\documentclass[12pt]{article}
\usepackage{epsfig, epsf, graphicx, subfigure}
\usepackage{pstricks, pst-node, psfrag}
\usepackage{amssymb,amsmath,bm}
\usepackage{verbatim,enumerate}
\usepackage{rotating, lscape}
\usepackage{setspace}
\usepackage[english]{babel}
\usepackage[sort&compress]{natbib}
\usepackage{multirow}
\usepackage{theorem}
\usepackage{authblk}

\usepackage{hyperref}
\hypersetup{
    colorlinks=true,
    citecolor=blue,
    linkcolor=red,
    urlcolor=blue,
    }

\def\d{\mathrm{d}}

\def\p{\partial}

\def\00{\mathrm{0}}
\def\UU{\mathbf{U}}

\def\XX{\mathbf{X}}
\def\xx{\mathbf{x}}
\def\ZZ{\mathbf{Z}}

\def\cc{\mathcal{c}}

\def\SS{\boldsymbol{\Sigma}}
\def\AA{\mathcal{A}}
\def\BB{\mathcal{B}}
\def\XA{\mathcal{X}}
\def\YA{\mathcal{Y}}

\def\II{\mathcal{I}}
\def\ww{\boldsymbol{w}}
\def\uu{\boldsymbol{u}}
\def\zz{\boldsymbol{z}}
\def\tht{\boldsymbol{\theta}}

\def\PP{\mathbf{P}}

\def\cc{\mathbf{c}}

\def\uu{\mathbf{u}}

\def\aa{\boldsymbol{\alpha}}

\def\SS{\boldsymbol{\Sigma}}

\def\ss{\boldsymbol{s}}
\def\gg{\boldsymbol{g}}

\newtheorem{prop}{Proposition}
\newtheorem{corol}{Corollary}

\begin{document}	
\title{\bf Parsimonious Factor Models for Asymmetric Dependence in Multivariate Extremes}
\thispagestyle{empty} \baselineskip=28pt \vskip 5mm

\author{
Pavel Krupskii\footnote{School of Mathematics and Statistics, University of Melbourne. \url{pavel.krupskiy@unimelb.edu.au}} \  
 and Boris B\'{e}ranger\footnote{School of Mathematics and Statistics, UNSW Sydney. \url{b.beranger@unsw.edu.au}}
}

\maketitle

\begin{abstract}

    Modelling multivariate extreme events is essential when extrapolating beyond the range of observed data. Parametric models that are suitable for real-world extremes must be flexible—particularly in their ability to capture asymmetric dependence structures—while also remaining parsimonious for interpretability and computationally scalable in high dimensions.
    Although many models have been proposed, it is rare for any single construction to satisfy all of these requirements. For instance, the popular H\"{u}sler-Reiss model \citep{Husler.Reiss1989} is limited to symmetric dependence structures.
    In this manuscript, we introduce a class of additive factor models and derive their extreme-value limits. This leads to a broad and tractable family of models characterised by a manageable number of parameters. These models naturally accommodate asymmetric tail dependence and allow for non-stationary behaviour.
    We present the limiting models from both the componentwise-maxima and Peaks-over-Thresholds perspectives, via the multivariate extreme value and multivariate generalized Pareto distributions, respectively. Simulation studies illustrate identifiability properties based on existing inference methodologies.
    Finally, applications to summer temperature maxima in Melbourne, Australia, and to weekly negative returns from four major UK banks demonstrate improved fit compared with the H\"{u}sler-Reiss model.
\end{abstract}

\section{Introduction}

The modelling of multivariate extreme events is a critical task in many disciplines --- including finance, insurance, hydrology, climate science, and engineering --- where a sound understanding of the probability of such events is highly consequential. Extreme Value Theory \citep{deHaan.Ferreira.2006, Resnick1987} provides a robust theoretical framework for addressing this class of problems. The most common approach is to specify a parametric model and use it to extrapolate beyond the range of the observed data in order to quantify the risk of extreme events.\\
While the theory for univariate random variables is well established, numerous challenges remain when extending these methods to two or more variables. The extremeness of a random vector can be defined via the componentwise maxima --- commonly referred to as the block maxima approach~--- or, alternatively, by considering any joint observation that exceeds a high threshold, known as the multivariate Peaks-over-Threshold approach. Both approaches are introduced in detail in Section~\ref{sec:background}, where their connections are also highlighted.\\
Since the marginal distributions can be readily modelled using univariate extreme value theory, we assume throughout this manuscript that the margins have been standardised, allowing us to concentrate solely on the extremal dependence structure. Although several mechanisms exist for constructing parametric models of extremal dependence, much of the literature focuses on the logistic \citep{Gumbel1960b} and H\"{u}sler–Reiss \citep{Husler.Reiss1989} models. The logistic model employs a single dependence parameter, making it too rigid for many real-world applications, while the H\"{u}sler–Reiss model cannot capture asymmetric dependence structures—in particular, situations where some components are more strongly dependent than others, as commonly observed in rainfall, temperature, and financial data. The extremal-$t$ \citep{Opitz2013} model uses elliptical distributions to extend the H\"{u}sler–Reiss model but possesses the same drawbacks. \citet{padoan2011} and \citet{beranger2019b} proposed asymmetric extensions of the H\"{u}sler–Reiss and extremal-$t$ models but do require some sum-to-zero conditions on the skewness parameters. Non-exhaustive overviews of existing models can be found in \citet{beirlant2004, Dey.Yan.2016, Naveau.Segers.2024}. The goal of this manuscript is to develop a new class of models capable of representing asymmetric extremal dependence structures. \\
Factor copula models \citep{Krupskii.Joe2013} provide a flexible framework for modelling data that exhibit tail dependence --- situations in which two or more variables can simultaneously take extreme values --- as well as tail asymmetry. To ensure computational scalability in high dimensions, structural assumptions on how the latent factors relate to the observed variables are often imposed, leading to more parsimonious model formulations \citep{Krupskii.Joe2015b}. When applied to finite-dimensional distributions, these constructions can also be used to model spatial data. \\
\citet{Krupskii.Huser.ea2016} introduced a common factor copula model tailored for spatial applications that is interpretable and scales well to high dimensions; however, it does not accommodate permutation asymmetry, a feature frequently observed in practice. \citet{Lee.Joe.2018} proposed factor copulas for multivariate extremes under a tree structure, and \citet{Krupskii.Genton2021} further extended this framework to conditional normal models. Yet, it remains unclear whether these approaches can adequately represent substantial permutation asymmetry. Moreover, they rely on numerical integration for likelihood evaluation, which becomes computationally demanding in high dimensions. More recently, \citet{Kiriliouk.Lee.ea.2025} adapted vine constructions to create highly flexible models for multivariate extremes. Despite their versatility, these models lack parsimony, and inference becomes prohibitive in very high-dimensional settings. \\
In this manuscript, we develop a broad class of additive factor copula models that extends the construction of \citet{Krupskii.Huser.ea2016}. We derive the corresponding limiting distributions and show that they permit permutation asymmetry between any pair of variables. We also identify parsimonious sub-models that are particularly well suited for capturing asymmetric tail dependence. The corresponding multivariate extreme value and multivariate generalized Pareto distributions are derived such that the modelling of block maxima and threshold exceedances is possible. The proposed class offers different types of extremal dependence structure for the variable of interests through models that can assume zero or non-zero mass at one or more of the boundaries of the support, meaning that joint extremes of a vector of interest may not always occur simultaneously \citep[see e.g.][]{Goix.Sabourin.ea2017, Simpson.Wadsworth.ea2020}. Furthermore, given that common factor ($V_0$, see Section~\ref{sec:factor_model}) affects all the variables and thus introduces extremal dependence in a small area under the influence of this factor, strengthens the interpretatbility of the proposed models.\\
The remainder of the paper is organised as follows. Section~\ref{sec:background} reviews essential background on multivariate extremes from the perspectives of Block Maxima and Peaks-over-Thresholds, and introduces the H\"{u}sler–Reiss model, which serves as our primary benchmark. Section~\ref{sec:factor_model} presents the main contributions of the paper: we define a class of additive factor models, derive their limiting distributions, and investigate their tail properties. Section~\ref{sec:factor_model} summarises inference methods for both componentwise maxima and threshold exceedances, and includes simulation studies demonstrating that accurate parameter estimation is achievable even in higher dimensions. Section~\ref{sec-empstudy} provides two empirical analyses that highlight the increased flexibility of the proposed models compared with the H\"{u}sler–Reiss model. Finally, Section~\ref{sec:conclusion} offers concluding remarks. All technical proofs are deferred to the Appendix~\ref{sec:appx}.

\section{Background}
\label{sec:background}

Let $\bm X_i = (X_{i,1}, \ldots, X_{i,d})^\top, i \in \left\{1, \ldots, n \right\}$ be $n$ independent and identically distributed (i.i.d.) replicates of a random vector $\bm X \in \mathbb{R}^d$ with continuous cumulative distribution function (cdf) $F$ (in short $\bm X \sim F$) with margins $F_1, \ldots, F_d$. By \citet{Sklar1959}'s representation theorem, $F$ can be represented as 
$F(\bm x) = C(F_1(x_1), \ldots, F_d(x_d))$, 
where $\bm x = (x_1, \ldots, x_d) \in \mathbb{R}^d$ and $C:[0,1]^d \rightarrow [0,1]$ is a copula. By continuity of the marginal distribution, the above can be re-written as
$$
C(\bm u) = F\left(F_1^{-1}(u_1), \ldots, F_d^{-1}(u_d)\right), \quad \bm u = (u_1, \ldots, u_d) \in [0,1]^d,
$$
highlighting that a copula is a restriction of a multivariate distribution with standard uniform margins to the unit hypercube.

\subsection{Extremes as Block Maxima}
\label{ssec:BM}

Let $\bm M_n = (M_{n, 1}, \ldots, M_{n,d})^\top$ denote the vector component-wise maxima, where $M_{n,j} = \max_{i=1, \ldots, n} X_{i,j}$. Assume there exist sequences of normalizing constants $\bm a_n = (a_{n, 1}, \ldots, a_{n,d}) \in \mathbb{R}^d_+$ and $\bm b_n = (b_{n, 1}, \ldots, b_{n,d}) \in \mathbb{R}^d$ such that, as $n \rightarrow \infty$, the location-scale standardized vector of maxima, $\bm a_n^{-1}(\bm M_n - \bm b_n)$ converges in distribution to a random vector $\bm Y$ with joint distribution $G$ and non-degenerate margins, $G_j, j=1, \ldots, d$. Let $\bm \mu \in \mathbb{R}^d, \bm \sigma \in \mathbb{R}^d_+ \backslash \{ \bm 0 \}$ and $\bm \xi \in \mathbb{R}^d$ denote the vectors of marginal location, scale and shape parameters. We then say that the random vector $\bm X$ lies in the maximum domain of attraction (MDA) of $\bm Y$. The limiting distribution $G$ is a multivariate extreme value distribution whose margins belong to the family of generalized extreme value (GEV) distributions \citep[see, e.g.,][Chap.~6]{deHaan.Ferreira.2006}. Consequently, the marginal distributions of $\bm X$ are in the MDA of the GEV family, which encompasses the Weibull, Gumbel, and Fr\'echet classes. The limiting distribution of the rescaled component-wise maxima can be expressed as
\begin{align}
\label{eq:G_copula}
G(\bm x) = C_{\textrm{EV}}\left( G_1(x_1), \ldots, G_d(x_d) \right), \quad \bm x \in \mathbb{R}^d
\end{align}
where $C_{\textrm{EV}}$ is the limiting copula defined as
\begin{align}
\label{eq:ev_copula}
C_{\textrm{EV}}(\bm u) = \lim_{n \rightarrow \infty} C\left( u_1^{1/n}, \ldots, u_d^{1/n}\right)^n, 
\quad \bm u = (u_1, \ldots, u_d) \in [0,1]^d,
\end{align}
and called an extreme value (EV) copula \citep[][Chap.~8]{beirlant2004}. The copula $C$ is said to be in the domain of attraction of $C_{\textrm{EV}}$. Extreme value copulas possess the max-stability property meaning that for any integer $k \geq 1$, $C_{\textrm{EV}}\left( u_1^{1/k}, \ldots, u_d^{1/k}\right)^k = C_{\textrm{EV}}\left( u_1, \ldots, u_d\right)$ \citep[see e.g.,][]{Gudendorf.Segers2010b}.
When studying multivariate extremes from the perspective of copulas, it is convenient to characterise the EV copula through the stable tail dependence function $\ell:\mathbb{R}^d_+ \rightarrow \mathbb{R}_+$, defined as
\begin{align}
\label{eq:tail_dep_fun}
\lim_{n \rightarrow \infty} n \left\{ 1 - C\left( 1 - \frac{x_1}{n}, \ldots, 1 - \frac{x_d}{n} \right)\right\}
= -\log C_{\textrm{EV}}\left( e^{-x_1}, \ldots, e^{-x_d}\right) 
\equiv \ell(\bm x), 
\end{align}
where $\bm x = (x_1, \ldots, x_d) \in \mathbb{R}^d_+$ \citep{huang1992, drees1998}. It can be seen that \eqref{eq:tail_dep_fun} is equivalent to \eqref{eq:ev_copula} after taking logarithms and applying a linear expansion. This yields the representation
\begin{align}
\label{eq:Gstdf}
G(\bm x) = \exp \left\{ - \ell\left( -\log G_1(x_1), \ldots, - \log G_d(x_d) \right) \right\}. 
\end{align}
The tail dependence function is homogeneous and convex, and satisfies
$$\max(\bm x) \leq \ell(\bm x) \leq \sum_{i=1}^d x_i,$$ with the lower and upper bounds corresponding respectively to complete dependence (i.e., $G(\bm x) = \min \left\{ G_1(x_1), \ldots, G_d(x_d) \right\}$) and asymptotic independence (i.e., $G(\bm x) = \prod_{i=1}^d G_i(x_i)$). At the copula level, this is equivalent to $$\prod_{i=1}^d u_i \leq C_{\textrm{EV}}(\bm u) \leq \max(u_1, \ldots, u_d).$$ By homogeneity, it is sufficient to restrict attention to the unit simplex $$\mathcal{S} = \left\{ \bm w \in [0,1]^d: \|\bm w \|_1 = 1\right\},$$ on which the Pickands dependence function $\textrm{A}: \mathcal{S} \rightarrow [1/d, 1]$ is defined. The tail dependence function then admits the representation
\begin{align*}
\ell(\bm x) = \sum_{i=1}^d x_i \textrm{A}(\bm w), \quad w_i = \frac{x_i}{\|\bm x \|_1}, \bm x \in \mathbb{R}^d_+ \backslash\{ \bm 0\}.
\end{align*}
By properties of the tail dependence function $\ell$, the Pickands function A is also homogeneous and convex, and satisfies  $$\max(w_1, \ldots, w_d) \leq A(\bm w) \leq 1, \quad \bm w \in \mathcal{S}.$$ Extremal dependence can also be quantified through the coefficient of upper tail dependence \citep{Li2009, Joe1997}, defined by
$$
\chi_{\textrm{U}} = \lim_{u \rightarrow 0} \Pr \left\{ X_i > F_i^{-1}(1-u), \forall i\neq j | X_j > F_j^{-1}(1-u) \right\}
= \lim_{u \rightarrow 0} \frac{\bar{C}(1-u, \ldots, 1-u)}{u},
$$
where $\bar{C}$ denotes the survival copula. This quantity links directly to the tail dependence and Pickands functions via $$\chi_{\textrm{U}} = d - \ell(1, \ldots, 1) = d \left\{ 1-A(1/d, \ldots, 1/d)\right\}.$$ Independence in the tails corresponds to $\chi_{\textrm{U}} = 0$, whereas complete dependence occurs when $\chi_{\textrm{U}} = d-1$. The quantity $\vartheta =  \ell(1, \ldots, 1) = d A(1/d, \ldots, 1/d)$ is known as the extremal coefficient, and ranges between 1 (complete dependence) and $d$ (asymptotic independence).

\subsection{Extremes as Peaks-over-Thresholds}
\label{ssec:POT}

\noindent Another approach to modelling multivariate extremes characterises extreme events through threshold exceedances rather than through componentwise maxima. More precisely, assume the convergence in distribution of $\bm a_n^{-1}(\bm M_n - \bm b_n)$ to $\bm Y$ as introduced above.  Using the notation $\bm X \nleq \bm b_n$ to indicate that $\bm X \leq \bm b_n$ does not hold --- i.e., there exists at least one $j \in \II = \{1, \ldots, d\}$ such that $X_j > b_{n,j}$ --- we have
$$
\frac{\bm X - \bm b_n}{\bm a_n} \bigvee \bm \eta  \Big| \bm X \nleq \bm b_n \overset{d}{\rightarrow} \tilde{\bm Y}, \quad \textrm{as } n \rightarrow \infty,
$$
where $\tilde{\bm Y}$ follows a multivariate Generalized Pareto distribution, $\vee$ denotes the componentwise maximum operator, and $\bm \eta$ is the vector of lower endpoints of the marginal distributions $G_1, \ldots, G_d$ \citep{Rootzen.etal:2018a}. Let $H$ denote the distribution function of the multivariate Generalized Pareto distribution associated with $G$, defined through
$$
H(\bm x) = \frac{1}{\log G(\bm 0)} \log \left\{ \frac{G(\bm x \wedge \bm 0)}{G(\bm x)}\right\}, \quad 0 < G(\bm 0) < 1,
$$
for $\bm x > \bm \eta$; otherwise $H(\bm x) = 0$ whenever $x_i < \eta_i$ for at least one $i \in \II$ \citep{RootzenTajvidi2006}. Here, $\wedge$ denotes the componentwise minimum operator. Using the representation \eqref{eq:Gstdf}, we obtain
$$
H(\bm x) = \ell\left[ \bm \pi \left\{ 1 + \frac{\bm \xi}{\tilde{\bm \sigma}} \left( \bm x \wedge \bm 0\right) \right\}^{-1/\bm \xi}\right]
- \ell\left\{ \bm \pi \left( 1 + \frac{\bm \xi}{\tilde{\bm \sigma}} \bm x \right)^{-1/\bm \xi}\right\},
$$
where $\tilde{\bm \sigma} = \bm \sigma - \bm \xi \bm \mu $, $\bm \pi = \bm \tau / \ell(\bm \tau)$, and $\tau_j = \left( 1+ \xi_j \mu_j / \sigma_j\right)^{-1/\xi_j}, j \in \II$ \citep[Prop.~2][]{Rootzen.etal:2018b}.
When $\bm \xi = \bm 0$, this representation simplifies to
$$
H(\bm x) = \ell\left\{ \bm \pi \exp \left( - \frac{\bm x \wedge \bm 0}{\bm \sigma}\right)\right\} - 
\ell\left\{ \bm \pi \exp \left( - \frac{\bm x}{\bm \sigma}\right)\right\},
$$
with $\tau_j = \exp \{ \mu_j / \sigma_j\}$.
By construction, each $j$-th conditional marginal $X_j - u_j | X_j > u_j$ follows a univariate Generalized Pareto distribution with parameters $\tilde{\sigma}_j, \xi_j$. A transformation to unit Pareto margins then yields
$H(\bm x) = \{\ell\left( \bm x^{-1} \wedge 1\right) - \ell(\bm x^{-1})\}\ell(\bm 1)^{-1}$.
These results imply that, for a sufficiently high threshold $\bm u$, the distribution of threshold exceedances $\bm X - \bm u | \bm X \nleq \bm u$ can be approximated by a multivariate Generalized Pareto distribution, which is fully characterised by the stable tail dependence function $\ell$. Since the support of $H$ is contained in the set of all $\bm x$ such that $x_j > \eta_j$ for all $j$ and $x_j >0$ for at least one $j$, the density of $\tilde{\bm Y}$ is given by 
$$
h(\bm x) = -\frac{\partial^d}{\partial x_1 \cdots \partial x_d} \ell\left\{ \bm \pi \left( 1 + \frac{\bm \gamma}{\tilde{\bm \sigma}} \bm x \right)^{-1/\bm \gamma}\right\}, \quad \bm x \in (\bm \eta, \bm \infty).
$$ 
The distribution $H$ may also place mass on lower-dimensional faces of $(\bm \eta, \bm \infty)$. For instance, the probability mass assigned to the hyperplane $\{ \bm x: x_j = \eta_j \}, j \in \mathcal{I}$ is 
\begin{align}
\label{eq:mass}
H_j(\{ \eta_j \}) = \lim_{x_j \rightarrow \infty} \ell\left( \pi_1, \ldots, \pi_{j-1}, \pi_j x_j, \pi_{j+1}, \ldots, \pi_d \right) - \pi_j x_j = \sum_{k \in \mathcal{I}_{-j}} \pi_k \frac{\p \ell}{\p x_k} (\pi_j \bm e_j),
\end{align}
where $\mathcal{I}_{-j} = \mathcal{I} \backslash \{j\}$ and $\p \ell / \p x_k$ denotes the first-order derivative of $\ell$ with respect to the $k$-th component and $\bm e_j$ is the $j$-th $d$-dimensional canonical vector \citep[see][Proposition~17]{Rootzen.etal:2018b}. \\
In the case of unit Pareto margins, $\pi_j = 1 / \ell(\bm 1)$ and \eqref{eq:mass} simplifies to 
\begin{align*}
H_j(\{ \eta_j \}) &= \lim_{x_j \rightarrow \infty} \left\{\ell\left( 1, \ldots, 1, x_j, 1, \ldots, 1 \right) - x_j\right\} \ell(\bm 1)^{-1} 
= \lim_{v \rightarrow 0} \frac{\partial^{d-1}}{\partial \bm x_{-j}}\ell \left( v, \ldots, v, 1, v, \ldots, v \right) \ell(\bm 1)^{-1},
\end{align*}
where the second equality follows from the homogeneity property of $\ell$. For a comprehensive treatment of the multivariate Peaks-over-Thresholds approach, the reader is referred to \citet{Naveau.Segers.2024}.

\subsection{The H\"{u}sler-Reiss distribution, \citet{Husler.Reiss1989}}
\label{ssec:HR}

\begin{figure}[t]
    \centering
    \includegraphics[width=0.32\linewidth]{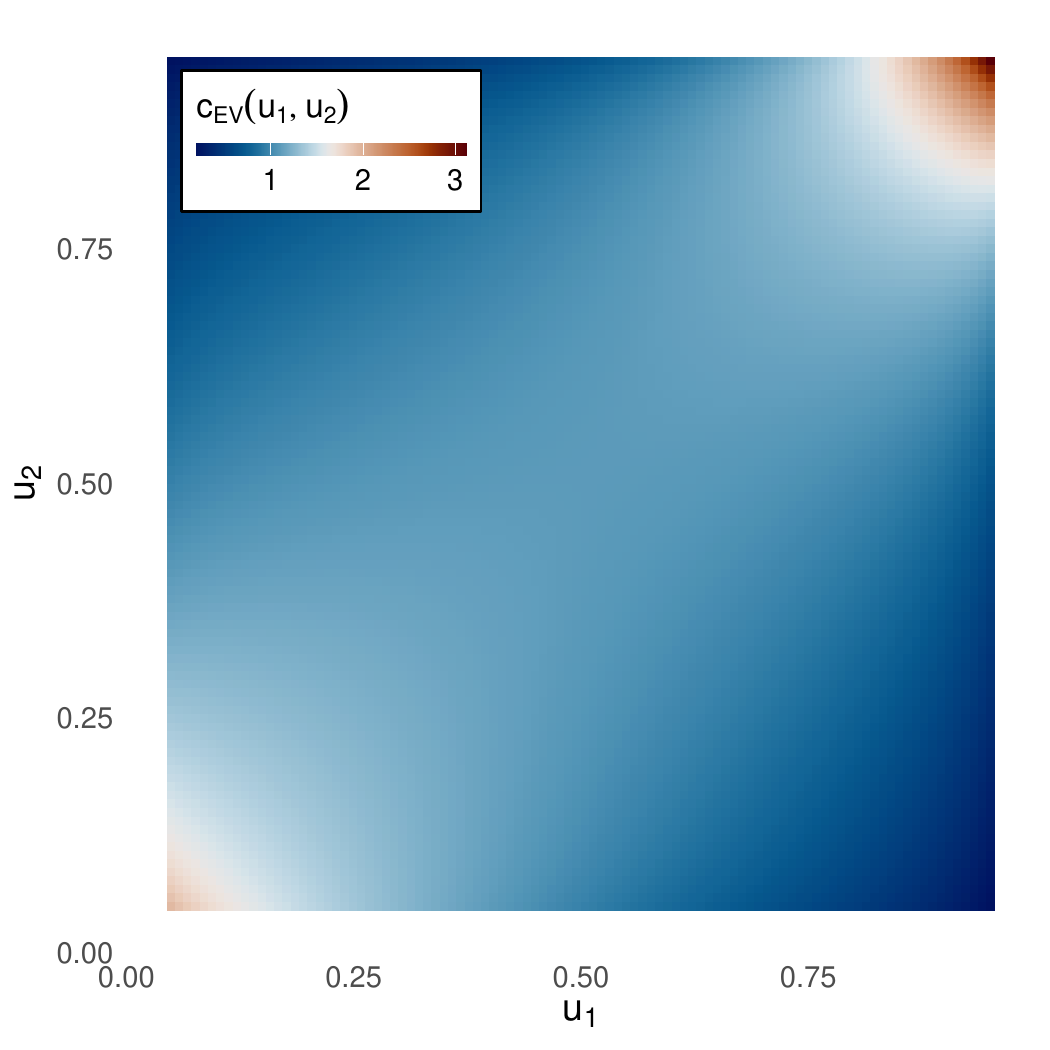}
    \includegraphics[width=0.32\linewidth]{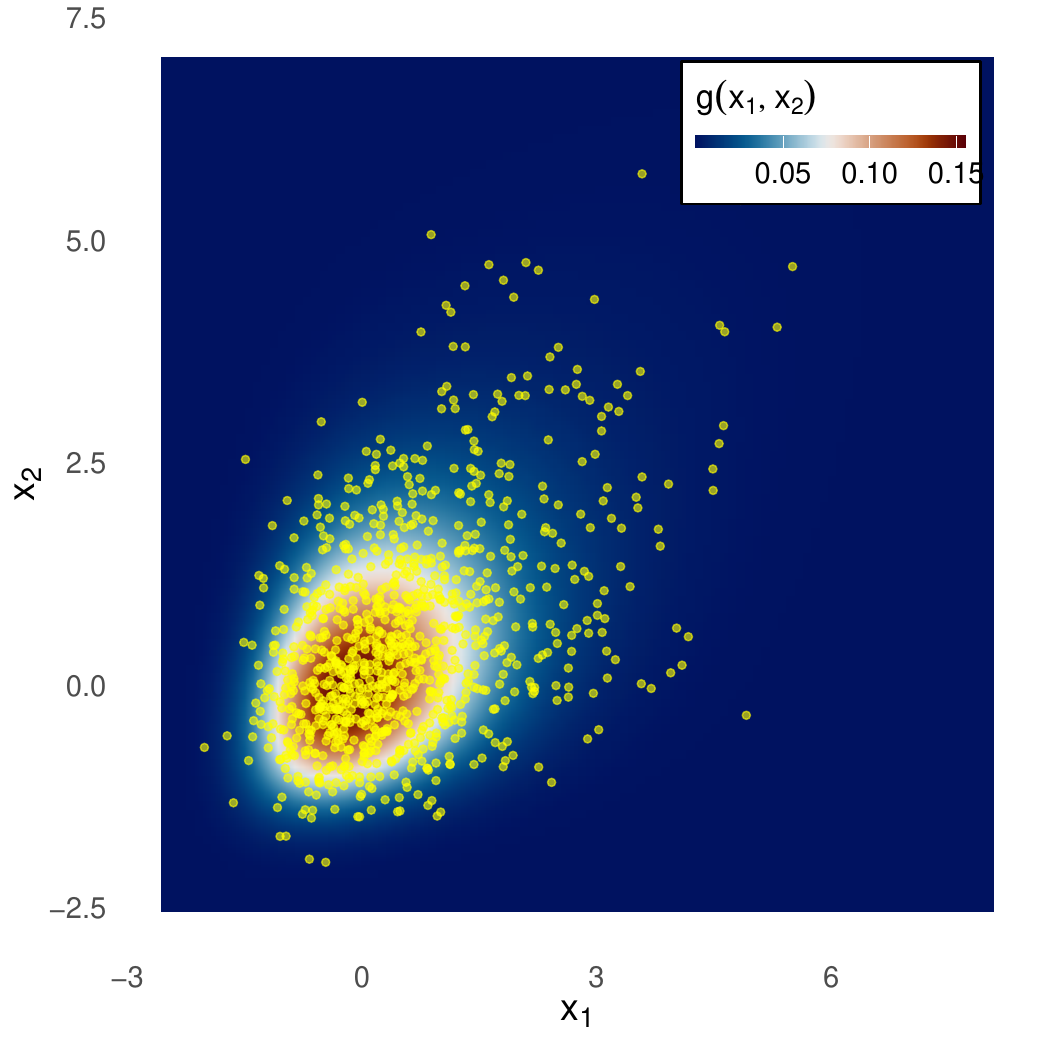}
    \includegraphics[width=0.32\linewidth]{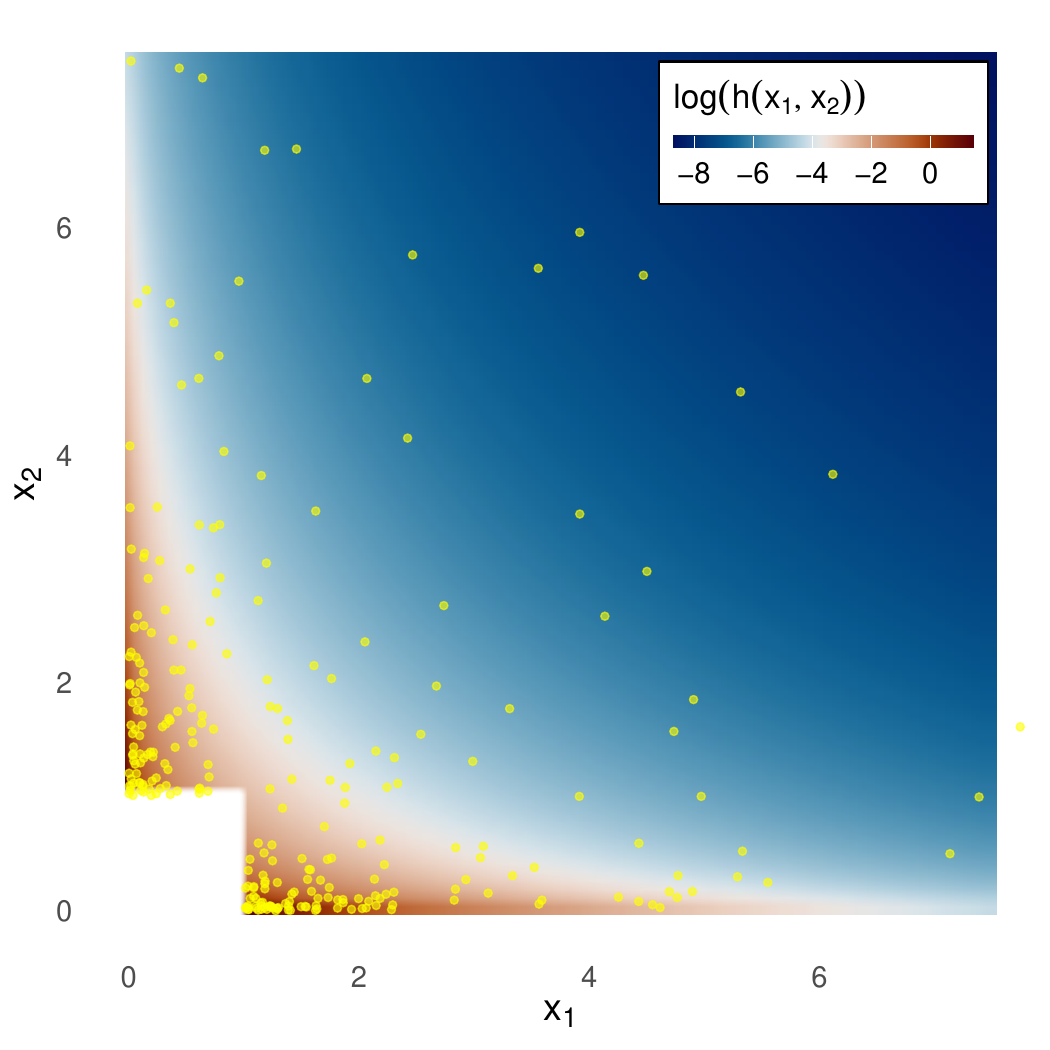} \\
    \includegraphics[width=0.32\linewidth]{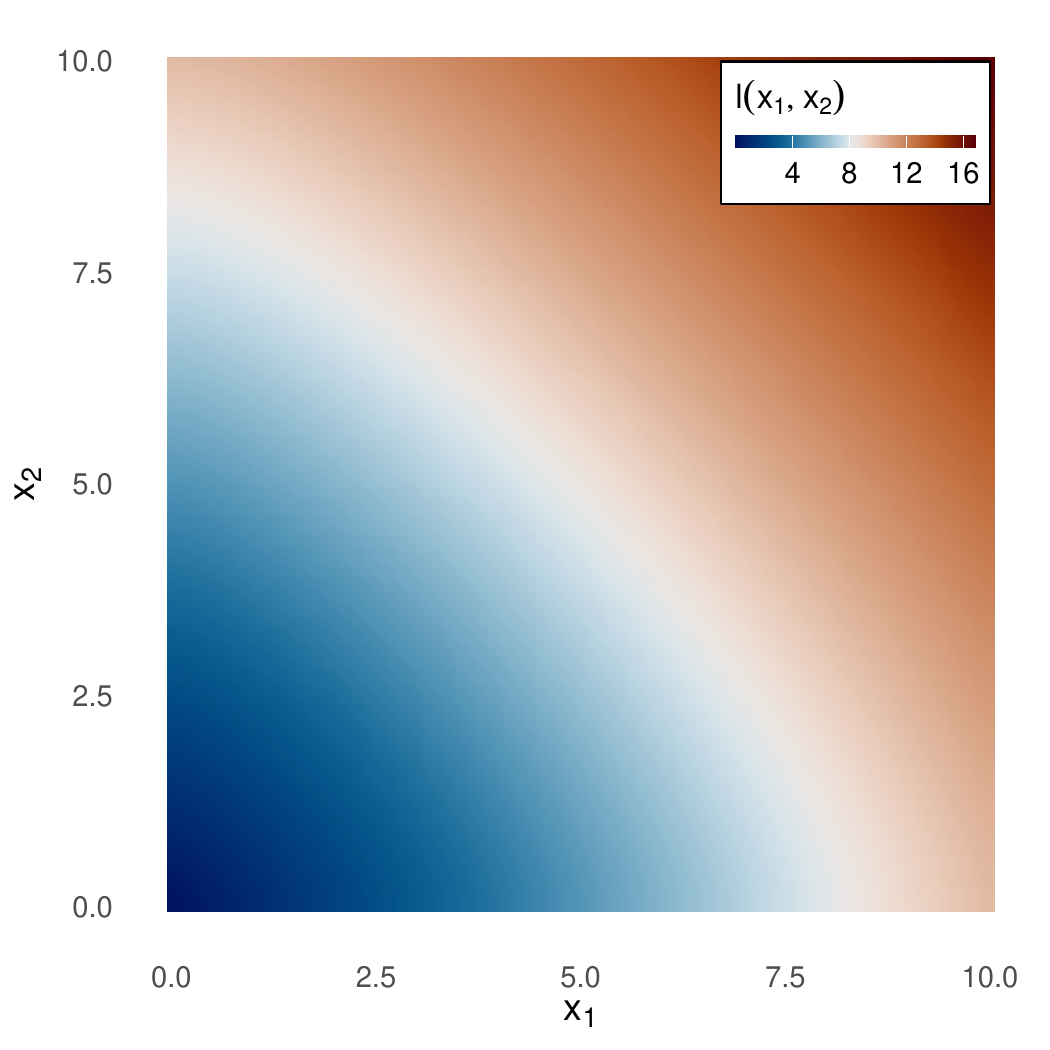}
    \includegraphics[width=0.32\linewidth]{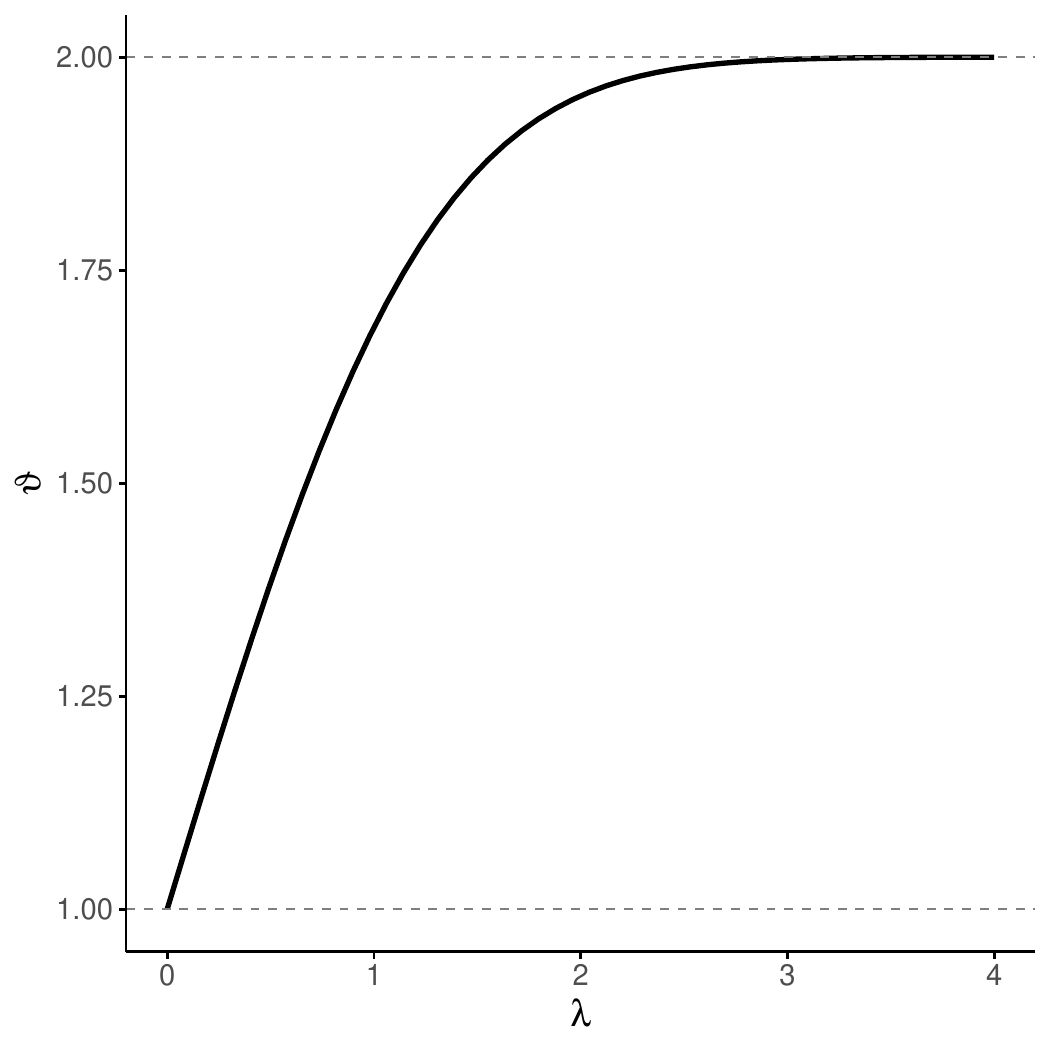}
    \includegraphics[width=0.32\linewidth]{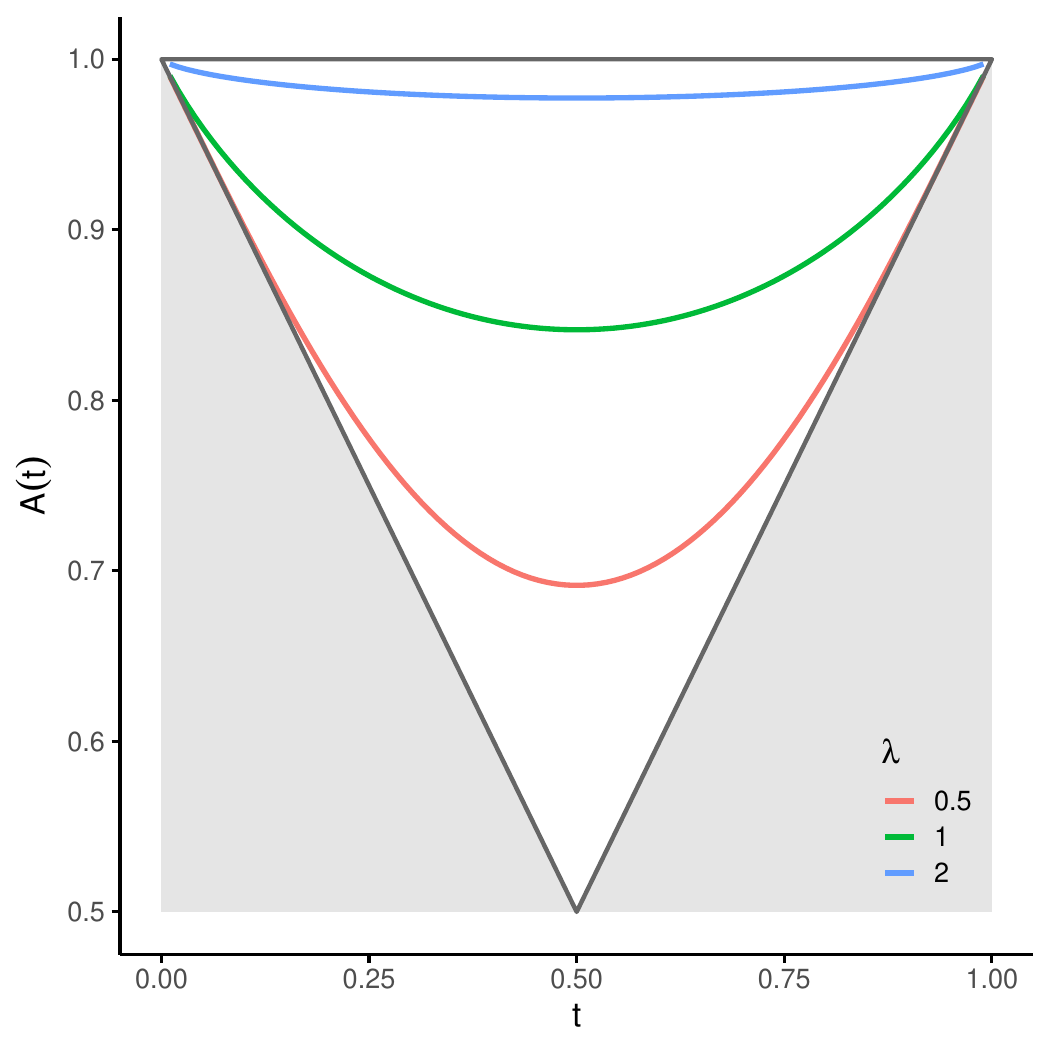}
    \caption{Left to right, top to bottom: density of the extreme value copula $c_{EV}$, density of the bivariate extreme value distribution $g(\bm x)$, density of the bivariate Generalized Pareto distribution with unit Pareto margins (on log scale, $\log h(\bm x)$) and stable tail dependence function $\ell(\bm x)$ of the H\"{u}sler-Reiss model with $\lambda = 1$. The remaining panels display the extremal coefficient $\vartheta$ as function of $\lambda$ and the Pickands dependence function for $\lambda = 0.5, 1$ and $2$.}
    \label{fig:summryHR}
\end{figure}

Let $\bm X$ be a zero-mean, unit-variance Gaussian random vector with correlations satisfying $(1-\rho_{ij}(n)) \log n \rightarrow \lambda_{ij}^2$ as $n \rightarrow \infty$. The limiting marginal distributions are $$G_j(x) = \exp\left\{ - \exp \{-x\}\right\}, \quad x \in \mathbb{R}, \quad j \in \II$$ and the corresponding EV copula is given by
$$
C_{\textrm{EV}}(\bm u) = \exp \left( \sum_{j=1}^d \log u_j \Phi_{d-1} \left[ \left\{ \lambda_{ij} + \log \left( \frac{\log u_j}{\log u_i} \right) / \left(2\lambda_{ij} \right)\right\}_{i \in \mathcal{I}_{-j}}; \bar{\Sigma}^{(j)}\right] \right), \quad \bm u \in [0,1]^d,
$$
where $\Phi_d(\cdot; \Sigma)$ denotes the cumulative distribution function (cdf) of a 
$d$-dimensional normal distribution with covariance (or correlation) matrix $\Sigma$
and $\bar{\Sigma}^{(j)}$ is the partial correlation matrix with entries $(\bar{\Sigma}^{(j)})_{i,k} = \left( \lambda_{k,j}^2 + \lambda_{i,j}^2 - \lambda_{k,i}^2 \right) / \left( 2 \lambda_{k,j} \lambda_{i,j}\right), \ i \neq k$. The stable tail dependence function is then
$$
\ell(\bm x) =\sum_{j=1}^d x_j \Phi_{d-1} \left[ \left\{ \lambda_{ij} + \log \left(x_j/x_i \right) / \left( 2\lambda_{ij} \right)\right\}_{i \in \mathcal{I}_{-j}}; \bar{\Sigma}^{(j)}\right], \quad \bm x \in \mathbb{R}^d_+,
$$
from which the multivariate extreme value distribution $G(\bm x)$ and the extremal coefficient $\vartheta = \sum_{j=1}^d \Phi_{d-1} \left( \left( \lambda_{ij} \right)_{i \in \mathcal{I}_{-j}}; \bar{\Sigma}^{(j)}\right)$ are obtained. In particular, $\lambda_{i,j} \rightarrow 0$  for all $i,j \in \mathcal{I}, i \neq j$  implies $\vartheta \rightarrow 1$ (complete dependence), whereas $\lambda_{ij} \rightarrow \infty$ implies $\vartheta \rightarrow d$ (independence).\\
Assuming unit Pareto marginals, the density of the multivariate Generalized Pareto distribution is given by \citep{EngelkeEtAl2015}
$$
h(\bm x) = \phi_{d-1} \left[ \left\{ \lambda_{ik} + \log(x_i/x_k)/(2\lambda_{ik}) \right\}_{i \in \mathcal{I}_{-k}}; \bar{\Sigma}^{(k)} \right] \left\{ \ell(\bm 1) x_k^2 \prod_{i \in \mathcal{I}_{-k}} x_i 2\lambda_{ik}\right\}^{-1},
$$
where $\phi_{d}(\cdot; \Sigma)$ denotes the probability density function (pdf) of a $d$-dimensional normal distribution with covariance (or correlation) matrix $\Sigma$.
Interestingly, this expression does not depend on $k$. \\
In the bivariate case ($d=2$), it can be shown that $$\dot{\ell}_k(x_1, x_2) = \Phi \left\{ \lambda + \log\left( \frac{x_k}{x_j}\right) / (2\lambda)\right\}, \quad j = \mathcal{I}_k.$$ Since the unit Pareto margins imply $\eta_1 = \eta_2 = 0$, from \eqref{eq:mass} we have $H_1(\{0\}) = H_2(\{0\}) = 0$, meaning that, for the H\"{u}sler-Reiss distribution, $H$ does not place mass on the boundaries of its support.\\
Figure~\ref{fig:summryHR} illustrates various aspects of the bivariate H\"{u}sler–Reiss distribution. In particular, the top and bottom left panels show the density of the EV copula $c_{EV}(\bm u) = \partial^d / \partial \bm u C_{EV}(\bm u)$,  and the stable tail dependence function $\ell(\bm x)$ for $\lambda =1$. For the same parameter setting, the top middle panel displays the density of the multivariate EV distribution $g(\bm x)$, with Gumbel margins, with yellow dots representing 200 random draws from the component-wise maxima. \\
Considering $200,000$ observations drawn from a bivariate normal distribution whose margins have been standardised to unit Pareto scale, and setting a threshold $\bm u$ at the $0.999$ marginal quantiles, the top right panel shows approximately $300$ observations of $\bm X / \bm u$ with at least one component above $1$ (yellow dots), together with the log-density of the multivariate Generalized Pareto distribution, $\log H(x_1, x_2)$.\\
Graphical summaries of the dependence are provided in the bottom middle and right panels of Figure~\ref{fig:summryHR}, showing the extremal coefficient $\vartheta$ as a function of $\lambda$ and the Pickands dependence function for $\lambda = 0.5, 1$ and $2$. These plots clearly illustrate that the level of dependence decreases as $\lambda$ increases. Note that the extremal coefficient, Pickands dependence function, and extreme value density $g$ were computed using the \texttt{R} package \citep{ExtremalDep2025, Beranger.Padoan2025}.

\section{A broad class of additive factor models for modelling extremes}
\label{sec:factor_model}

Consider the $d$-dimensional random vectors $\ZZ = (Z_1, \ldots, Z_d)^{\top}$ and $\ZZ^* = (Z_1^*, \ldots, Z_d^*)^{\top}$ defined such that $\{\ZZ^{\top}, (\ZZ^*)^{\top}\}^{\top}$ follows a multivariate normal distribution with zero mean, unit variance, and correlation matrix $\boldsymbol{\mathrm{P}}$. We construct the random vector $\XX = (X_1, \ldots, X_d)^{\top}$ by defining each marginal random variable $X_i$ through the following additive factor models
\begin{equation}
    \label{main-model}
    X_i = Z_i + (1/\alpha_i)\cdot V_0 \cdot \mathbf{1}(Z_i^* > c_i), \quad i \in \mathcal{I},
\end{equation}
where $\bm \alpha = (\alpha_1, \ldots, \alpha_d)^{\top} \in \mathbb{R}^d_+$, the common factor $V_0$ is introduced to model upper tail dependence and $\cc = (c_1, \ldots, c_d)^{\top}$ is a vector of real-valued truncation parameters. If $\cc = -\infty$ (no truncation), then the model \eqref{main-model} simplifies to the factor copula model for spatial data \citep{Krupskii.Huser.ea2016}. For such model, considering an exponential factor $V_0$ yields the H\"{u}sler-Reiss distribution for which the non-trivial EV limiting distribution can be obtained. However, the associated bivariate marginal distributions are permutation symmetric which can be a restrictive assumption in many applications. We now study the EV limiting distributions for the model \eqref{main-model}, and show that this model can generate more flexible dependence structures including permutation asymmetry.

\subsection{Extreme-value limiting distributions}
\label{ssec:EVlimits}

In this section, we consider the additive factor model defined in \eqref{main-model} and examine its limiting distributions under various assumptions for the common factor $V_0$ and under different dependence or independence regimes between $\ZZ$ and $\ZZ^*$.

\begin{prop}
    \label{prop0}
    Consider the random vector $\XX = (X_1, \ldots, X_d)^{\top}$ defined in \eqref{main-model}. If the common factor $V_0$ follows a continuous distribution with cdf $F_0$ such that for any constants $M \in \mathbb{R}$, $ k > 0$ and some $\alpha_0 \in (0, 0.5)$, $\gamma < 0$ we have
    $$ F_0\left\{F_0^{-1}(1-k/n) + M(\ln n)^{\alpha_0} \right\} = 1 - k/n + o(1/n), \quad n \to \infty,$$ 
    and
    $$ F_0\left\{F_0^{-1}(1-k/n) + y \right\} \geq  1 - (k/n)\cdot \exp(\gamma y), \quad \text{ if } \quad   -|M| \ln n < y < -|M|(\ln n)^{\alpha_0}, \quad n \to \infty,$$ 
    then the stable tail dependence function associated with $\XX$ is given by
    $$\ell_{\XX}(\xx) = \sum_{\AA \subset \mathcal{I}, \BB = \mathcal{I}_{-\AA}} \left(\max_{i \in \AA}\frac{x_i}{\zeta_i}\right) \cdot\Pr(\ZZ^*_{\AA} > \cc_{\AA}, \ZZ^*_{\BB} < \cc_{\BB})
    $$
    where we recall that $\mathcal{I}_{-\AA} = \mathcal{I} \backslash \{ \AA \}$, the indices $\AA$ and $\BB$ indicate the corresponding subvectors, and $\zeta_i = \Phi(-c_i), i \in \mathcal{I}$.
\end{prop}

\noindent \textbf{Proof:} See Appendix \ref{appx-prop0}. \hfill $\Box$

\begin{corol}
\label{corol00}
The assumptions on the cdf of the common factor $V_0$ stated above in Proposition~\ref{prop0} are fulfilled by the Pareto distribution with $F_0(x) = 1 - x^{-\beta}$, $\beta > 0$ and $x > 1$, and the Weibull distribution with $F_0(x) = 1 - \exp(-x^{\beta})$, $\beta \in (0,1)$ and $x > 0$.
\end{corol}

\noindent \textbf{Proof}: Let $|M|(\ln n)^{\alpha_0} < |y| < |M|\ln n$, where $\alpha_0 \in (0, 0.5)$. For the Pareto distribution, $F_0^{-1}(1-k/n) = (n/k)^{1/\beta}$ and
\begin{align*}
F_0\left\{F_0^{-1}(1-k/n) + y \right\} &= 1 - \frac{k}{n} \cdot \left\{1 + y \left(\frac{k}{n}\right)^{1/\beta}\right\}^{-\beta} = 1 - \frac{k}{n} + \beta y \cdot \left(\frac{k}{n}\right)^{1+1/\beta} + o(n^{-1-1/\beta})\\
&= 1 - \frac{k}{n} + o(1/n).
\end{align*}
The second condition holds for any $\gamma < 0$ provided that $n$ is large enough:
$$
1 - \frac{k}{n}\exp(\gamma y) \leq 1 - \frac{k}{n} - \frac{\gamma k}{n} y \leq 1 - \frac{k}{n} + \frac{k \gamma}{n}\,.
$$

For the Weibull distribution, $F_0(1-k/n) =\{ \ln (n/k) \}^{1/\beta}$ and
\begin{align*}
F_0\left\{F_0^{-1}(1-k/n) + y \right\} &= 1 - \exp\left[\left(\ln\frac{k}{n}\right)\cdot\left\{1 + y\left(\ln \frac{n}{k}\right)^{-1/\beta}\right\}^{\beta} \right]\\
&= 1 - \frac{k}{n}\exp\left[- \beta y \left(\ln \frac{n}{k}\right)^{1-1/\beta} + o\left\{y \left(\ln \frac{n}{k}\right)^{1-1/\beta}\right\}\right]\,.
\end{align*}
If $y = M(\ln n)^{\alpha_0}$, then $$y \left(\ln \frac{n}{k}\right)^{1-1/\beta} = o(1) \quad \text{and} \quad F_0\left\{F_0^{-1}(1-k/n) + y \right\} = 1 - \frac{k}{n} + o(1/n) \quad \text{as} \quad n \to \infty,$$ provided that $\alpha_0 < 1/\beta - 1$. On the other hand, if $-|M|(\ln n)^{\alpha_0} < y < -|M|\ln n$, then
$$
F_0\left\{F_0^{-1}(1-k/n) + y \right\} \geq 
1 - \frac{k}{n}\exp(-\beta y) \geq 1 - \frac{k}{n}\exp(-\gamma y),$$
provided that $\beta < \gamma$ and $n$ is large enough. \hfill $\Box$

\begin{corol}
\label{corol0}
Let $F_i$ be the cdf of $X_i$ as defined in \eqref{main-model}, then
\begin{enumerate}[i)]
    \item If $V_0 \sim \textrm{Pareto}(\beta)$, $\beta > 0$ then $F_i$ belongs to the MDA of the Fr\'echet distribution with shape $\beta$.
    \item If $V_0 \sim \textrm{Weibull}(\beta)$, $\beta \in (0,1)$, then $F_i$ belongs to the MDA of the Gumbel distribution.
\end{enumerate}
%
\end{corol}

\noindent \textbf{Proof}: Consider $X_{i1}, \ldots, X_{in} \sim_{\text{i.i.d.}} F_i$ and from the proof of Proposition \ref{prop0}, recall that
\begin{equation}
\label{eq-marg-asymp}
\Pr\left\{X_i < \alpha_i^{-1}F_0^{-1}\left(1- \frac{x_i}{n\zeta_i}\right)\right\} = 1 - \frac{x_i}{n} + o(1/n).
\end{equation}

\begin{enumerate}[i)]
    \item With the Pareto cdf $F_0(x) = 1 - x^{-\beta}$, \eqref{eq-marg-asymp} can be rewritten as
    $$
    \Pr(X_i < x) = 1 - \frac{\zeta_i}{(\alpha_i x)^{\beta}} +  o\left(x^{-\beta}\right)\,,
    $$
    and setting $a_n = n^{1/\beta} \zeta_i^{1/\beta}/\alpha_i$ yields
    $$
    \Pr\left(\max_{j=1,\ldots,n} X_{ij}/a_n < x\right) =\left(1 - \frac{1}{x^{\beta}n} \right)^n + o(1).
    $$
    Consequently $\lim_{n\to\infty} \Pr(\max_{j=1,\ldots,n} X_{ij}/a_n < x) = \exp(-x^{-\beta})$ and the distribution of $X_i$ belongs to the MDA of the Fr\'echet distribution. 
    \item With the Weibull cdf $F_0(x) = 1 - \exp(-x^{\beta})$, \eqref{eq-marg-asymp} can be rewritten as
    $$
    \Pr(X_i < x) = 1 - \zeta_i \exp\left\{-(\alpha_i x)^{\beta}\right\} + o\left[\exp\left\{-(\alpha_i x)^{\beta}\right\}\right],
    $$
    and setting $a_n = (\alpha_i \beta)^{-1} \left\{ \ln (\zeta_i n)\right\}^{1/\beta - 1}$ and $b_n = \alpha_i^{-1} \left\{\ln (\zeta_i n)\right\}^{1/\beta}$ yields
    \begin{align*}
    \Pr\left\{\max_{j=1,\ldots,n} (X_{ij} - b_n)/a_n < x\right\} &=\left\{1 - \left(\frac{1}{n}\right)^{\left( 1 + x a_n/b_n\right )^{\beta}} \right\}^n   + o(1)\\ &= \left(1 - \frac{1}{n} \exp\left[-x \cdot \frac{\ln n}{\ln(\zeta_i n)} + o\left\{\frac{\ln n}{\ln(\zeta_i n)}\right\}\right]\right)^n + o(1).
    \end{align*}
    Consequently, $\lim_{n\to\infty} \Pr\left\{\max_{j=1,\ldots,n} (X_{ij}-b_n)/a_n < x\right\} = \exp\{-\exp(-x)\}$ and the distribution of $X_i$ belongs to the MDA of the Gumbel distribution. \hfill $\Box$    
\end{enumerate}

The following result extends Proposition \ref{prop0} to the case where $V_0$ is exponentially distributed, under the assumption that $\ZZ$ and $\ZZ^*$ are independent.

\begin{prop}
    \label{prop1a}
    Consider the random vector $\XX = (X_1, \ldots, X_d)^{\top}$ defined in \eqref{main-model}. If $V_0$ follows a unit exponential distribution and the vectors $\ZZ$ and $\ZZ^*$ are independent, then the stable tail dependence function associated with $\XX$ is given by
    $$\ell_{\XX}(\xx) = \sum_{\AA \subset \mathcal{I}, \BB = \mathcal{I}_{-\AA}} \ell_{\AA}(\tilde\xx_{\AA}) \cdot \Pr(\ZZ^*_{\AA} > \cc_{\AA}, \ZZ^*_{\BB} < \cc_{\BB})
    $$
    where $\tilde x_i = x_i/\zeta_i$ and $\zeta_i = \Phi(-c_i), i \in \mathcal{I}$, and $\ell_{\AA}$ is the stable tail dependence function of the lower-dimensional H\"usler-Reiss distribution which can be obtained as an EV limiting distribution of the vector $\XX_{\AA} = \ZZ_{\AA} + (1/\aa_{\AA}) \cdot V_0$; see \cite{Krupskii.Joe.ea2018}. 
\end{prop}

\noindent \textbf{Proof:} See Appendix \ref{appx-prop1a}. \hfill $\Box$

A generalization of the above result can be obtained by assuming $\ZZ$ and $\ZZ^*$ are dependent, but at the cost of more involved formula for $\ell_{\XX}$. Such scenario is considered in the following proposition where, for simplicity, the focus is restricted to the stable tail dependence function for the $(i,j)$-margin, denoted $\ell_{X_i,X_j}$, as it admits a simpler form.

\begin{prop}
    \label{prop1b}
    Consider the random vector $\XX = (X_1, \ldots, X_d)^{\top}$ defined in \eqref{main-model}. If $V_0$ follows a unit exponential distribution and the vectors $\ZZ$ and $\ZZ^*$ are dependent, then the stable tail dependence function associated with the $(X_i,X_j), i,j \in \mathcal{I}, i \neq j$, is given by    
    \begin{align*}
    \ell_{X_i,X_j}(x_i, x_j) &= 
    x_i \left\{1 -  \frac{
        \Phi_{3}\left( \left(-\frac{\lambda_{ij}}{2} - \frac{1}{\lambda_{ij}} \ln \frac{\tilde x_i}{\tilde x_j}, \alpha_i \rho_{j^*,i}-c_j, \alpha_i\rho_{i^*,i}-c_i\right)^\top; \SS_{i|j}\right)}
        {\Phi(\alpha_i\rho_{i^*,i} - c_i)}\right\} \\
    & \quad + x_j \left\{1 -  \frac{
        \Phi_{3}\left( \left( -\frac{\lambda_{ij}}{2} - \frac{1}{\lambda_{ij}} \ln \frac{\tilde x_j}{\tilde x_i}, \alpha_j \rho_{i^*,j}-c_i, \alpha_j\rho_{j^*,j}-c_j\right)^\top; \SS_{j|i} \right)}
        {\Phi(\alpha_j\rho_{j^*,j} - c_j)}\right\}\,,
    \end{align*}
    where $\tilde x_l = x_l/\Phi(\alpha_l\rho_{l^*,l} - c_l), l \in \{i,j\}$, $\lambda_{ij} = (\alpha_i^2 -2\rho_{i,j}\alpha_i\alpha_j + \alpha_j^2)^{1/2}$, and
    $$
    \SS_{i|j} = \left(\begin{tabular}{ccc}
            $1$ & $\frac{1}{\lambda_{i,j}}(\alpha_j \rho_{j^*,j} - \alpha_i \rho_{j^*,i})$ &  $\frac{1}{\lambda_{i,j}}(\alpha_j \rho_{i^*,j} - \alpha_i \rho_{i^*,i})$ \\
             $\frac{1}{\lambda_{i,j}}(\alpha_j \rho_{j^*,j} - \alpha_i \rho_{j^*,i})$ & $1$ & $\rho_{i^*, j^*}$ \\
             $\frac{1}{\lambda_{i,j}}(\alpha_j \rho_{i^*,j} - \alpha_i \rho_{i^*,i})$ & $\rho_{i^*, j^*}$ & $1$
        \end{tabular}\right)\,.
    $$
    Note that the above uses the notation $\rho_{i,j} = \PP_{i,j} = \mathrm{Cor}(Z_i,Z_j)$, $\rho_{j^*,i} = \PP_{i,j+d} = \mathrm{Cor}(Z_i,Z_j^*)$ and $\rho_{i^*,j^*} = \PP_{i+d,j+d} = \mathrm{Cor}(Z_i^*,Z_j^*)$.
\end{prop}

\noindent \textbf{Proof:} See Appendix \ref{appx-prop1b}. \hfill $\Box$



The limiting distribution characterized by the above tail dependence function is an extension of the H\"{u}sler-Reiss distribution (see Section~\ref{ssec:HR}) that allows for asymmetric dependence structures and we therefore refer to it as the skew-H\"usler-Reiss distribution. It is seen that the standard H\"usler-Reiss distribution is recovered when either $c_i =  c_j = -\infty$ or $c_i = c_j$ combined with $\rho_{i^*,j^*} = 1$ and independence of $(Z_i, Z_j)^{\top}$ and $(Z_i^*, Z_j^*)^{\top}$. The asymmetric extension of the H\"{u}sler-Reiss distribution is not unique \citep[see e.g.,][]{padoan2011, beranger2019b, zhong2024} however the proposed construction doesn't not require additional assumptions on the (skewness) parameters.

\noindent \textbf{Remark 1:} Distributions $F_0$ of the common factor $V_0$ with lighter tails than exponential, such as the Weibull distribution with the shape parameter $\beta > 1$ result in asymptotic independence of the components of $\XX$ \citep{Krupskii.Huser.ea2016}, and hence $\ell(\xx) = \sum_{i=1}^d x_i$ in this case.

\noindent \textbf{Remark 2:} Following from Proposition \ref{prop1b}, for a pair $(X_i, X_j)^{\top}$,
$$
\xi_{i:j} := \lim_{x_i \to 0} \frac{\partial \ell_{\XX_{i:j}}((x_i,x_j)^{\top})}{\partial x_i} = 1 - \frac{\Phi_2((\alpha_i\rho_{j^*,i}-c_j,\alpha_i\rho_{i^*,i}-c_i)^{\top}; \rho_{i^*,j^*})}{\Phi(\alpha_i\rho_{i^*,i}-c_i)}\,,
$$
which is a strictly positive quantity unless $\rho_{i^*,j^*} = 1$ and $\alpha_i\rho_{i^*,i}-c_i \leq \alpha_i\rho_{j^*,i}-c_j$. From \eqref{eq:mass}, the bivariate generalized Pareto distribution has a point mass at one or both boundaries.  On the other hand, $\xi_{i:j} = 0$ for any $i \neq j$ if $c_1 = c_2 = \cdots = c_d = c_0$ and $Z_1^* = Z_2^* = \cdots = Z_d^*$. In this case, the stable tail dependence function simplifies to
\begin{align*}
\ell_{X_i,X_j}(x_i, x_j) =\ & x_i\frac{
        \Phi_{2}\left( \left(\frac{\lambda_{ij}}{2} + \frac{1}{\lambda_{ij}} \ln \frac{\tilde x_i}{\tilde x_j}, \alpha_i \rho_{i,0}-c_0\right)^\top; \frac{1}{\lambda_{ij}}(\alpha_i\rho_{i,0} - \alpha_j\rho_{j,0})\right)}
        {\Phi(\alpha_i\rho_{i,0} - c_0)}\\
        &+ x_j\frac{
        \Phi_{2}\left( \left(\frac{\lambda_{ij}}{2} + \frac{1}{\lambda_{ij}} \ln \frac{\tilde x_j}{\tilde x_i}, \alpha_j \rho_{j,0}-c_0\right)^\top; \frac{1}{\lambda_{ij}}(\alpha_j\rho_{j,0} - \alpha_i\rho_{i,0})\right)}
        {\Phi(\alpha_j\rho_{j,0} - c_0)}\,,
\end{align*}
where $\rho_{k,0} = \PP_{k,d+1}$. This is a different parameterization of the asymmetric extension of the H\"usler-Reiss distribution by \cite{beranger2019b} with the advantage of not requiring any constraints on the skewness parameters. The next proposition gives the full formula for $\ell_{\XX}$ in this special case.

\begin{prop}
    \label{prop1c} Consider the random vector $\XX = (X_1, \ldots, X_d)^{\top}$ defined in \eqref{main-model}. Assume $V_0$ follows a unit exponential distribution, $c_1 = c_2 = \cdots = c_d = c_0$, $Z^*_1 = Z^*_2 = \cdots = Z_d^* = Z_0^*$, and $\rho_{k,0} = \mathrm{Cor}(Z_k,Z_0^*)$, $k \in \II$. Then the stable tail dependence function associated with $\XX$ is given by
    $$
    \ell_{\XX}(\xx) = \sum_{j = 1}^d x_j  \Phi_{d-1}^{\textrm{ESN}} \left\{ 
    \left(\frac{\lambda_{ij}}{2} + \frac{1}{\lambda_{ij}}\log \frac{\tilde x_j}{\tilde x_i}\right)_{i \in \II_{-j}}; 
    \bar{\Sigma}^{(j)},
    \frac{\bar{\Sigma}^{(j)-1} \bm \delta_j}{\sqrt{1-\bm \delta^\top_j \bar{\Sigma}^{(j)-1} \bm \delta_j}},
    \frac{\alpha_j \rho_{j,0} - c_0}{\sqrt{1-\bm \delta^\top_j \bar{\Sigma}^{(j)-1} \bm \delta_j}}
    \right\} 
    $$
    where $\Phi^{\textrm{ESN}}_{d-1}(\cdot; \Sigma, \bm \alpha, \tau)$ is the cdf of the centered extended Skew-Normal (ESN) distribution with covariance matrix $\Sigma$, skewness vector $\bm \alpha \in \mathbb{R}^{d-1}$ and extension parameter $\tau \in \mathbb{R}$. Note that here $\bar{\Sigma}^{(j)}$ is identical to the H\"{u}sler-Reiss model and in addition $\tilde x_j = x_j/\Phi(\alpha_j\rho_{j,0} - c_0)$, $\lambda_{i,j} = (\alpha_i^2 - 2\rho_{i,j}\alpha_i\alpha_j + \alpha_j^2)^{1/2}$ and 
    $\bm \delta_j = \left\{ (\alpha_i \rho_{i,0} - \alpha_j \rho_{j,0})/\lambda_{ij} \right\}_{i \in \II_{-j}}$.
\end{prop}

\noindent \textbf{Proof:} See Appendix \ref{appx-prop1c}. \hfill $\Box$

\noindent As noted in Section \ref{sec:background}, the stable tail dependence function fully characterizes extremal dependence, since the EV copula and the multivariate EV distribution with unit Fr\'echet margins are given respectively by $C_{\textrm{EV}}(\bm u) = \exp \left\{ - \ell(- \log \bm u)\right\}$ and $G(\bm x) = \exp \left\{ - \ell(1/\bm x)\right\}$.
Moreover, as discussed above, the corresponding multivariate Pareto model is completely determined by its density on the interior of the support, which is given by the formula:
\begin{align}
\label{eq:h_prop4}
    h(\bm x) &= \phi_{d-1}^{\textrm{ESN}} \left[ 
    \left\{\frac{\lambda_{ik}}{2} + \frac{1}{\lambda_{ik}}\log \frac{x_i \Phi(\alpha_i \rho_{i,0}-c_0)}{x_k \Phi(\alpha_k \rho_{k,0} - c_0)}\right\}_{i \in \II_{-k}}; 
    \bar{\Sigma}^{(k)},
    \frac{\bar{\Sigma}^{(k)-1} \bm \delta_k}{\mathcal{S}_k},
    \frac{\alpha_k \rho_{k,0} - c_0}{\mathcal{S}_k}
    \right] \nonumber \\
    & \quad \times \left\{ \ell(\bm 1) x_k^2 \prod_{i \in \mathcal{I}_{-k}} x_i \lambda_{ik} \right\}^{-1},
\end{align}
where $\mathcal{S}_k = \sqrt{1-\bm \delta^\top_k \bar{\Sigma}^{(k)-1} \bm \delta_k}$.
The extremal coefficient and the Pickands dependence function are also determined by the stable tail dependence function, via $\vartheta = \ell (\bm 1)$ and $A(\bm w) = \ell \left( \bm x / \|\bm x \|_1 \right)$.

\noindent \textbf{Remark 3:} A parsimonious parameterization of this model can be obtained by assuming that the covariance matrix of $(Z_0^*,Z_1,\ldots,Z_d)^{\top}$ is determined by a covariance function, such that $\rho_{i,0}$ depends on the distance between the corresponding locations $\ss_i$ and $\ss_0$. In practical spatial applications, the locations $\ss_1, \ldots, \ss_d$ are typically known, while the coordinate vector $\ss_0$ can be treated as an additional model parameter to be estimated jointly with the other parameters. Even when the underlying covariance function is stationary, the extremal dependence between any two components of the vector $\XX$ is influenced by their respective distances to $\ss_0$. This induces skewness and renders the model non-stationary. The proximity of the locations $\ss_1, \ldots, \ss_d$ to $\ss_0$ governs the degree of skewness: larger distances lead to smaller values of $\bm \delta_j$ and hence weaker skewness, as implied by Proposition \ref{prop1c}.

\begin{figure}[t]
    \centering
    \includegraphics[width=0.32\linewidth]{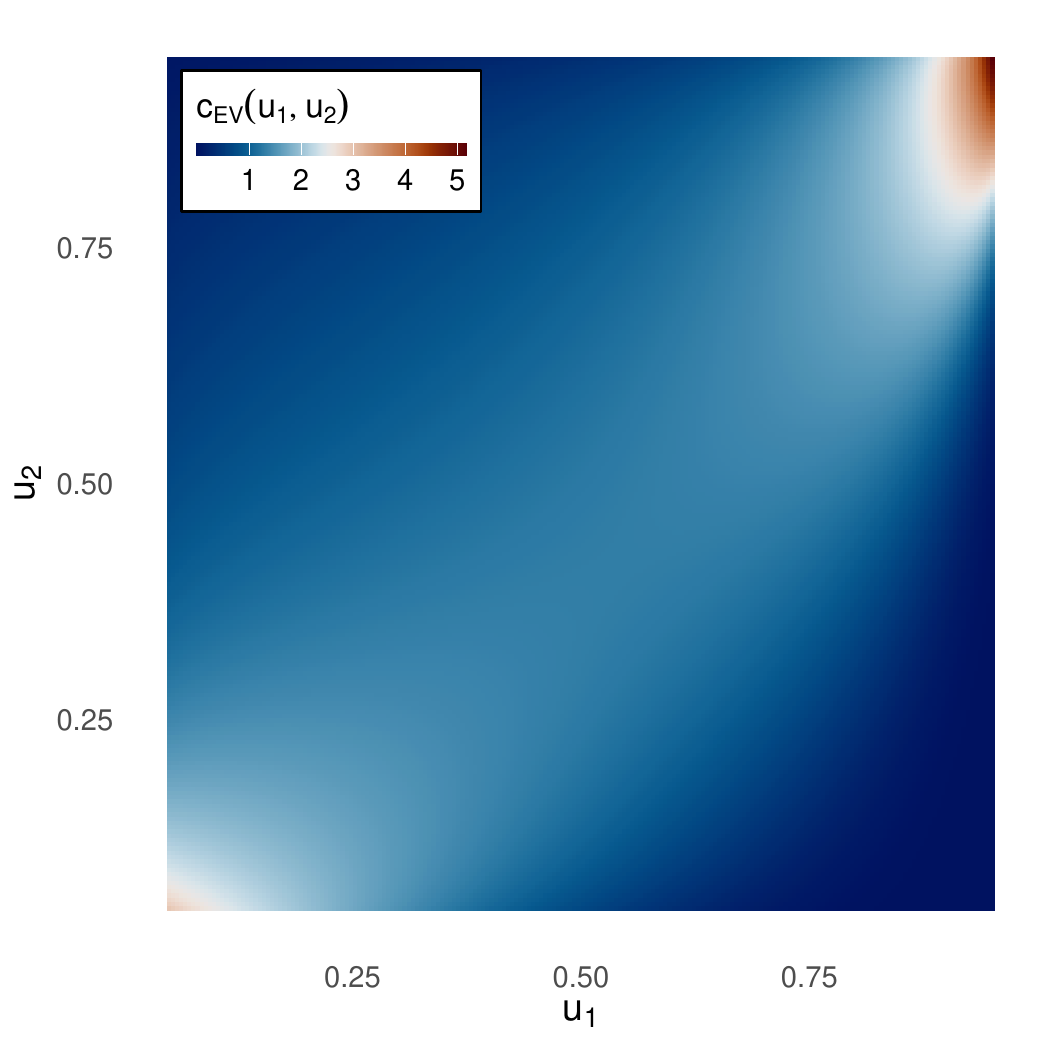}
    \includegraphics[width=0.32\linewidth]{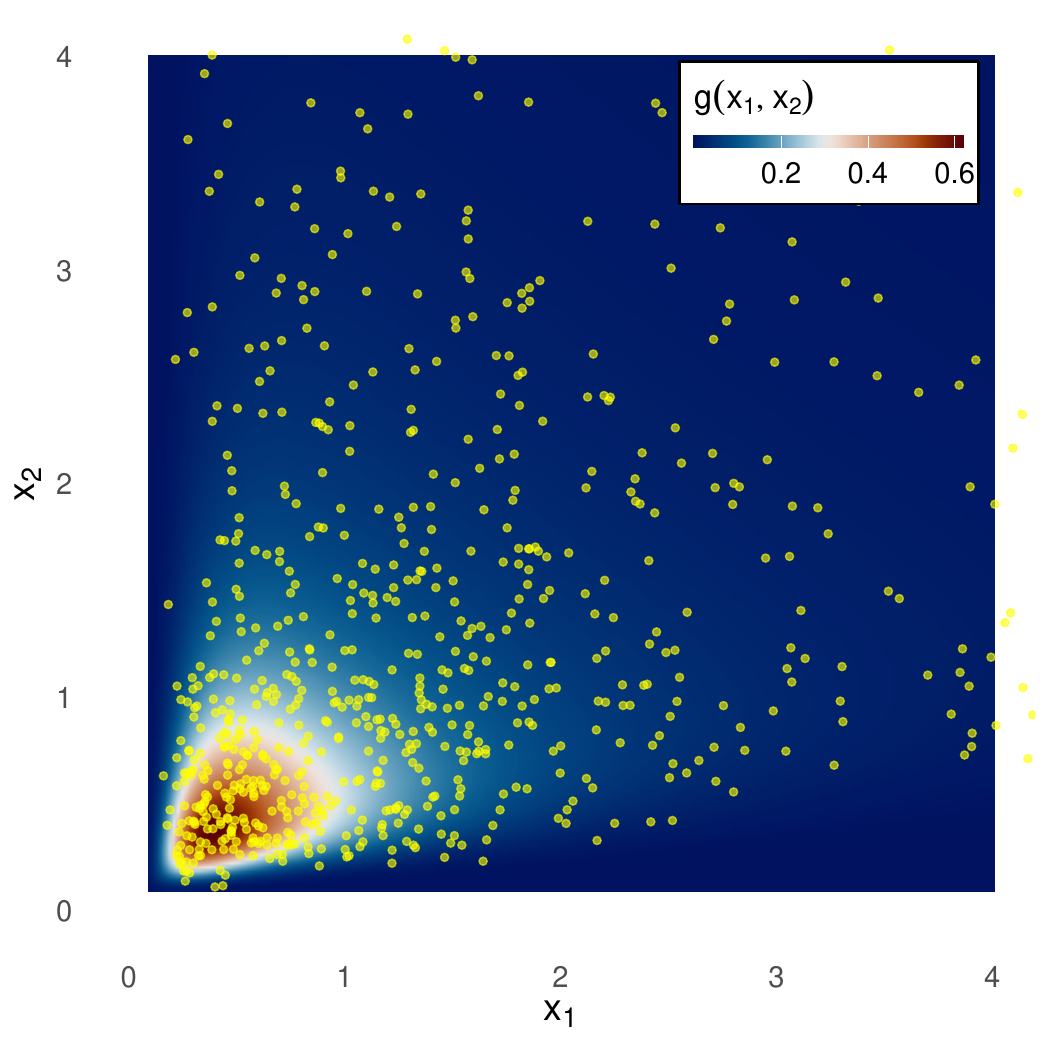}
    \includegraphics[width=0.32\linewidth]{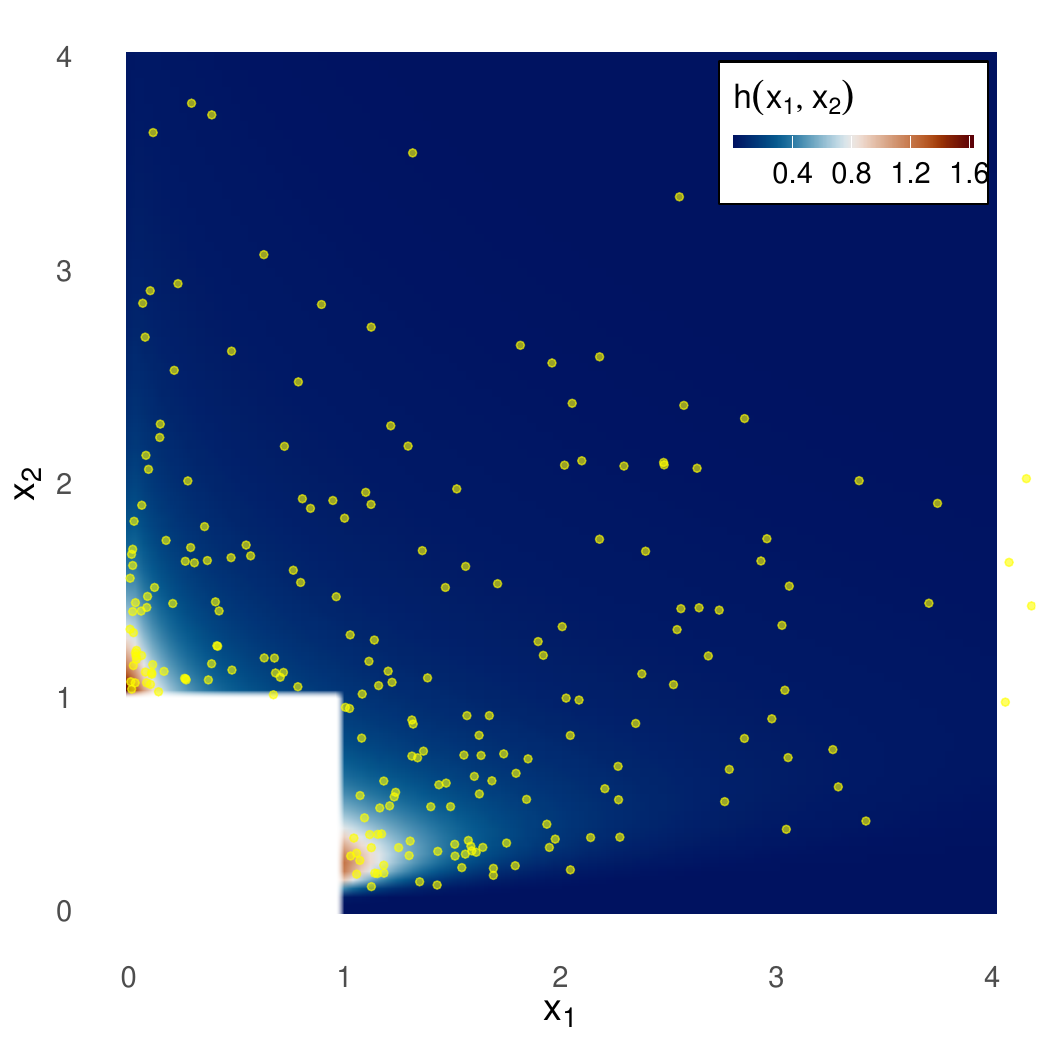} \\
    \includegraphics[width=0.32\linewidth]{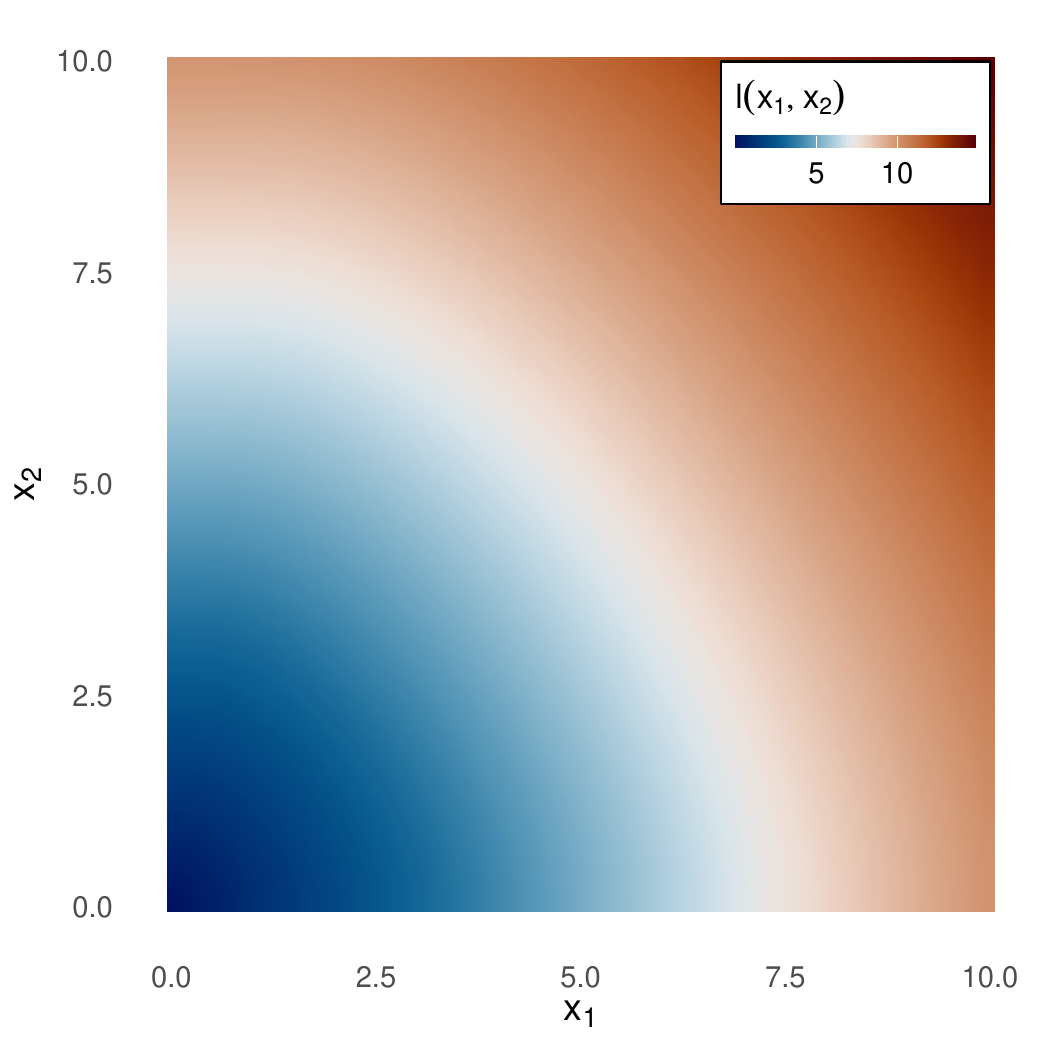}
    \includegraphics[width=0.32\linewidth]{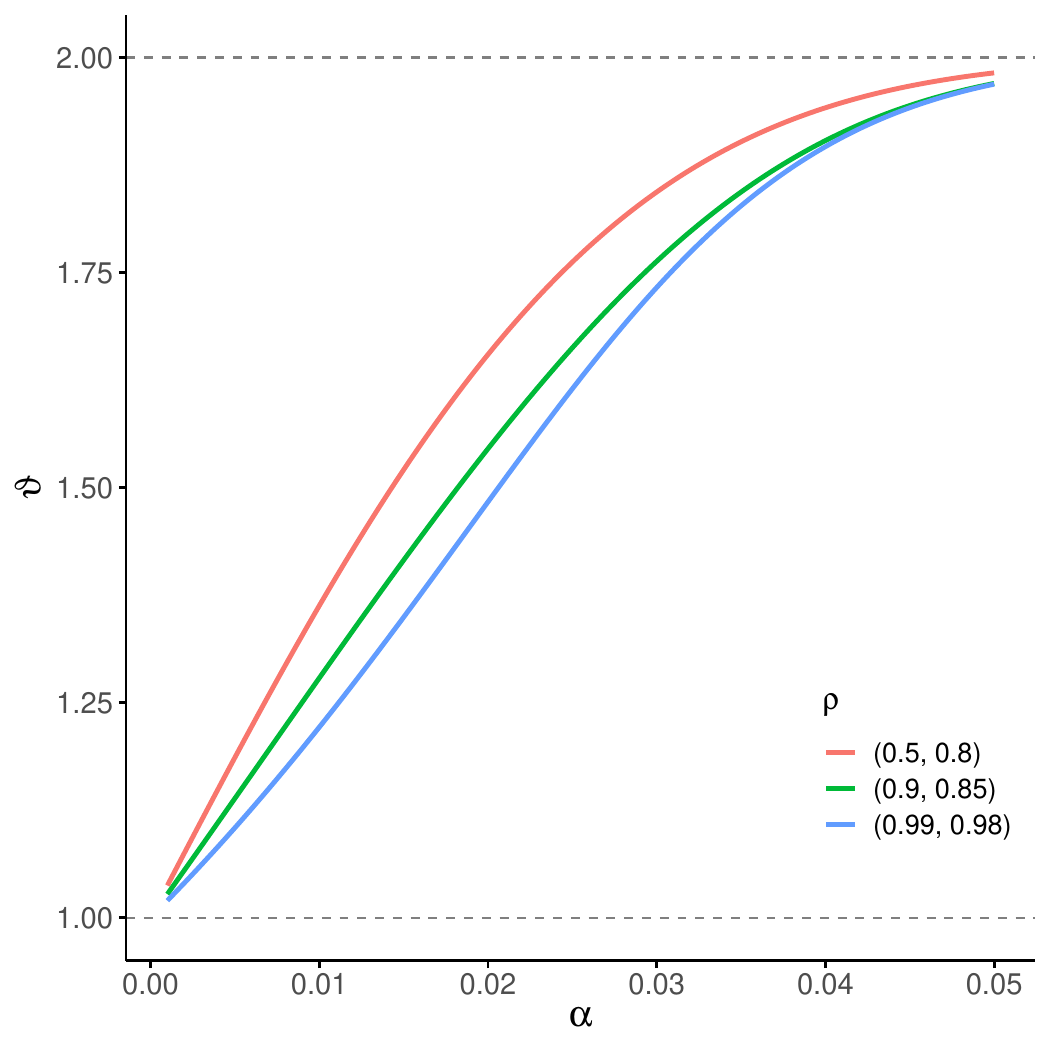}
    \includegraphics[width=0.32\linewidth]{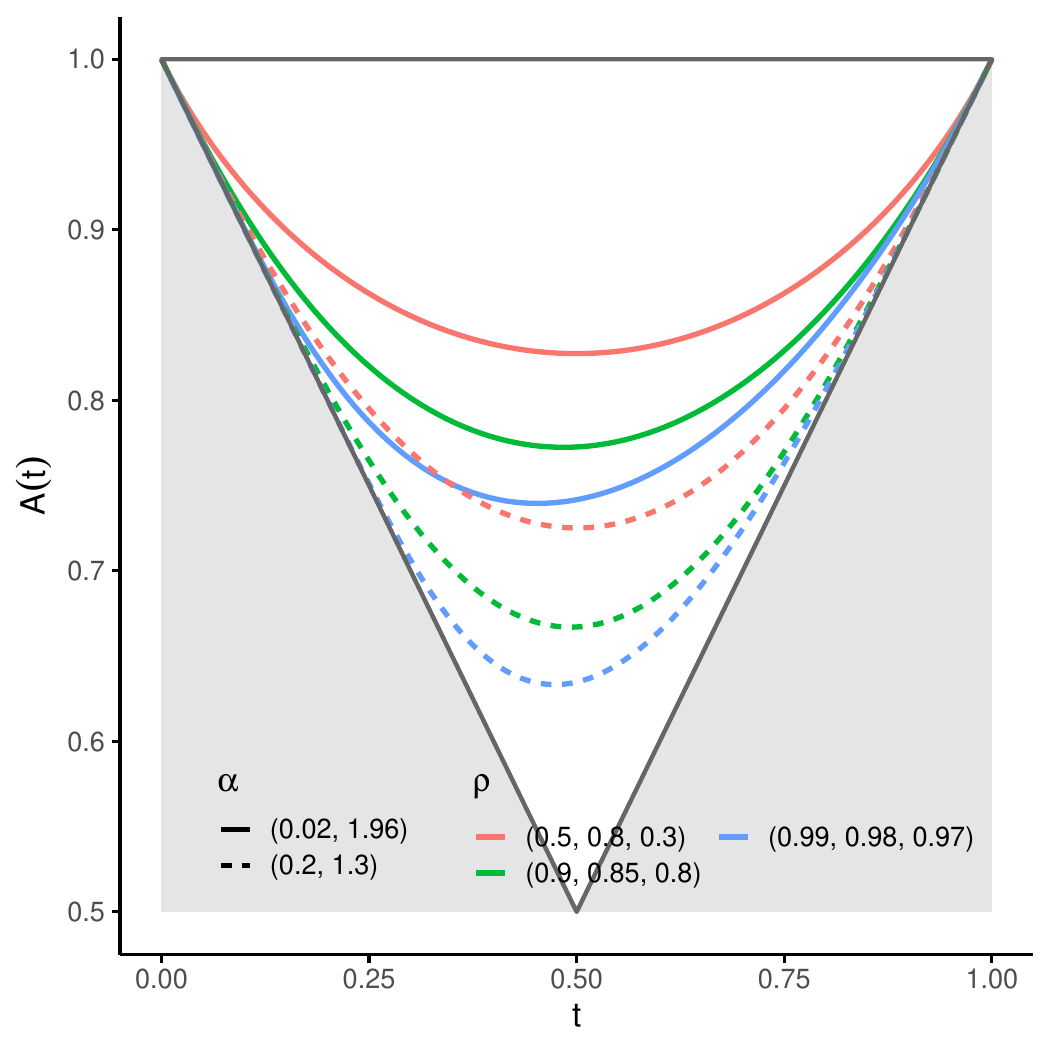}
    \caption{Left to right, top to bottom: density of the extreme value copula $c_{EV}$, density of the bivariate extreme value distribution $g(\bm x)$, density of the bivariate Generalized Pareto distribution with unit Pareto margins $\log h(\bm x)$ and stable tail dependence function $\ell(\bm x)$ of the Proposition~\ref{prop1c} model with $\bm c_0 = 0.8$, $\bm \alpha = (\alpha_1, \alpha_2)$ and correlation $\bm \rho =(\rho_{1,2}, \rho_{1,0}, \rho_{2,0}) = (0.99, 0.98, 0.97)$. The remaining panels display for several choices of $\bm \rho$ and $\bm \alpha$, the extremal coefficient $\vartheta$ as function of $\alpha$ with $\bm \alpha = (\alpha, 98 \alpha)$ and the Pickands dependence function with $c_0 = 0.8$.}
    \label{fig:summrySHR}
\end{figure}

\subsection{Tail properties}
\label{ssec:tail}

We now delve into the tail properties of the models defined in Section~\ref{ssec:EVlimits} and derive insights about their characteristics and potential flexibility.

In the context of Proposition~\ref{prop0}, the limiting EV copula corresponds to the Marshall-Olkin distribution \citep{Marshall.Olkin1967}, and it depends on the correlation matrix of the vector $\ZZ^*$ and the vector of truncation parameters $\mathbf{c}$. Indeed, the correlation matrix of $\ZZ$ has no impact on the copula since the heavy-tailed factor $V_0$ dominates the Gaussian component $\ZZ$ in the upper tail. With $\mathbf{c} = -\infty$, the EV copula simplifies to the comonotonicity copula, i.e. $C_{\textrm{EV}}(\bm u) = \min(\bm u)$. For example, when $d=2$, the stable tail dependence function of $\XX = (X_1, X_2)$ is
$$
\ell_{\XX}(\xx) = (1-\alpha) x_1 + (1-\beta) x_2 + \max (\alpha x_1, \beta x_2),
$$
where $\alpha = \Pr(Z_2^* >c_2 | Z_1^* > c_1) \in [0,1]$ and $\beta = \Pr(Z_1^* >c_1 | Z_2^* > c_2) \in [0,1]$. The corresponding EV copula is therefore $C_{\textrm{EV}}(u_1, u_2) = \min \left( u_1^{1-\alpha} u_2, u_1 u_2^{1-\beta}\right)$ and its density $c_{\textrm{EV}}(u_1, u_2) = (1-\alpha) u_1^\alpha$ if $u_1 > u_2^{\beta/\alpha}$ and $(1-\beta) u_2^{\beta}$ otherwise. The multivariate EV density is given by $g(\bm x) = \exp \{ -\ell(1/\bm x) \} (1-\alpha)(1-\beta) / (x_1 x_2)^2$ while the bivariate generalized Pareto distribution does not admit a density (since $\partial^2\ell_\XX/\partial x_1 \partial x_2  =0 $) but admits mass $(\alpha \wedge \beta) / (2 - (\alpha \vee \beta) )$ along the line $x_2 = \tfrac{\alpha}{\beta}x_1$ and at the boundaries of the support $H_1(\{0\}) = (1-\beta) / \{2 - (\alpha \vee \beta) \}$,  $H_2(\{0\}) = (1-\alpha) / \{2 - (\alpha \vee \beta) \}$. Illustrations are provided in Appendix~\ref{appx-MO}, Figure~\ref{fig:summryMO}.

Models with an exponential factor preserve substantial flexibility in the tail behaviour, unlike those based on a heavy-tailed distribution for $V_0$ considered in Proposition~\ref{prop0}. However, Proposition~\ref{prop1b} shows that the resulting bivariate stable tail dependence function requires evaluating a trivariate normal cdf, making extensions to higher dimensions computationally demanding. By contrast, Proposition~\ref{prop1a} demonstrates that assuming independence between $\ZZ$ and $\ZZ^*$ yields models that are far more tractable, as they rely only on lower-dimensional Hüsler–Reiss distributions while still offering substantial tail flexibility, including the ability to capture asymmetry. We now turn to the following submodel.

\noindent \textbf{Submodel 1:} Assume that $Z_1^* = Z_2^* = \cdots = Z_d^* = Z^*$ and $(Z_1, \ldots, Z_d)^{\top}$ is independent of $Z^*$. From Proposition \ref{prop1a}, for $i,j \in \mathcal{I}, i \neq j$,
\begin{align*}
\ell_{X_i,X_j}(x_i,x_j) 
&= x_i\left\{1 - \Phi\left(-\frac{\lambda_{ij}}{2} - \frac{1}{\lambda_{ij}}\log \frac{\tilde x_i}{\tilde x_j}\right) \cdot \min (\zeta_j/\zeta_i, 1)\right\} \\ 
& \quad + x_j\left\{1 - \Phi\left(-\frac{\lambda_{ij}}{2} - \frac{1}{\lambda_{ij}}\log \frac{\tilde x_j}{\tilde x_j}\right) \cdot \min (\zeta_i/\zeta_j, 1)\right\}\,,
\end{align*}
where $\tilde x_l = x_l/\zeta_l$ and $\zeta_l = \Phi(-c_l)$ for $l \in \{i,j\}$. 
\begin{figure}[t]
    \centering
    \includegraphics[width=0.32\linewidth]{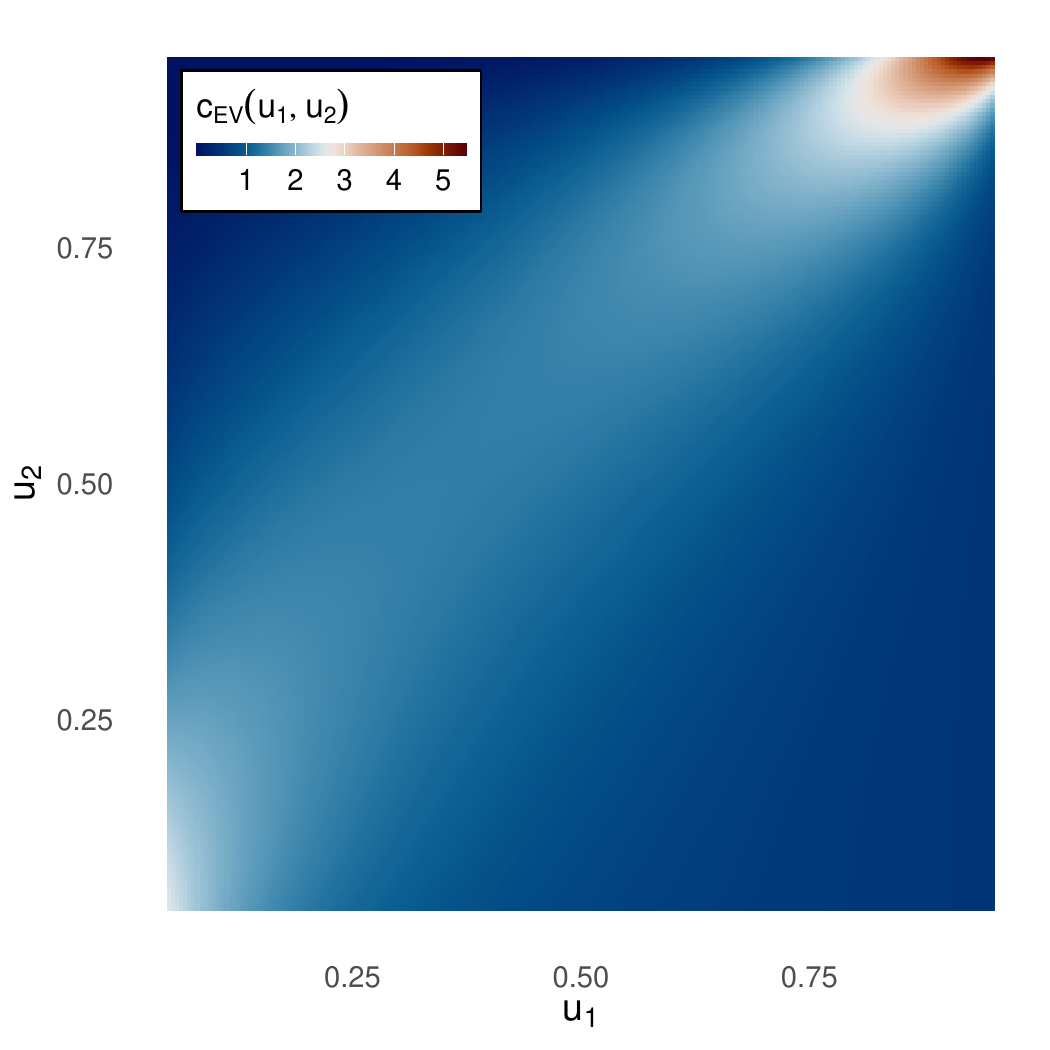}
    \includegraphics[width=0.32\linewidth]{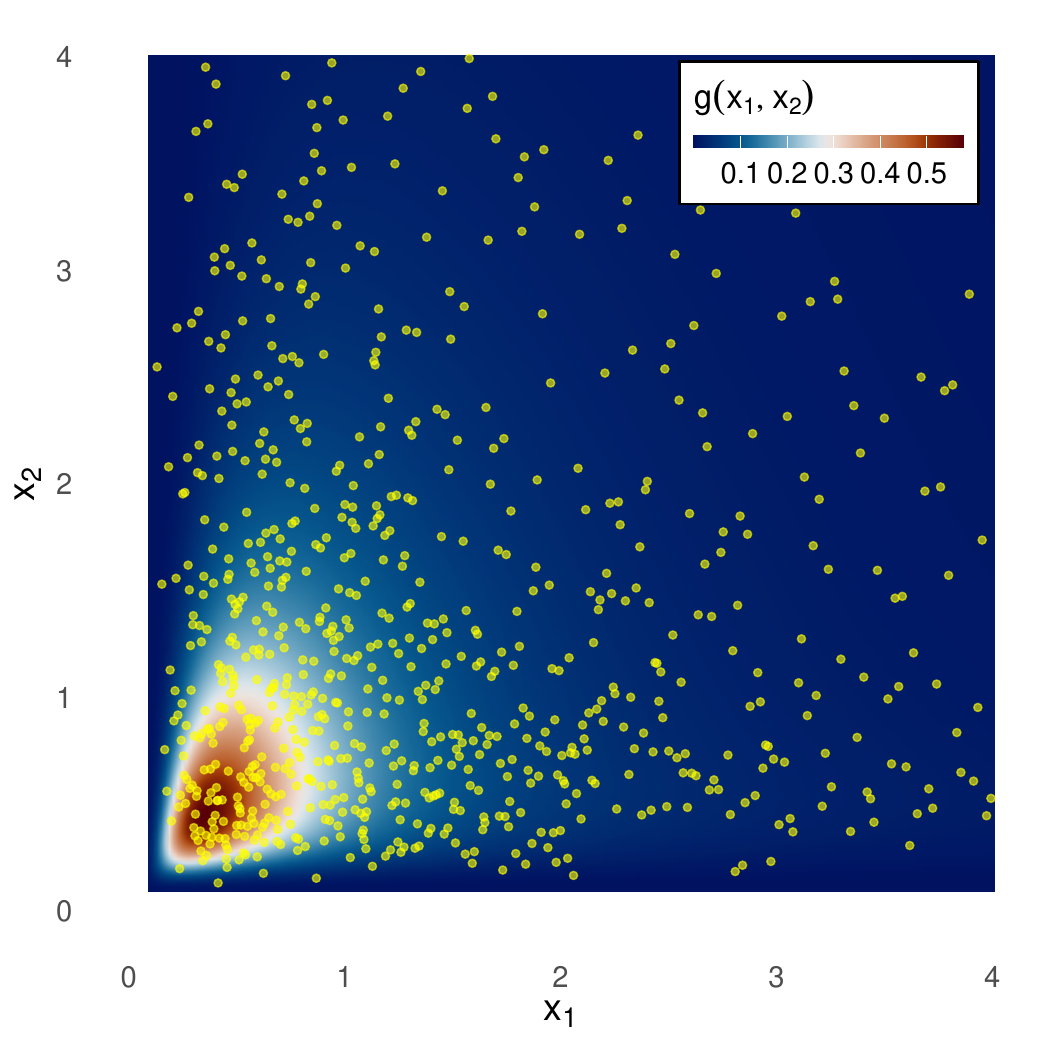}
    \includegraphics[width=0.32\linewidth]{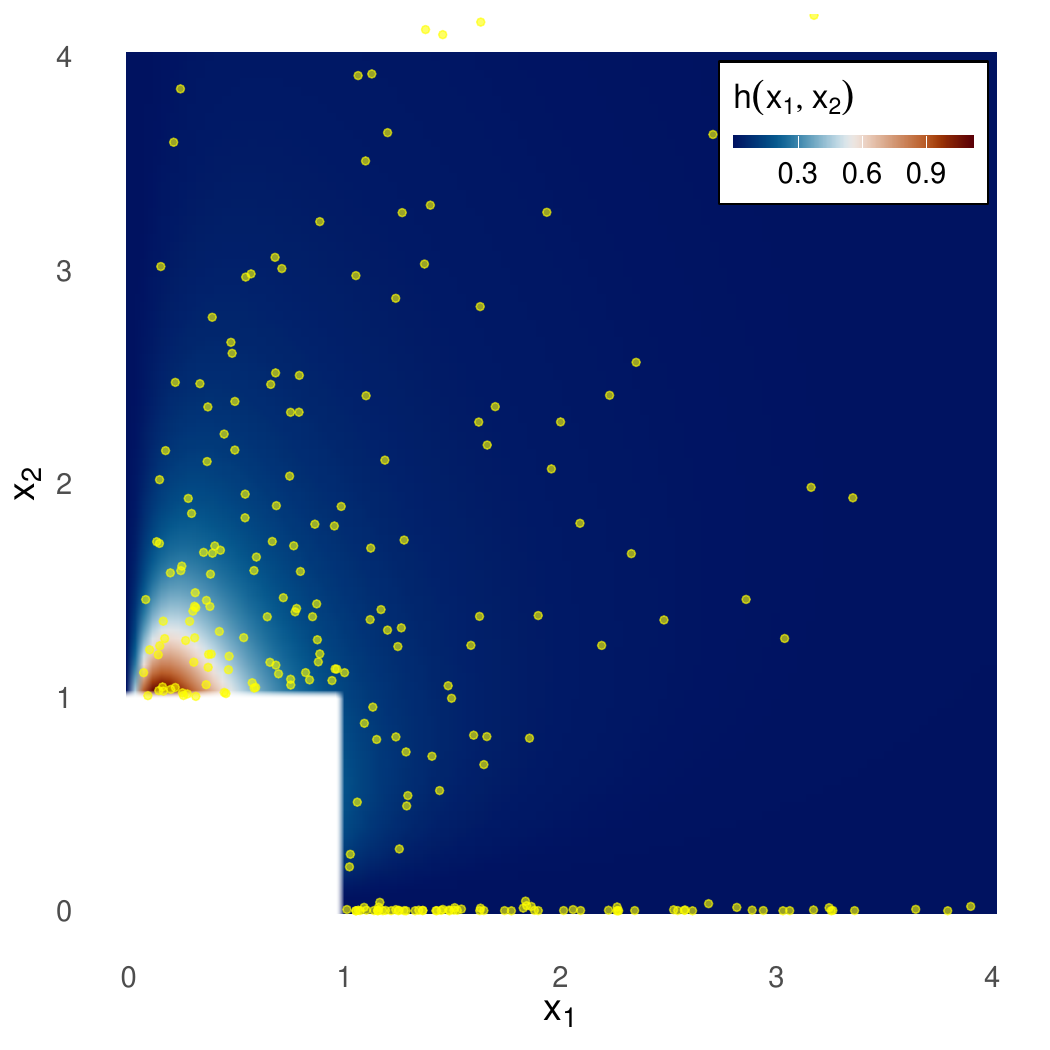} \\
    \includegraphics[width=0.32\linewidth]{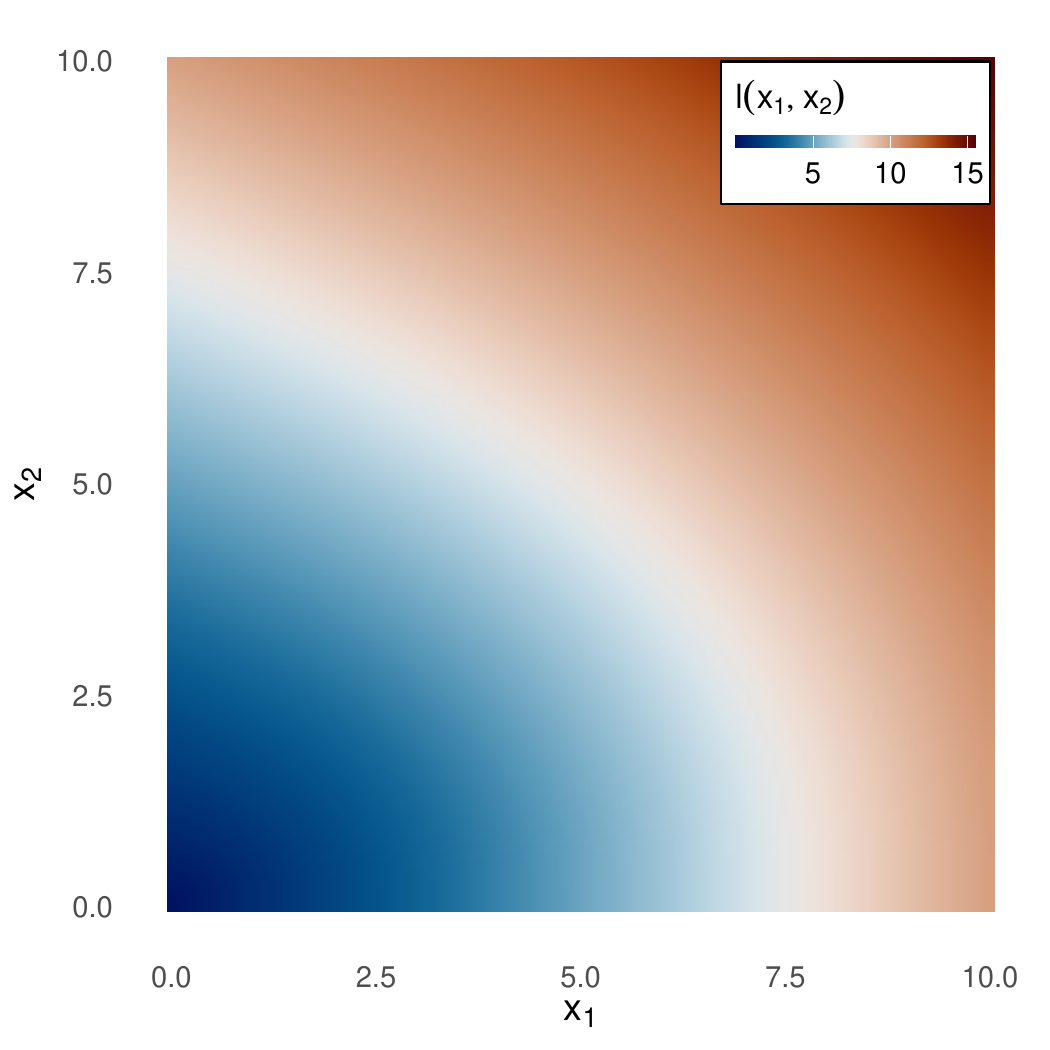}
    \includegraphics[width=0.32\linewidth]{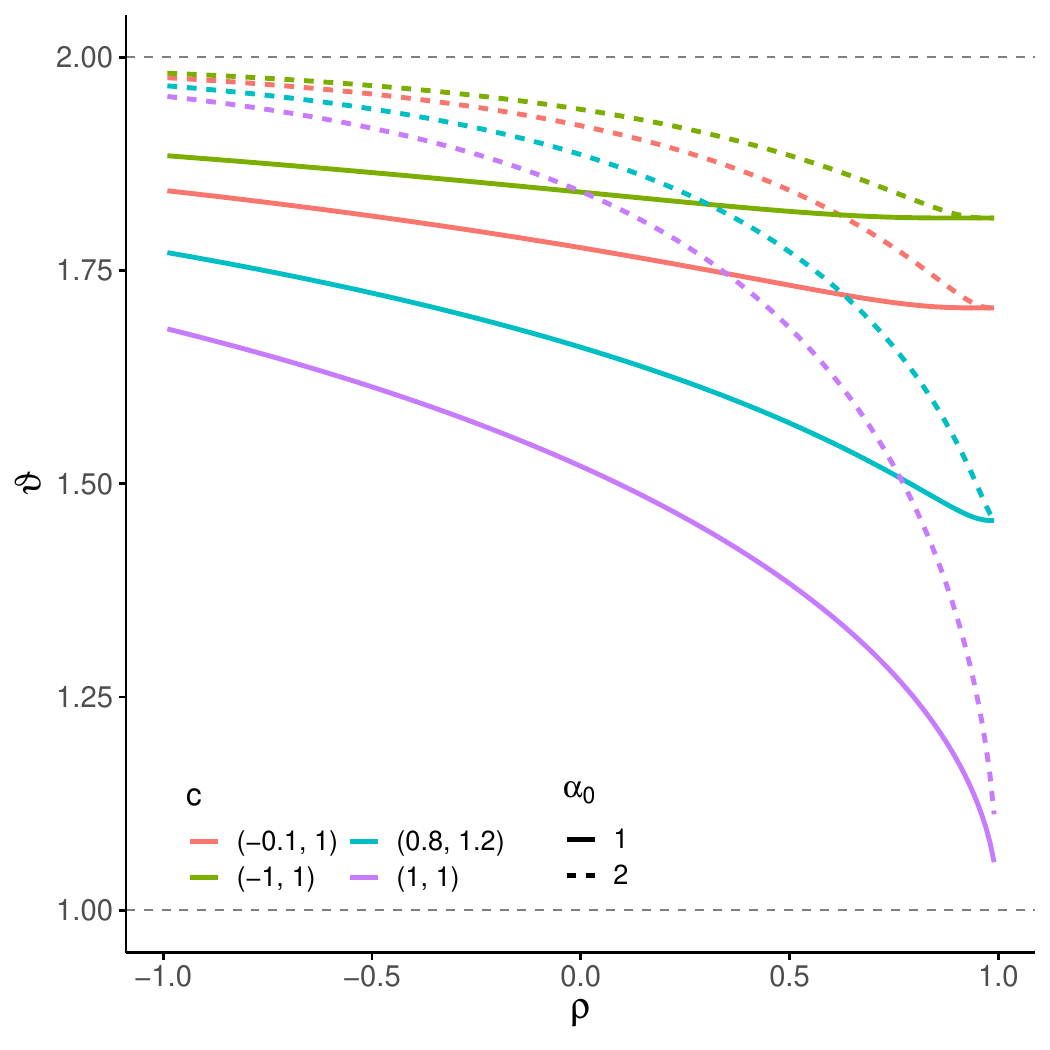}
    \includegraphics[width=0.32\linewidth]{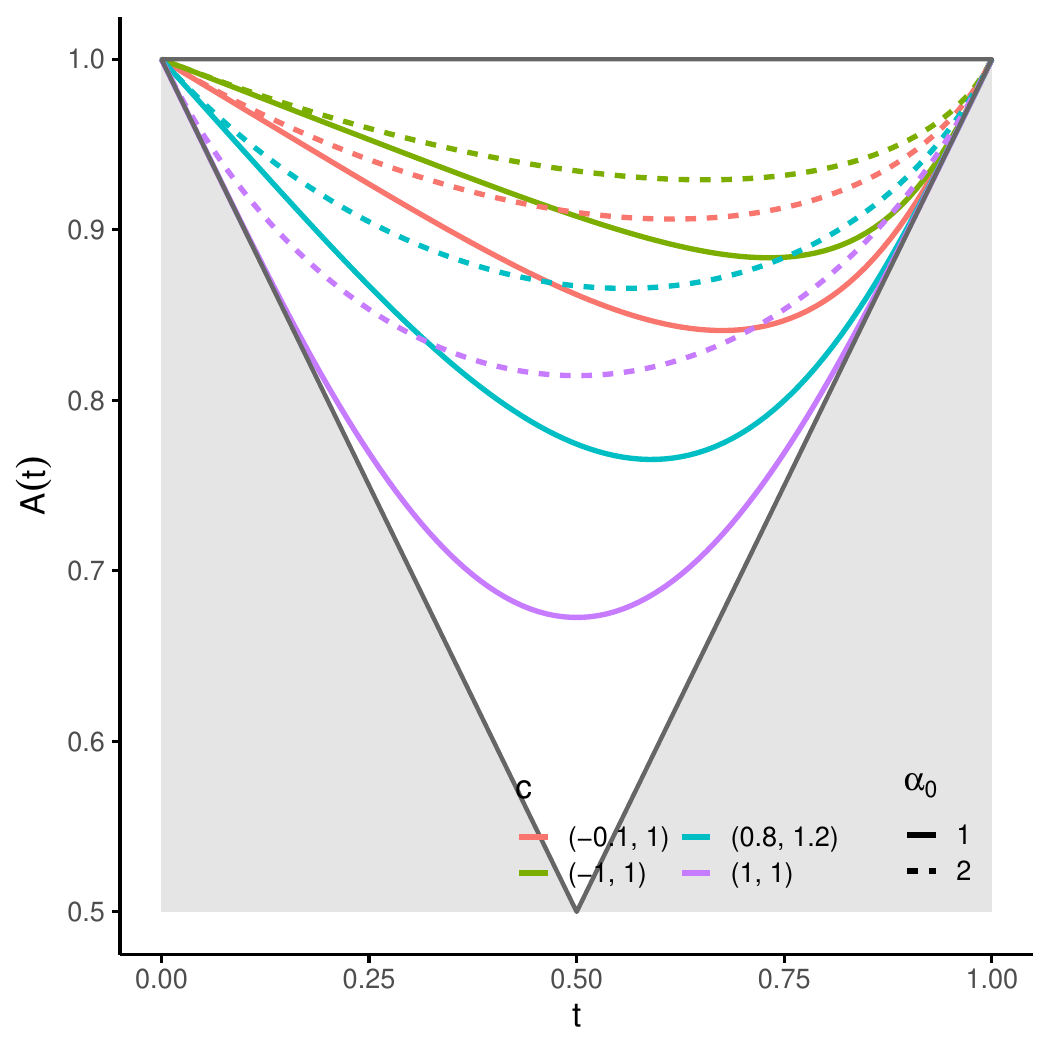}
    \caption{Left to right, top to bottom: density of the extreme value copula $c_{EV}$, density of the bivariate extreme value distribution $g(\bm x)$, density of the bivariate Generalized Pareto distribution with unit Pareto margins $\log h(\bm x)$ and stable tail dependence function $\ell(\bm x)$ of Submodel 1 with $\bm c = (0.8, 1.2)$, $\alpha_0 = 1$ and correlation $\rho_{1,2} = 0.6$. The remaining panels display for several choices of $\bm c$ and $\bm \alpha$, the extremal coefficient $\vartheta$ as function of $\rho_{1,2}$ and the Pickands dependence function with $\rho = 0.6$.}
    \label{fig:summryS1}
\end{figure}
Note that $\ell_{X_i,X_j}$ only involves univariare normal cdfs thus making this model computationally tractable (see Section~\ref{sec:inf}). As previously mentioned, the EV copula and the multivariate EV distribution with unit Fr\'{e}chet margins can be easily recovered from the stable tail dependence function while its partial derivatives can be used to compute the corresponding densities as well as the density of the multivariate Generalized Pareto distribution. Regarding the latter, we have
$$
\frac{\p}{\p x_k}\ell_{X_i,X_j}(x_i,x_j) = 1 - \Phi\left(-\frac{\lambda_{ij}}{2} - \frac{1}{\lambda_{ij}}\log \frac{\tilde x_k}{\tilde x_l}\right) \cdot \min (\zeta_l/\zeta_k, 1), \quad k \in \{i,j\}, \ l  \neq k,
$$
which, from \eqref{eq:mass}, yields $H_i(\{0\}) = \left[ 1 - \min \left( \frac{\zeta_i}{\zeta_j}, 1\right) \right] / \ell_{X_i,X_j}(1,1)$.  This implies that $H$ places mass on a single boundary determined by the ratio $\zeta_i/\zeta_j$, and therefore by the values of $c_i$ and $c_j$; in particular,  when $c_i = c_j$, no boundary mass arises. The density in the interior of the support $(\bm \eta, \bm \infty)$ is then
$$
h(x_i, x_j) = \frac{\phi\left(-\frac{\lambda_{ij}}{2} - \frac{1}{\lambda_{ij}}\log \frac{\tilde x_j}{\tilde x_i}\right) \cdot \min (\zeta_j/\zeta_i, 1)}{\lambda_{ij} x_i^2 x_j \ell_{X_i,X_j}(1,1)}.
$$
For simplicity and to avoid over parametrisation, from now on we will consider $\alpha_1 = \cdots = \alpha_d = \alpha_0$.
Reading from left to right and top to bottom, Figure~\ref{fig:summryS1} illustrates the EV copula density $c_{\textrm{EV}}(u_1, u_2)$, the multivariate EV density $g(x_1, x_2)$, the multivariate generalized Pareto density $h(x_1, x_2)$, the stable tail dependence function $\ell_\XX(x_1, x_2)$, the extremal coefficient $\vartheta$, and the Pickands dependence function $A(t)$ for Submodel 1 with $\bm c = (0.8, 1.2)$, $\alpha_0 = 1$, and correlation $\rho_{1,2} = 0.6$. Since $c_1 < c_2$, we have $\zeta_1 > \zeta_2$, implying that $H_1({0})=0$ (no mass along $x_1=0$) and $H_1({0})\neq 0$ (mass along $x_2=0$). The resulting asymmetry is clearly reflected in the Pickands dependence function, which is directly influenced by the parameters $\bm c$.

Instead of imposing $Z_1^* = \cdots = Z_d^*$ as in Submodel 1, an etxension is to allow the components of $\ZZ^*$ to be equicorrelated, leading to the following submodel.

\noindent \textbf{Submodel 2:} Assume that $\ZZ^* = (Z_1^*, Z_2^*, \ldots Z_d^*)^{\top}$ has equicorrelation structure with $\rho_{i^*,j^*} = \rho^*$ for $i^* \neq j^*$ and $\ZZ= (Z_1, \ldots, Z_d)^{\top}$ is independent of $\ZZ^*$. This model extends Submodel~1, which arises as the special case $\rho^* = 1$. From Proposition \ref{prop1a}, for $i,j \in \mathcal{I}, i \neq j$,
\begin{align*}
\ell_{X_i,X_j}(x_i,x_j) 
&= x_i\left[1 - \Phi\left(-\frac{\lambda_{ij}}{2} - \frac{1}{\lambda_{ij}}\log \frac{\tilde x_i}{\tilde x_j}\right) \cdot \Phi_{2}\left\{\left(-c_i,-c_j\right)^\top; \rho^*\right\}/\zeta_i, \right]\\ 
& \quad + x_j\left[1 - \Phi\left(-\frac{\lambda_{ij}}{2} - \frac{1}{\lambda_{ij}}\log \frac{\tilde x_j}{\tilde x_j}\right) \cdot \Phi_{2}\left\{\left(-c_i,-c_j\right)^\top;\rho^*\right\}/\zeta_j\right]\,,
\end{align*}
where $\tilde x_l = x_l/\zeta_l$, $\zeta_l = \Phi(-c_l)$ for $l \in \{i,j\}$ as before. 
Similarly to Submodel~1, all functions associated with this EV distribution can be derived. In particular, for the generalized Pareto distribution, we obtain
$$
\frac{\p}{\p x_k}\ell_{X_i,X_j}(x_i,x_j) = 1 - \Phi\left(-\frac{\lambda_{ij}}{2} - \frac{1}{\lambda_{ij}}\log \frac{\tilde x_k}{\tilde x_l}\right) \cdot \Phi_{2}\left\{\left(-c_i,-c_j\right)^\top; \rho^*\right\}/\zeta_k, \quad k \in \mathcal{I}, l =\mathcal{I}_{-k},
$$
which, from \eqref{eq:mass}, yields $H_i(\{0\}) = \left[ 1 - \Phi_{2}\left\{\left(-c_i,-c_j\right)^\top; \rho^*\right\}/\zeta_i \right] / \ell_{X_i,X_j}(1,1)$. This implies that $H$ places mass on both boundaries. These masses are equal when $c_i = c_j$ and vanish when $c_i = c_j = -\infty$. Finally, the density in the interior of the support $(\bm 0, \bm \infty)$ is given by
$$
h(x_i, x_j) = \frac{\phi\left(-\frac{\lambda_{ij}}{2} - \frac{1}{\lambda_{ij}}\log \frac{\tilde x_j}{\tilde x_i}\right) \cdot \Phi_{2}\left\{\left(-c_i,-c_j\right)^\top; \rho^*\right\}}{\lambda_{ij} \zeta_i x_i^2 x_j \ell_{X_i,X_j}(1,1)}.
$$
Illustrations for Submodel~2 are shown in Figure~\ref{fig:summryS2} of Appendix~\ref{appx-S2}, where a higher degree of independence among the extremes is observed when $\rho^* = 0$ compared to $\rho^* = 1$ (Submodel~1, Figure~\ref{fig:summryS1}). Additionally, the multivariate generalized Pareto distribution now assigns mass to both boundaries ($x_1=0$ and $x_2=0$).

\section{Inference}
\label{sec:inf}

The explicit form of the stable tail dependence function can be used to estimate the parameters of the limiting EV distribution of the vector $\XX$ in \eqref{main-model} for the general case. Depending on whether extremes are defined via component-wise maxima or threshold exceedances, the limiting distribution may be $G$ or $H$, respectively. In this section, we briefly review some inference approaches and demonstrate how they can be applied to estimate the parameters of Submodels~1 and 2.

\subsection{Inference for Block Maxima data}
\label{ssec:inferenceBM}

Consider a sample $\boldsymbol{\XA}_n \equiv \{ (\XA_{k1}, \ldots, \XA_{kd})^{\top}\}_{k=1}^n$ from the distribution $G$ with univariate marginal cdfs $G_1, \ldots, G_d$ and pdfs $g_1,\ldots, g_d$. Following Section~\ref{ssec:BM}, we define the EV copula cdf of the $(i,j)$-th marginal as $C_{\textrm{EV}}^{(i,j)}(u_i, u_j) = \exp\left\{-\ell_{X_i,X_j}(x_i, x_j)\right\}$, where $x_i = -\log u_i$ and $x_j = -\log u_j$, and its density by
\begin{align*}
    c_{\textrm{EV}}^{(i,j)}(u_i, u_j) &= \frac{\exp\left\{-\ell_{X_i,X_j}(x_i, x_j)\right\}}{u_iu_j} \cdot\left\{\frac{\p \ell_{X_i,X_j}(x_i, x_j)}{\p x_i}\frac{\p \ell_{X_i,X_j}(x_i, x_j)}{\p x_j} - \frac{\p^2 \ell_{X_i,X_j}(x_i, x_j)}{\p x_i \p x_j}\right\}\,.
\end{align*}
From \eqref{eq:G_copula}, the distribution of the $(i,j)$-th marginal is $G^{(i,j)}(x_i, x_j) = C_{\textrm{EV}}^{(i,j)}\left\{ G_i(x_i), G_j(x_j)\right\}$ and its density 
$$
g^{(i,j)}(x_i, x_j) = \frac{\p^2}{\p x_i \p x_j} G^{(i,j)}(x_i, x_j) = c_{\textrm{EV}}^{(i,j)}\left\{G_i(x_i), G_j(x_j)\right\} \cdot g_i(x_i) g_j(x_j),
$$
so that the pairwise log-likelihood for the sample $\boldsymbol{\XA_n}$ is
$$
\mathcal{L}(\boldsymbol{\XA_n}; \tht) = 
\sum_{k=1}^n\sum_{i < j} \log c_{\textrm{EV}}^{(i,j)}\left\{G_i(\XA_{ki};\tht_M),G_j(\XA_{kj};\tht_M);\tht_D\right\} 
+ (d-1)\sum_{k=1}^n\sum_{i=1}^d \log g_i(\XA_{ki};\tht_M),
$$
where $\tht = (\tht_M^{\top},\tht_D^{\top})^{\top}$ with $\tht_M$ and $\tht_D$ respectively being the vectors of marginal and dependence parameters. The composite maximum likelihood estimator of $\tht$, $\widehat\tht_{CMLE}$, is obtained by maximizing $\ell(\boldsymbol{\XA_n};\tht)$. This estimator is consistent and asymptotically normal under standard regularity conditions \citep{Lindsay.1998}. Alternatively, $\tht_M$ can be estimated using the marginal likelihood:
$$
\mathcal{L}_M(\boldsymbol{\XA_n};\tht_M) = \sum_{k=1}^n\sum_{i=1}^d \log g_i(\XA_{ki};\tht_M),
$$
and the dependence parameters can be obtained by maximizing the joint likelihood:
$$
\mathcal{L}_D(\boldsymbol{\XA_n}; \tht_D) = \sum_{k=1}^n\sum_{i < j} \log c_{\textrm{EV}}^{(i,j)}\left\{G_i(\XA_{ki};\widehat\tht_M),G_j(\XA_{kj};\widehat\tht_M);\tht_D\right\},
$$
where $\widehat\tht_M$ is the estimator of $\tht_M$ obtained in the first step. This two-step estimator of $\tht$ is more computationally efficient, especially for large $d$, and is consistent and asymptotically normal \citep{Joe.Xu1996}. 

In the general case, higher-dimensional marginals can be incorporated to construct a composite maximum likelihood estimator. While this approach is more computationally demanding, it can provide greater efficiency than the pairwise maximum likelihood estimator introduced above.

We now conduct a simulation study to assess the finite sample performance of the pairwise likelihood estimator. We generate $N = 500$ datasets made of $n = 250$ and $n=1000$ component-wise maxima drawn from the EV limiting distribution of Submodel 1 considering a parsimonious parameterization with spatial covariance structure for the vector $(Z_1, \ldots, Z_d)^{\top}$. More precisely, we use $\alpha_1 = \cdots = \alpha_d = \alpha = 1/3$, $c_i = \beta_0 + \beta_1 \ss_{i1} + \beta_2 \ss_{i2} = 0.5 + 1.5 \ss_{i1} - 0.5 \ss_{i2}$ where $\{(\ss_{i1},\ss_{i2})\}_{i=1}^d$ is a set of $d = 20, 30, 50$ coordinates randomly selected in $(0,1)^2$ for each generated dataset, and $\rho_{i,j} = \exp\left\{-(\mathrm{dst}_{i,j}/p_1)^{p_2}\right\} = \exp\left\{-(\mathrm{dst}_{i,j}/1.5)\right\}$ where $\mathrm{dst}_{i,j}$ is the Euclidean distance between two points with the coordinates $(\ss_{i1},\ss_{i2})$ and $(\ss_{j1},\ss_{j2})$. Assuming the marginals are known, we estimate $\tht_C = (p_0, p_1, \alpha, \beta_0, \beta_1, \beta_2)^{\top}$ and report the root mean squared errors (RMSEs) for the proposed pairwise likelihood estimator in Table \ref{tab1}.

\begin{table}[ht!]
    \centering
    \begin{tabular}{c|cccccc}
     & $p_0$ & $p_1$ & $\alpha$ & $\beta_0$ & $\beta_1$ & $\beta_2$  \\
     \hline
     $d = 20$ & 1.62/0.95 &  0.12/0.03 
 & 0.07/0.06 & 0.11/0.05 & 0.06/0.03 & 0.02/0.01\\ 
     $d = 30$ & 1.52/0.74 &  0.11/0.09 
 & 0.07/0.05 & 0.09/0.05 & 0.06/0.03 & 0.02/0.01\\
     $d = 50$ & 1.38/0.63 &  0.06/0.03 
 & 0.07/0.05 & 0.09/0.05 & 0.05/0.03 & 0.02/0.01\\
    \end{tabular}
    \caption{RMSEs of the copula parameter estimates of the EV distribution of Submodel 1 obtained using the pairwise likelihood approach. The true values are $p_0 = 1.5, p_1 = 1$, $\alpha = 1/3$, $\beta_0 = 0.5$, $\beta_1 = 1.5, \beta_2 = -0.5$. The results are based on 500 simulated datasets with $N = 250$/$N = 1000$ component-wise maxima replicates and $d=20, 30, 50$ randomly selected locations in $(0,1)^2$.}
    \label{tab1}
\end{table}

It is seen that the pairwise likelihood approach yields pretty accurate parameter estimates, except for $p_0$, and more accurate estimates can be obtained with larger sample size, as expected.

We now conduct a simulation study to assess the finite sample performance of the pairwise likelihood estimator for Submodel 2. We generate $N = 500$ datasets made of $n = 250$ and $n=1000$ component-wise maxima drawn from the EV limiting distribution of submodel 2, and we use the same parameter values as for Submodel 1 above with the addition of $\rho = 0.8$. For comparison, we also assess the finite-sample performance of the triplewise likelihood estimator for Submodel 2 under the same simulation settings. The stable tail dependence function for this model has a simple analytical form in the trivariate case, rendering the triplewise likelihood estimator computationally tractable.  Assuming again the marginals to be known, we estimate $\tht_C = (p_0, p_1, \alpha, \beta_0, \beta_1, \beta_2, \rho)^{\top}$ using the proposed pairwise and triplewise likelihood estimators. The corresponding RMSEs for both estimators, along with their relative efficiency, are reported in Table \ref{tab2}.

\begin{table}[ht!]
    \centering
    \begin{tabular}{c|ccccccc}
     & $p_0$ & $p_1$ & $\alpha$ & $\beta_0$ & $\beta_1$ & $\beta_2$ & $\rho$ \\
     \hline
      & 1.78/1.25 &  0.27/0.19 
 & 0.35/0.35 & 0.13/0.10 & 0.09/0.05 & 0.05/0.04 & 0.03/0.02\\
      $d=20$ & 1.83/1.20 & 0.15/0.12 & 0.10/0.06 & 0.11/0.05 & 0.07/0.03 & 0.02/0.01 & 0.03/0.01\\
      & 1.03/0.97 & 0.54/0.63 & 0.28/0.17 & 0.84/0.49 & 0.74/0.57 & 0.53/0.30 & 0.85/0.40 \\
      \hline
      & 1.63/0.92 &  0.17/0.11 
 & 0.25/0.23 & 0.09/0.04 & 0.06/0.04 & 0.03/0.01 & 0.02/0.01\\
     $d = 30$ & 1.48/0.96 & 0.13/0.10 & 0.08/0.05 & 0.08/0.04 & 0.05/0.02 & 0.02/0.01 & 0.02/0.01 \\
     & 0.91/1.05 & 0.78/0.97 & 0.31/0.23 & 0.90/0.79 & 0.81/0.63 & 0.77/0.81 & 0.71/0.63\\
     \hline
      & 1.41/0.73 &  0.21/0.17 
 & 0.08/0.06 & 0.09/0.04 & 0.05/0.02 & 0.02/0.01 & 0.02/0.01\\
    $d = 50$ & 1.30/0.64 & 0.13/0.02 & 0.07/0.04 & 0.09/0.03 & 0.05/0.02 & 0.02/0.01 & 0.01/0.01 \\
    & 0.92/0.88 & 0.61/0.12 & 0.95/0.74 & 0.99/0.90 & 0.97/0.95 & 0.97/0.94 & 0.83/0.98\\
    \end{tabular}
    \caption{RMSEs of the copula parameter estimates of the EV distribution of Submodel 2 obtained using the pairwise likelihood approach (first line), triplewise likelihood approach (second line), and relative efficiency of the two estimators (third line). The true values are $p_0 = 1.5, p_1 = 1$, $\alpha = 1/3$, $\beta_0 = 0.5$, $\beta_1 = 1.5, \beta_2 = -0.5, \rho = 0.8$. The results are based on 500 simulated datasets with $N = 250$/$N = 1000$ component-wise maxima replicates and $d=20, 30, 50$ randomly selected locations in $(0,1)^2$.}
    \label{tab2}
\end{table}

The pairwise likelihood approach produces relatively poor estimates of 
$p_0$, $p_1$, and $\alpha$ in this setting, suggesting that these parameters may be weakly identifiable, particularly for small $d$. The estimation accuracy improves as 
$d$ increases, and the parameters $\beta_0, \beta_1, \beta_2$, which govern skewness, as well as $\rho$, are estimated much more precisely. In contrast, the triplewise likelihood method provides substantially more accurate estimates for all parameters except $p_0$, with notable improvements for $p_1$ and $\alpha$. However, the relative improvement diminishes as 
$d$ becomes larger.

\subsection{Inference for Peaks-over-Thresholds data}

The aim of this section is to introduce an inference methodology for the proposed multivariate Generalized Pareto distributions of Proposition~\ref{prop1c}, Submodels~1 and 2. Since the focus is on the estimation accuracy of the dependence structure, the marginal distributions are assumed to be unit Pareto. \\ 
Let us consider a sample $\boldsymbol{\XA}_{1:n} \equiv \{ \bm \XA_k = (\XA_{k1}, \ldots, \XA_{kd})^{\top}\}_{k=1}^n$ drawn from the original distribution $F$. After transforming the margins to the unit Pareto scale and selecting a high threshold $\bm u = (u_1, \ldots, u_d)$, we retain the subset of samples $\bm \XA_l$ for which there exists at least one $j \in \mathcal{I}$ such that $ \XA_{lj} > u_j$, with $l \in \{1, \ldots, m\}$ and $m\ll n$. The resulting sample, $\boldsymbol{\YA}_{1:m}\equiv \{ \bm \YA_l = (\YA_{l1}, \ldots, \YA_{ld})^{\top}\}_{l=1}^m$ is therefore assumed to be drawn from the distribution $H$.  

\noindent A first approach for performing inference on multivariate Generalized Pareto models is to directly use the known density $h$ derived in Section~\ref{ssec:POT}, yielding the (full) likelihood:
\begin{align}
\label{eq:lik1MGPD}
\mathcal{L}(\bm \YA_{1:m}; \tht_D) = \sum_{l=1}^m \log h(\bm \YA_l; \tht_D),
\end{align}
which enables simpler and more efficient inference than for Block Maxima models, which, as discussed in the previous section, lack a tractable analytic expression for their densities, particularly in high dimensions. This approach is especially useful for models that allow all variables to be simultaneously extreme and place no mass on the boundaries of the support, as in the model derived in Proposition~\ref{prop1c}. \\
A censored likelihood approach is sometimes employed when the multivariate Pareto distribution may not be reliable over the full set of threshold exceedances \citep{Rootzen.etal:2018a}. Since the goal of this section is primarily to demonstrate that model parameters can be accurately recovered, the censored likelihood is not considered here, although in some cases it can help reduce bias \citep{Huser.Davison.Genton.2026}. This is partly motivated by the fact that certain likelihood contributions would otherwise require numerical integration.

\noindent Focusing on the model of Proposition~\ref{prop1c}, we first generate $d=5$ random locations within $(0,1)^2$ and set $\bm s_0$ at the center of the study region, i.e., $(0.5, 0.5)$. We then generate $n = 200{,}000$ observations from the model using the parsimonious parametrization suggested in Remark~3, where the correlations are defined as $\rho_{i,j} = \exp\left\{-(\mathrm{dst}_{i,j}/p_1)^{p_2}\right\}$, with $p_1 = 0.5$, $p_2 = 1.5$ (including for $j=0$), $\alpha_1 = \cdots = \alpha_5 = \alpha_0 = 1.2$, and $c_0 = 0.8$.

For simplicity, we assume $\bm s_0$ is known and maximize the likelihood \eqref{eq:lik1MGPD}, with $h$ given by \eqref{eq:h_prop4}, with respect to $\bm \theta_D = (p_1, p_2, \alpha_0, c_0)$. The estimation procedure is performed for threshold levels corresponding to the $99\%$, $99.5\%$, and $99.9\%$ empirical quantiles and is repeated $500$ times.
\begin{figure}[t!]
    \centering
    \includegraphics[width=0.9\linewidth]{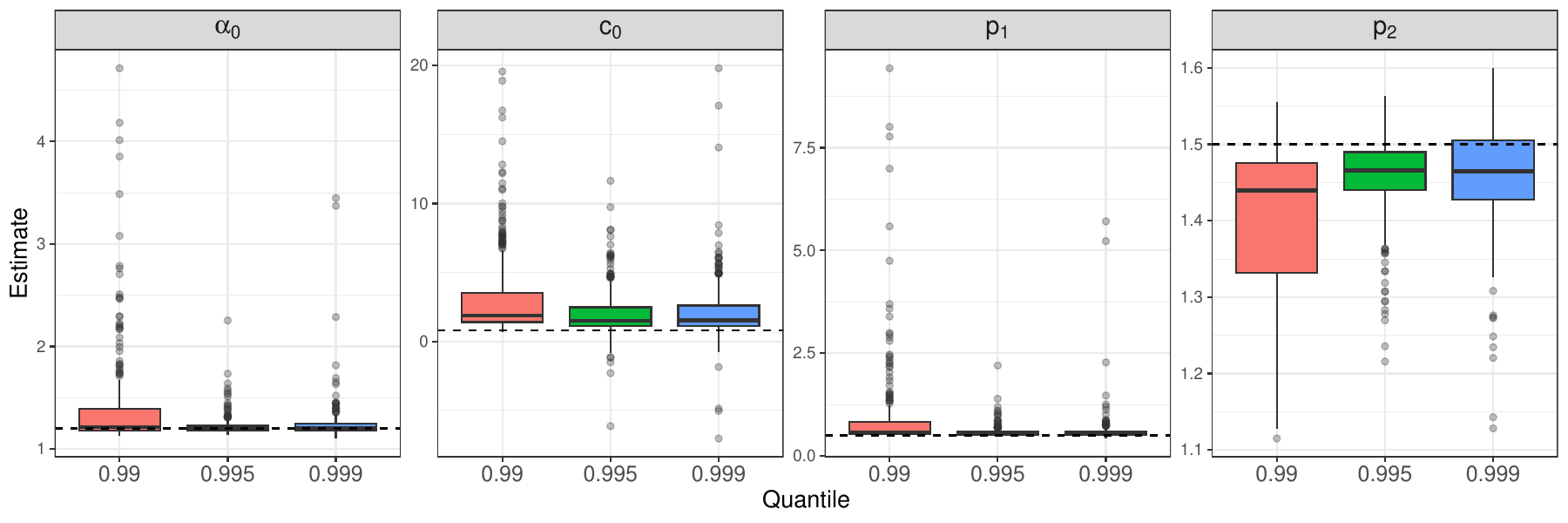} 
    \caption{Boxplots of the parameter estimates $\hat{\alpha}_0, \hat{c}_0, \hat{p}_1$ and $\hat{p}_2$ (left to right) from the Proposition~\ref{prop1c} model with $d=5$ and where $\bm s_0 = (0.5,0.5)$ is assumed fixed and the exceedance thresholds are set at the $99\%, 99.5\%$ and $99.9\%$ empirical quantiles. The true parameter values are indicated by dashed horizontal lines.}
    \label{fig:simProp4}
\end{figure}
Figure~\ref{fig:simProp4} presents boxplots summarizing the distribution of the estimates for each parameter and threshold level. As the threshold increases, the bias in the estimates decreases, which is consistent with the fact that the multivariate generalized Pareto is the limiting distribution of threshold exceedances. Increasing the threshold from the $99.5\%$ to the $99.9\%$ quantile also increases the variability of the estimates, although both thresholds yield approximately unbiased estimates. This suggests that the $99.9\%$ quantile is too high, providing insufficient data to accurately estimate the model parameters. Overall, the $99.5\%$ threshold appears to offer a good balance between bias and variance.

For models such as Submodels 1 and 2, the likelihood \eqref{eq:lik1MGPD} is, however, unsuitable since there is positive mass on the boundaries and omitting it would lead to overestimation of the probability of joint extremes \citep{Thibaud.Opitz.2015}. We therefore adapt the strategy of \citet{Beranger.Padoan.Sisson.2017}, which was originally developed for the unit simplex. Since this approach does not scale well with the dimension $d$, we restrict our attention to the bivariate case ($d=2$). In this context, the density $h(\bm x)$ is defined on the interior $(\bm 0, \bm \infty)^2$, while point masses $H_1({0})$ and $H_2({0})$ are present on the boundaries $\{\bm x \in [\bm 0, \bm \infty)^2: x_1=0 \}$ and $\{\bm x \in [\bm 0, \bm \infty)^2: x_2=0 \}$, respectively. Given that data are never exactly located on the boundary, define
$$
h^\epsilon(x_1, x_2) = \left\{
\begin{array}{ll}
h(x_1, x_2) & \textrm{if } (x_1, x_2) \in (\bm \epsilon, \infty)^2 \\
H_1(\{0\}) & \textrm{if } x_1 < \epsilon \\
H_2(\{0\}) & \textrm{if } x_2 < \epsilon \\
\end{array}
\right.,
$$
for some small $\epsilon$ and the likelihood is then 
\begin{align}
\label{eq:lik2MGPD}
\mathcal{L}^\epsilon(\bm \YA_{1:m}; \tht_D) 
&= \sum_{l=1}^m \log h^\epsilon(\bm \YA_l; \tht_D) \nonumber \\
&= \sum_{l=1}^{m_{12}} \log \left(h(\bm \YA_l^\epsilon; \tht_D)\right) + m_1 H_1(\{0\}; \tht_D) + m_2 H_2(\{0\}; \tht_D),
\end{align}
where $\{\bm \YA_l^\epsilon\}_{l=1}^{m_{12}}$ is the restriction of $\bm \YA_{1:m}$ to $(\bm \epsilon, \infty)^2$ and $m_{12} + m_1 + m_2 = m$.
For a review and comparison of estimation methods for multivariate extremes, we refer the reader to \citet{Huser.Davison.Genton.2026}. We note also that \citet{Goix.Sabourin.ea2017} and \citet{Simpson.Wadsworth.ea2020} have proposed methodologies to estimate extremal mass at the boundaries.

\begin{figure}[tbp]
    \centering
    \includegraphics[width=0.9\linewidth]{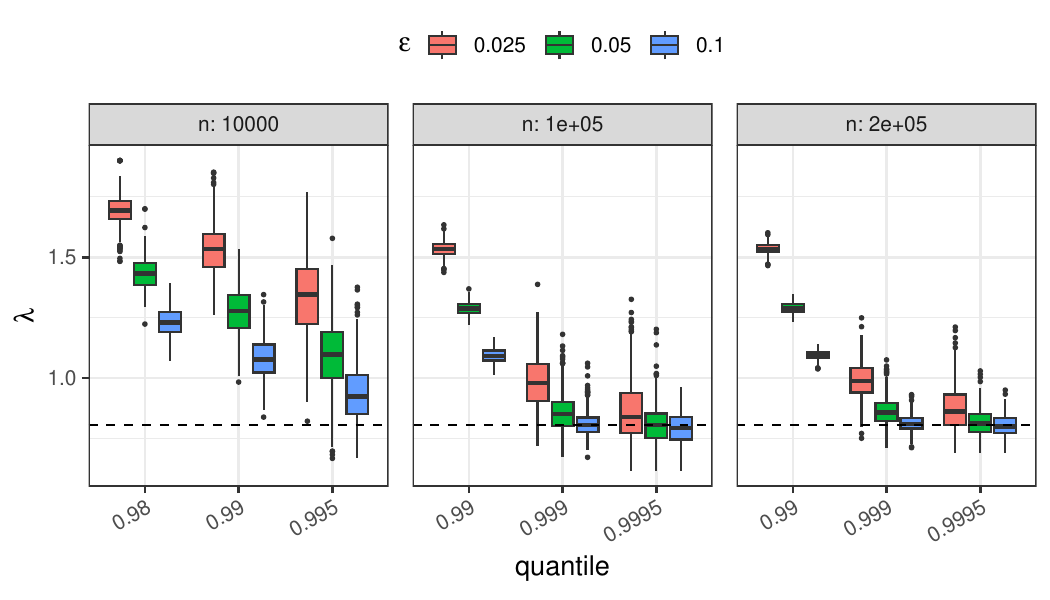} \\
    \includegraphics[width=0.9\linewidth]{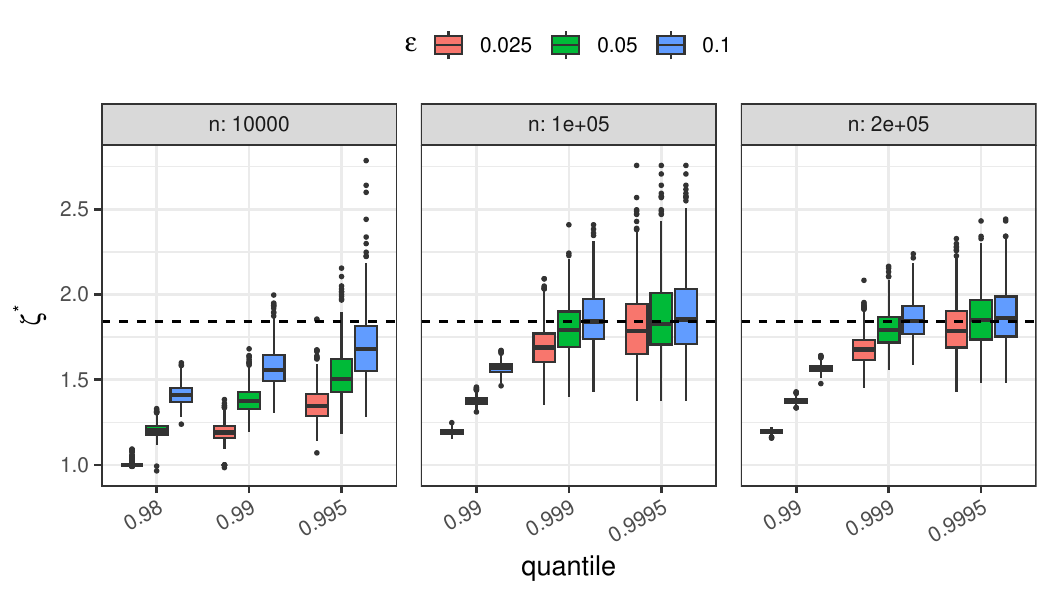}
    \caption{Boxplots of the parameter estimates $\hat{\lambda}$ (top) and $\hat{\zeta}^*$ (bottom) for Submodel~1 in dimension $d=2$, obtained using the likelihood \eqref{eq:lik2MGPD} with $\epsilon=0.025, 0.05$ and $0.1$.  Exceedance thresholds set at the $98\%, 99\%$ and $99.5\%$ empirical quantiles for $n=10000$ (left panels) and at the $99\%, 99.9\%$ and $99.95\%$ quantiles for $n=100000$ and $200000$ (middle and right panels). The true parameter values are indicated by dashed horizontal lines.}
    \label{fig:simS1}
\end{figure}

We now investigate the performance of the inference approach described above, focusing on Submodel~1, where the true parameter values are the same as in Figure~\ref{fig:summryS1}, i.e., $c=(0.8, 1.2)$, $\alpha_0 = 1$, and $\rho_{1,2} = 0.6$. In its current form, the parameters of this bivariate model are not identifiable; we therefore reduce it to a two-parameter model with $\lambda = \sqrt{\alpha_0^2 (1-\rho_{1,2})} \approx 0.8050$ and $\zeta^* = \zeta_1 / \zeta_2 \approx 1.8411$. We generate $n = 10{,}000$, $100{,}000$, and $200{,}000$ observations from $\bm X$ and consider exceedances over the $98\%$, $99\%$, and $99.5\%$ empirical quantiles for $n=10{,}000$, and $99\%$, $99.9\%$, and $99.95\%$ for the larger sample sizes. The model is fitted using \eqref{eq:lik2MGPD} with $\epsilon = 0.025$, $0.05$, and $0.1$, with each experiment repeated $500$ times. Boxplots of the estimates of $\lambda$ and $\zeta^*$ are reported in Figure~\ref{fig:simS1}. \\
It is observed that, as the quantile threshold increases, the multivariate Generalized Pareto model provides a better approximation, resulting in more accurate estimates. Additionally, since the model places mass on one boundary, a reasonably high value of $\epsilon$ (around $0.05$–$0.1$) is necessary to ensure that observations are appropriately allocated to each subspace. For $n = 10{,}000$, the threshold cannot be set too high, as this would result in too few exceedances (on average $m \approx 77$ for the $99.5\%$ quantile), leading to slight bias. For larger sample sizes, a threshold at the $99.9\%$ quantile appears sufficient, as the estimates are stable. 
%

\section{Illustrations on real datasets}
\label{sec-empstudy}

In this section, we demonstrate that asymmetric models offer substantial flexibility gains over symmetric models, such as H\"{u}sler-Reiss, when modeling extremal dependence. In particular, we apply the models introduced in Section~\ref{ssec:EVlimits} to the analysis of two datasets: the first requires a block maxima approach in relatively high dimensions, while the second uses the Peaks-over-Thresholds approach in more moderate dimensions.

\subsection{Summer temperature maxima in Melbourne, Australia}

This dataset comprises $50$ summer temperature maxima in Melbourne, Australia, recorded over the extended summer period from August to April (August 1961 to April 2011) across a study region of 90 locations arranged on a $0.15^{\circ}$ grid in a 9-by-10 layout. The data are available in the \texttt{R} package \texttt{ExtremalDep} \citep{ExtremalDep2025} as the object \texttt{heat}, and an analysis using max-stable processes is provided by \citet{Beranger.Stephenson.Sisson.2019}. \\
In this high-dimensional setting ($d = 90$ spatial locations), we consider four models corresponding to different extreme-value limiting distributions of the random vector $\XX$, as defined in \eqref{main-model}. In all cases, the covariance matrix of the vector $\ZZ$ is specified using the powered-exponential covariance function. The models are defined as follows:
\begin{enumerate}
    \item[1] Submodel 1 from Section~\ref{ssec:tail}, with $c_i = 0$ for $i \in \II$, corresponding to the H\"usler-Reiss distribution.
    \item[2] Submodel 1 from Section \ref{ssec:tail}, with $c_i = \beta_0 + \beta_1 \ss_{i1} + \beta_2 \ss_{i2}$, $i \in \II$, where $\{(\ss_{i1}, \ss_{i2})\}_{i=1}^d$ denote the spatial coordinates of the $d=90$ locations. 
    \item[3] Submodel 2 from Section \ref{ssec:tail}, with $c_i = \beta_0 + \beta_1 \ss_{i1} + \beta_2 \ss_{i2}$, $i \in \II$, where $\{(\ss_{i1}, \ss_{i2})\}_{i=1}^d$ are the coordinates of the $d=90$ locations.
    \item[4] Model from Proposition \ref{prop1c}, in which the covariance matrix of $(Z_0^*, Z_1, \ldots, Z_d)^{\top}$ is also determined by the powered-exponential covariance function.  
\end{enumerate}

\begin{figure}[h!]
    \centering
    \includegraphics[width=0.32\linewidth]{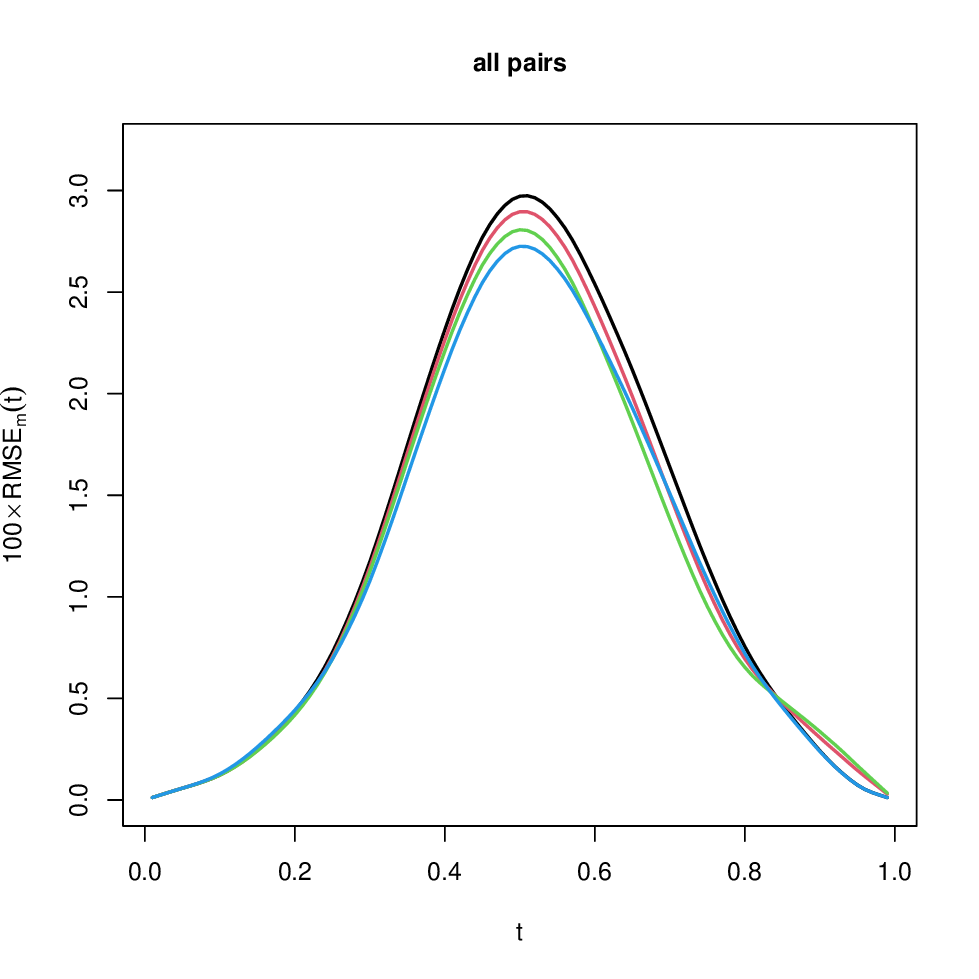} 
    \includegraphics[width=0.32\linewidth]{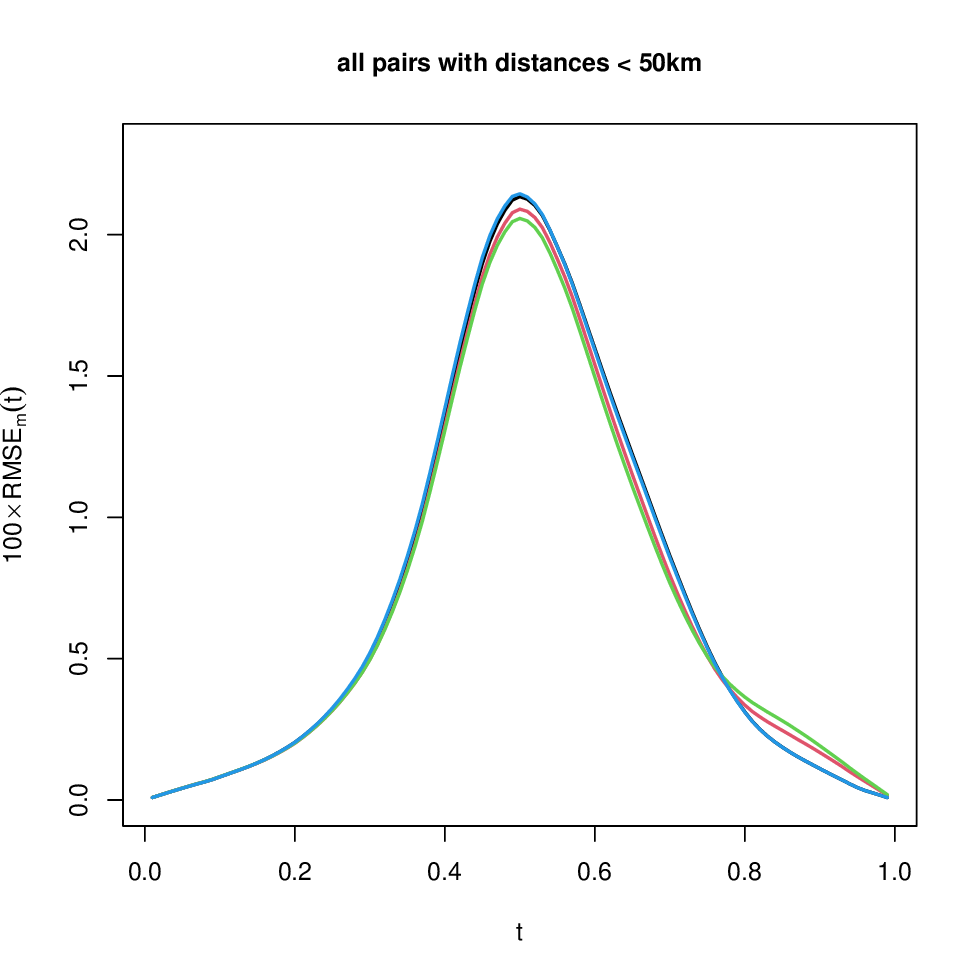} 
    \includegraphics[width=0.32\linewidth]{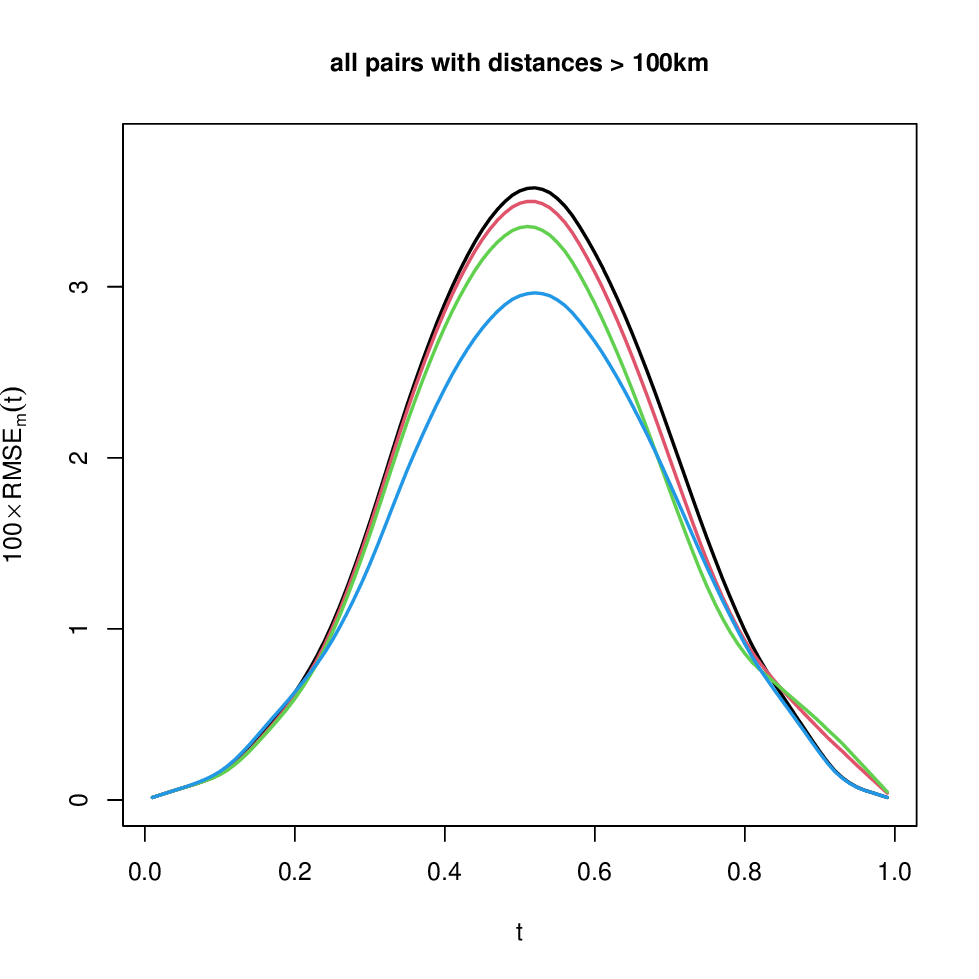} 
    \caption{$\mathrm{RMSE}_m(t)$ values for $t \in (0,1)$ for models 1-4 (black, red, green, and blue lines, respectively), computed for all pairs of variables (left), for pairs separated by less than 50 km (middle), and for pairs separated by more than 100 km (right). }
    \label{fig:GoF}
\end{figure}

Following insights drawn from the simulation studies in Section~\ref{ssec:inferenceBM}, these models are fitted using the pairwise likelihood estimator since no substantial loss of efficiency is aniticipated when $d=90$. The goodness of fit of model $m \in \{1,2,3,4\}$ is evaluated by computing the RMSE defined as
$$
\text{RMSE}_m(t) = \sqrt{\frac{1}{|S|}\sum_{(i,j) \in S}\{\widehat A_{X_i,X_j}(t) - \widehat A_{X_i,X_j}^m(t)\}^2},
$$
where $\widehat A_{X_i,X_j}^m(t)$ represents model-based estimates of the Pickands dependence function $A_{X_i, X_j}(t)$ for each pair of locations $\ss_i$ and $\ss_j$ and $\widehat A_{X_i,X_j}^{\text{emp}}(t)$ is its corresponding nonparametric estimate. \\
Figure~\ref{fig:GoF} shows the RMSE values for different sets of pairs, $S$.  
\noindent While all models perform similarly at short distances, Models 2–4 --- whose tail-dependence functions are permutation-asymmetric --- achieve smaller RMSEs than Model 1 at larger distances, indicating nonstationarity in the data. We also compute aggregated RMSEs defined by
$$
\mathrm{RMSE}_m = \int_0^1 \mathrm{RMSE}_m(t) dt
$$
as well as $\mathrm{RMSE}_m(0.5)$, which reflects the accuracy of the estimated upper tail-dependence coefficients. Table \ref{tab3} reports the RMSE values for the different sets of pairs, $S$.

\begin{table}[ht!]
    \centering
    \begin{tabular}{c|ccc}
     \multirow{2}{*}{Model} &  \multicolumn{3}{c}{$100 \times\mathrm{RMSE}_m$/$100 \times \mathrm{RMSE}_m(0.5)$} \\
     & (all pairs) & ($\mathrm{dst}<$ 50km) & ($\mathrm{dst}>$ 100km) \\
     \hline
     $m=1$ & 1.24/2.97 &  0.73/2.14 & 1.57/3.56 \\
     $m=2$ & 1.20/2.90 &  0.71/2.09 & 1.53/3.49 \\
     $m=3$ & 1.15/2.81 &  0.71/2.06 & 1.46/3.35 \\
     $m=4$ & 1.14/2.73 &  0.73/2.15 & 1.34/2.95 
     \end{tabular}
    \caption{$100\cdot\text{RMSE}_m/100 \cdot \mathrm{RMSE}_m(0.5)$ values for Models 1-4, computed for all pairs of variables (left), for pairs separated by less than 50 km (middle), and for pairs separated by more than 100 km (right).}
    \label{tab3}
\end{table}

Again, all models perform similarly at short distances, while Models 2–4 achieve smaller RMSEs than Model 1 at larger distances. Among these, Model 4 attains the lowest RMSEs, showing a 17–21\% reduction relative to Model 1 at larger distances, indicating a better overall fit to the data.

\subsection{UK Bank Returns}

This second illustration focuses on the analysis of weekly negative raw returns for stocks of four major UK banks (HSBC, Lloyds, RBS, and Barclays) recorded between 29 October 2007 and 17 October 2016, yielding $n=470$ observations. A multivariate Peaks-over-Thresholds approach was applied to this dataset by \citet{Kirilouk2019}, and we adopt their proposed marginal threshold selection at the $83\%$ quantile, leaving 149 observations for which at least one component exceeds the threshold. \\
Since our primary interest lies in modeling extremal dependence, the margins are standardized to unit Pareto, and we consider both the H\"usler-Reiss model and its asymmetric counterpart as defined in Proposition~\ref{prop1c}. As demonstrated earlier, these models are fully defined in the interior of the support, which allows inference to be conducted directly using \eqref{eq:lik1MGPD}, a simpler and more elegant approach than using \eqref{eq:lik2MGPD} when $d=4$. \\
For the model in Proposition~\ref{prop1c}, we assume $\alpha_1 = \cdots = \alpha_d = \alpha_0$, which leads to the following parametrization: $\lambda_{ij} = \alpha_0 \left( 2(1-\rho_{ij})\right)^{1/2}$ and $\tau_{j} = \alpha_0 \rho_{j,0} - c_0$. Thus, we estimate the following 10-dimensional parameter vector: $$\bm \theta_D = \left( \lambda_{12}, \lambda_{13}, \lambda_{14}, \lambda_{23}, \lambda_{24}, \lambda_{34}, \tau_1, \tau_2, \tau_3, \tau_4 \right)^{\top},$$ while the H\"usler-Reiss model is recovered by setting $\tau_1 = \tau_2 = \tau_3 = \tau_4 = 0$. 
\begin{table}[ht!]
    \centering
    \begin{tabular}{c|cccccccccc}
      Model & $\lambda_{12}$ & $\lambda_{13}$ & $\lambda_{14}$ & $\lambda_{23}$ & $\lambda_{24}$ & $\lambda_{34}$ & $\tau_1$ & $\tau_2$ & $\tau_3$ & $\tau_4$ \\
      \hline
      H\"{u}sler-Reiss & $1.10$ & $1.04$ & $1.03$ & $0.94$ & $0.96$ & $0.96$ & $-$ & $-$ & $-$ & $-$ \\
      & $(0.05)$ & $(0.04)$ & $(0.04)$ & $(0.05)$ & $(0.04)$ & $(0.05)$ & $-$ & $-$ & $-$ & $-$ \\
      Prop.~\ref{prop1c} & $1.34$ & $1.07$ & $1.10$ & $1.39$ & $1.06$ & $1.10$ & $0.30$ & $-0.59$ & $0.66$ & $-0.12$\\
      & $(0.14)$ & $(0.05)$ & $(0.12)$ & $(0.17)$ & $(0.08)$ & $(0.16)$ & $(0.47)$ & $(0.60)$ & $(0.47)$ & $(0.61)$ \\
     \hline
    \end{tabular}
    \caption{Parameter estimates, along with parametric bootstrap estimates of their standard deviations, for the Hüsler–Reiss and Proposition~\ref{prop1c} models fitted to the UK Bank dataset. The data were transformed to unit Pareto margins, with thresholds set at the 83\% marginal quantiles.}
    \label{tab_UK_Bank}
\end{table}
\noindent Table~\ref{tab_UK_Bank} presents the estimated parameters for both models, along with bootstrap estimates of their standard deviations. The $\lambda_{ij}$ estimates are broadly similar across models, whereas Model~\ref{prop1c} shows larger bootstrap standard deviations, as expected due to the additional parameters being estimated. \\
Although the individual confidence intervals for each $\tau_j$ appear to include zero, both the AIC and BIC are substantially lower when allowing for asymmetry ($1722$ vs. $1825$ for AIC and $1752$ vs. $1843$ for BIC), indicating a better fit for the Proposition~\ref{prop1c} model. Furthermore, since the models are nested, this conclusion is supported by a likelihood ratio test of the null hypothesis $\tau_1 = \tau_2 = \tau_3 = \tau_4 = 0$, which yields a $p$-value less than $0.001$.

\section{Conclusion}
\label{sec:conclusion}

In this paper, we introduced a new class of additive factor models, derived their corresponding extreme-value limiting distributions, and studied their tail behavior. By extending the common factor model of \citet{Krupskii.Huser.ea2016}, we were able to introduce permutation asymmetry in the dependence structure of extremes across multiple random variables, thereby increasing modeling flexibility compared to the popular (symmetric) H\"usler–Reiss model, which can be recovered as a special case. Extremes are modeled from both the Block Maxima and Peaks-over-Thresholds perspectives, making the approach applicable to a wide range of multivariate extremes problems. Several sub-models are derived from the proposed family --- some capturing joint extremes of the random vector of interest, and others not --- illustrating their modeling capabilities. All considered models share the appealing property of being able to accommodate non-stationarity. \\
%
Simulation experiments are included to demonstrate that the model parameters can be reliably recovered, while two analyses of real datasets ---consisting of Block Maxima and Peaks-over-Thresholds data --- showcase an improved quality of fit compared to the H\"usler–Reiss model. Although the primary focus of this manuscript is on multivariate extremes, some illustrations are also provided in the context of spatial data. 
Many of the proposed models assign positive mass to some or all of the boundaries of their support, which makes inference challenging in moderate to high dimensions. A direction for future research would therefore be to develop a class of parsimonious factor models that, by construction, are fully characterized within the interior of their support. In addition, multi-factor models can better capture dependence over large spatial domains and therefore extensions of the proposed factor models coudl also be a topic of future research.

\section*{Acknowledgements}

Boris B\'eranger was supported by the Australian Research Council (ARC) through the Discovery Project Scheme (DP220103269).

\appendix
\renewcommand{\thesection}{\Alph{section}}
\setcounter{section}{0}

\section{Appendix}
\label{sec:appx}

\subsection{Proof of Proposition \ref{prop0}}
\label{appx-prop0}

Consider a vector $\ww = (w_1, \ldots, w_d)^{\top}$ with $w_i = \alpha_i^{-1}F_0^{-1}\left\{1-x_i/(n\zeta_i)\right\}$ for $i=1,\ldots, d$. We find:
\begin{align*}
\Pr(\XX < \ww) =&  \sum_{\AA \in \II,\BB = \AA^c} \Pr\{\ZZ_{\AA} + (1/\aa_{\AA})\cdot V_0 < \ww_{\AA}, \ZZ_{\BB} < \ww_{\BB}, \ZZ^*_{\AA} > \cc_{\AA}, \ZZ^*_{\BB} < \cc_{\BB}\} \\
=& \sum_{\AA \in \II,\BB = \AA^c} \Pr\{\ZZ_{\AA} + (1/\aa_{\AA})\cdot V_0 < \ww_{\AA}, \ZZ^*_{\AA} > \cc_{\AA}, \ZZ^*_{\BB} < \cc_{\BB}\} + o(1/n).
\end{align*}

Denote $i^* = \mathrm{argmin}_{i\in \AA} (w_i\alpha_i)$. Note that
\begin{align*}
    P_{\AA} &:= \Pr\{\ZZ_{\AA} + (1/\aa_{\AA})\cdot V_0 < \ww_{\AA}, \ZZ^*_{\AA} > \cc_{\AA}, \ZZ^*_{\BB} < \cc_{\BB}\}\\
    &= \int_x \Pr(\ZZ_{\AA}  < \ww_{\AA} - y/\aa_{\AA}, \ZZ^*_{\AA} > \cc_{\AA}, \ZZ^*_{\BB} < \cc_{\BB}) dF_{0}(y) = P_{\AA,1} + P_{\AA,2} + P_{\AA,3},
\end{align*}
where, with $w_{L,n} = w_{i^*}\alpha_{i^*} - (\ln n)^{\alpha_0}$ and $w_{U,n} = w_{i^*}\alpha_{i^*} + (\ln n)^{\alpha_0}$, 
\begin{align*}
    P_{\AA,1} &= \int_{y < w_{L,n}} \Pr(\ZZ_{\AA}  < \ww_{\AA} - y/\aa_{\AA}, \ZZ^*_{\AA} > \cc_{\AA}, \ZZ^*_{\BB} < \cc_{\BB})dF_0(y)\\
    &= \int_{y < w_{L,n}} \Pr(\ZZ^*_{\AA} > \cc_{\AA}, \ZZ^*_{\BB} < \cc_{\BB})dF_0(y) - \upsilon_n\\
    &= F_0(w_{i^*}\alpha_{i^*}) \cdot\Pr(\ZZ^*_{\AA} > \cc_{\AA}, \ZZ^*_{\BB} < \cc_{\BB}) - \upsilon_n + o(1/n),\\
    P_{\AA,2} &= \int_{w_{L,n}}^{w_{U,n}} \Pr(\ZZ_{\AA}  < \ww_{\AA} - y/\aa_{\AA}, \ZZ^*_{\AA} > \cc_{\AA}, \ZZ^*_{\BB} < \cc_{\BB})dF_0(y)\\
    & < F_0(w_{U,n}) - F_0(w_{L,n}) = o(1/n),\\
    P_{\AA,3} &:= \int_{y > w_{U,n}}\Pr(\ZZ_{\AA} < \ww_{\AA} - y/\alpha_{\AA}, \ZZ^*_{\AA} > \cc_{\AA}, \ZZ^*_{\BB} < \cc_{\BB}) dF_0(y)\\
    & < \Phi\{-(\ln n)^{\alpha_0}/\alpha_{i^*}\}\bar F_0(w_{U,n})= o(1/n).
\end{align*}

Here, 
\begin{align*}
    \upsilon_n &= \int_{y < w_{L,n}} \Pr(\ZZ_{\AA}  > \ww_{\AA} - y/\aa_{\AA}, \ZZ^*_{\AA} > \cc_{\AA}, \ZZ^*_{\BB} < \cc_{\BB})dF_0(y)\\
    & < \int_{y < w_{L,n}} \Phi(y/\alpha_{i^*} - w_{i^*})dF_0(y) = \upsilon_{n,1}+ \upsilon_{n,2},
\end{align*}
where, using the integration by parts formula, we get:
\begin{align*}
\upsilon_{n,1} &= -\{1-F_0(w_{L,n})\}\Phi\left\{-(\ln n )^{\alpha_0}\right\} = o(1/n), \\
\upsilon_{n,2} & = \alpha_{i^*}^{-1} \int_{y<w_{L,n}} \{1 - F_0(y)\}\phi(y/\alpha_{i^*} - w_{i^*})dy
     = \int_{y < -(\ln n)^{\alpha_0}} \left\{1 - F_0(z \alpha_{i^*} + w_{i^*}\alpha_{i^*})\right\}\phi(z) dz\\
     & = \int_{ -(\ln n)^{\alpha^*}}^{ -(\ln n)^{\alpha_0}} \left\{1 - F_0(z \alpha_{i^*} + w_{i^*}\alpha_{i^*})\right\}\phi(z) dz + o(1/n)\\
     & \leq \frac{x_{i^*}}{n\zeta_{i^*}}\int_{ -(\ln n)^{\alpha^*}}^{ -(\ln n)^{\alpha_0}} \exp(\gamma z\alpha_{i^*})\phi(z) dz + o(1/n)\\
     & = \frac{x_{i^*}}{n\zeta_{i^*}} \exp(0.5\gamma^2\alpha_{i^*}^2)\left[\Phi\left\{-(\ln n)^{\alpha_0} - \gamma\alpha_{i^*}\right\} - \Phi\left\{-(\ln n)^{\alpha^*} - \gamma\alpha_{i^*}\right\}\right] + o(1/n) = o(1/n)\,,
\end{align*}
where $\alpha^* \in (0.5, 1)$ is a constant.
It follows that
\begin{align*}
\Pr(\XX < \ww) &= \sum_{\AA \in \II,\BB = \AA^c} F_0\{\min_{i\in\AA}(w_{i}\alpha_{i})\} \cdot\Pr(\ZZ^*_{\AA} > \cc_{\AA}, \ZZ^*_{\BB} < \cc_{\BB}) + o(1/n)\\
& = \left\{1 - \max_{i \in \AA} \frac{x_i}{n\zeta_i}\right\}\cdot\Pr(\ZZ^*_{\AA} > \cc_{\AA}, \ZZ^*_{\BB} < \cc_{\BB}) + o(1/n).
\end{align*}

In particular, $$\Pr(X_i < w_i) = \{1-x_i/(n\zeta_i)\}\Pr(Z_i^*  > c_i) + \Pr(Z_i^* < c_i) + o(1/n) = 1 - x_i/n + o(1/n),$$
 and
$$\ell_{\XX}(\xx) = \lim_{n \to \infty} n\{1 - \Pr(\XX < \ww)\} = \sum_{\AA \in \II, \BB = \AA^c} \left(\max_{i \in \AA}\frac{x_i}{\zeta_i}\right) \Pr(\ZZ^*_{\AA} > \cc_{\AA}, \ZZ^*_{\BB} < \cc_{\BB}).
$$
The proof is now complete. \hfill $\Box$

\subsection{Proof of Proposition \ref{prop1a}}
\label{appx-prop1a}

Let $F_i(z)$ be the cdf of $X_i$ as before. We can write
\begin{align*}
    F_i(z) &= \Pr(X_i < z) = \Pr(X_i < z, Z_i^* > c_i) + \Pr(X_i < z, Z_i^* < c_i)\\
    &= \int_{\mathbb{R}^2}\Pr\left\{V_0 < \alpha_l(z-Z_i), Z_i^* > c_i| Z_i = z_i, Z_i^* = z_i^*\right\}\phi_{\rho_{i,i^*}}(z_i,z_i^*)dz_idz_i^* + \Pr(Z_i < z, Z_i^* < c_i)\\
    &= \int_{c_i}^{\infty}\int_{-\infty}^z \left[1 - \exp\left\{-\alpha_i (z - z_i)\right\} \right]\phi_{\rho_{i,i^*}}(z_i, z_i^*)dz_i dz_i^* + \Phi_{\rho_{i,i^*}}(z,c_i)\\
    &= \Phi(z) - \exp(0.5\alpha_i^2 - \alpha_i z) \Phi_{-\rho_{i,i^*}}(\alpha_i\rho_{i,i^*} - c_i, z - \alpha_i).
\end{align*}

As $z \to \infty$, we have $\Phi(z) \to 1$ and $\Phi_{-\rho_{i,i^*}}(\alpha_i\rho_{i,i^*} - c_i, z - \alpha_i) \to \Phi(\alpha_i\rho_{i,i^*} - c_i)$. Solving 
$$
1 - \exp(0.5\alpha_i^2 - \alpha_i z) \Phi(\alpha_i\rho_{i,i^*} - c_i) = 1 - x/n,
$$
we obtain $z = z_i(x/n) = 0.5\alpha_i + \alpha_i^{-1} \ln (n/x) + \alpha_i^{-1}\ln \Phi(\alpha_i\rho_{i,i^*} - c_i) = 0.5\alpha_i + \alpha_i^{-1} \ln (n\zeta_i/x)$ since $\rho_{i,i^*} = 0$. It is easy to see that $F_i(z_i(x/n)) = 1 - x/n + o(1/n)$ as $n \to \infty$.

Now let $\tilde F_i$ be the cdf of $\tilde X_i = Z_i + (1/\alpha_i)\cdot V_0$, and let $w_i = z_i(x_i/n)$, then $\tilde F_i(w_i) = 1 - x_i/(\zeta_i n) + o(1/n)$. It follows that
\begin{align*}
\Pr(\XX < \ww) =&  \sum_{\AA \in \II,\BB = \AA^c} \Pr\{\ZZ_{\AA} + (1/\aa_{\AA})\cdot V_0 < \ww_{\AA}, \ZZ_{\BB} < \ww_{\BB}, \ZZ^*_{\AA} > \cc_{\AA}, \ZZ^*_{\BB} < \cc_{\BB}\} \\
=& \sum_{\AA \in \II,\BB = \AA^c} \Pr\{\tilde\XX_{\AA} = \ZZ_{\AA} + (1/\aa_{\AA})\cdot V_0 < \ww_{\AA}\}\cdot \Pr( \ZZ^*_{\AA} > \cc_{\AA}, \ZZ^*_{\BB} < \cc_{\BB}) + o(1/n),
\end{align*}
and
$$
n\{1 - \Pr(\XX < \ww)\} = \sum_{\AA \in \II,\BB = \AA^c} n\{1 - \Pr(\tilde\XX_{\AA} < \ww_{\AA})\}\cdot \Pr( \ZZ^*_{\AA} > \cc_{\AA}, \ZZ^*_{\BB} < \cc_{\BB}) + o(1).
$$
To conclude the proof, we notice that $n\{1 - \Pr(\tilde\XX_{\AA} < \ww_{\AA})\} \to \ell_{\AA}(\tilde \xx_{\AA})$ \citep{Krupskii.Joe.ea2018}. \hfill $\Box$

\subsection{Proof of Proposition \ref{prop1b}}
\label{appx-prop1b}

Without loss of generality, consider $i = 1$ and $j = 2$. Let $F_{12}$ be the joint cdf of $(X_1, X_2)^{\top}$ as defined in \eqref{main-model}, and $F_1$, $F_2$ be the respective marginal cdfs. From the proof of Proposition \ref{prop1a}, if we let $ z_l(x/n) = 0.5\alpha_l + \alpha_l^{-1} \ln (n/x) + \alpha_l^{-1}\ln \Phi(\alpha_l\rho_{l,l^*} - c_l)$, then $F_l(z_l(x/n)) = 1 - x/n + o(1/n)$ as $n \to \infty$.

%

Denote $w_l = z_l(x_l/n)$ and $\zz = (z_1, z_2, z_1^*, z_2^*)^{\top}$. Let $\Phi_{i.j,k^*}$ be the cdf of the vector $(Z_i,Z_j,Z_k^*)^{\top}$ and $\Phi_{i,j,k^*,l^*}$, $\phi_{i,j,k^*,l^*}$ be the cdf (pdf, respectively) of the vector $(Z_i, Z_j, Z_k^*, Z_l^*)^{\top}$. We find:
$$
F_{12}(w_1,w_2) = \Pr(X_1 < w_1, X_2 < w_2) = A_{12} + A_1 + A_2 + A_0,
$$
where
\begin{align*}
    A_{12} &:= \Pr(X_1 < w_1, X_2 < w_2, Z_1^* > c_1, Z_2^* > c_2)\\ &= \int_{S_{12}}\left(1 - \exp\left[\max\left\{\alpha_1(z_1-w_1), \alpha_2(z_2 - w_2)\right\}\right]\right)\phi_{1,2,1^*,2^*}(\zz)d\zz\\
    & = \Phi_{\rho_{1,2}}(w_1,w_2) - \Phi_{1,2,1^*}(w_1,w_2,c_1) - \Phi_{1,2,2^*}(w_1,w_2,c_2) + \Phi_{1,2,1^*,2^*}(w_1,w_2,c_1,c_2) \\ 
    &\ \ \ - \int_{S_{1|2}}\exp\left\{\alpha_1(z_1-w_1)\right\}\phi_{1,2,1^*,2^*}(\zz) d\zz - \int_{S_{2|1}}\exp\left\{\alpha_2(z_2-w_2)\right\}\phi_{1,2,1^*,2^*}(\zz) d\zz,
\end{align*}
\begin{align*}
    A_1 &:= \Pr(X_1 < w_1, X_2 < w_2, Z_1^* > c_1, Z_2^* < c_2)= \int_{S_{1}}\left[1 - \exp\left\{\alpha_1(z_1-w_1) \right\}\right]\phi_{1,2,1^*,2^*}(\zz)d\zz \\&= \Phi_{1,2,2^*}(w_1,w_2,c_2) - \Phi_{1,2,1^*,2^*}(w_1,w_2,c_1,c_2) - \int_{S_{1}}\exp\left\{\alpha_1(z_1-w_1) \right\}\phi_{1,2,1^*,2^*}(\zz)d\zz,\\
    A_2 &:= \Pr(X_1 < w_1, X_2 < w_2, Z_1^* < c_1, Z_2^* > c_2)= \int_{S_{2}}\left[1 - \exp\left\{\alpha_2(z_2-w_2) \right\}\right]\phi_{1,2,1^*,2^*}(\zz)d\zz \\&= \Phi_{1,2,1^*}(w_1,w_2,c_1) - \Phi_{1,2,1^*,2^*}(w_1,w_2,c_1,c_2) - \int_{S_{1}}\exp\left\{\alpha_1(z_1-w_1) \right\}\phi_{1,2,1^*,2^*}(\zz)d\zz,\\
    A_0 &:= \Pr(X_1 < w_1, X_2 < w_2, Z_1^* < c_1, Z_2^* < c_2) = \Phi_{1,2,1^*,2^*}(w_1,w_2,c_1,c_2),
\end{align*}
and $S_{12} = \{\zz: z_1 < w_1, z_2 < w_2, z_1^* > c_1, z_2^* > c_2\}$, $S_l = \{\zz: z_1 < w_1, z_2 < w_2, z_l^* > c_l, z_{-l}^* < c_{-l}\}$. Here, $S_{l|-l} = \{\zz \in S_{1,2}: \alpha_l(z_l - w_l) > \alpha_{-l}(z_{-l} - w_{-l})\} = \{\zz \in S_{1,2}: z_l > \alpha_{-l}z_{-l}/\alpha_l + \xi_{l,-l}\}$, where $$\xi_{l,-l} = \alpha_l^{-1}\ln (x_{-l}/x_l) + 0.5\alpha_l - 0.5\alpha_{-l}^2/\alpha_l + \alpha_l^{-1} \ln \left\{\zeta_l/\zeta_{-l}\right\}\,,$$
with $\zeta_l = \Phi(\alpha_l\rho_{l,l^*} - c_l)$, $l \in \{1,2\}$. It implies that 
\begin{align*}
1 - C_{X_1,X_2}(1-x_1/n, 1-x_2/n)& = \exp(-\alpha_1w_1)\int_{S_1^* \cup S_{1|2}^*}\exp(\alpha_1z_1)\phi_{1,2,1^*,2^*}(\zz)d\zz\\
&\ \ \ + \exp(-\alpha_2w_2)\int_{S_2^* \cup S_{2|1}^*}\exp(\alpha_2z_2)\phi_{1,2,1^*,2^*}(\zz)d\zz
 + o(1/n),
 \end{align*}
where $S_{l}^* = \{\zz: z_l^* > c_l,\  z_{-l}^* < c_{-l}\}$, $S_{l|-l}^* = \{\zz:  z_{l} > \alpha_{-l} z_{-l}/\alpha_{l} + \xi_{l,-l}, \ z_1^* > c_1, \ z_2^* > c_2\}$.

Note that $\exp(-\alpha_l w_l) = \frac{x_l}{n\zeta_l} \cdot \exp(-0.5\alpha_l^2)$, and
\begin{align*}
    I_l &:=\int_{S_l^* \cup S_{l|-l}^*}\exp(\alpha_lz_l)\phi_{1,2,1^*,2^*}(\zz)d\zz\\ &= \exp(0.5\alpha_l^2) \left\{\zeta_l - \Phi_{\SS_{l|-l}}\left(-\frac{\lambda_{12}}{2} - \frac{1}{\lambda_{12}}\ln \frac{\tilde x_l}{\tilde x_{-l}}, \alpha_l\rho_{-l^*,l} - c_{-l}, \alpha_l\rho_{l^*,l} - c_l \right)\right\}\,.
\end{align*}
The result of proposition now follows by taking the limit: $$\hspace{3.6cm} \ell_{X_1,X_2}(x_1,x_2) = \lim_{n \to \infty} n \left\{1 - C_{X_1,X_2}\left(1 - \frac{x_1}{n}, 1 - \frac{x_2}{n}\right)\right\}\,. \hspace{3.6cm} \Box$$ 

\subsection{Proof of Proposition \ref{prop1c}}
\label{appx-prop1c}

Consider a vector $\ww = (w_1,\ldots,w_d)^{\top}$ with $w_l = w_l(n) =  0.5\alpha_l + \alpha_l^{-1}\ln (n/x_l) + \alpha_l^{-1}\ln \Phi(\alpha_l \rho_{l.0} - c_0)$ for $l = 1,\ldots, d$, then from the proof of Proposition \ref{prop1a}, $F_l(w_l) = 1 - x_l/n + o(1/n)$, where $F_l$ is the cdf of $X_l$ as before. We can write: 
$$
\Pr(\XX < \ww) = I_1(\ww) + I_2(\ww),
$$
where 
$$
I_1(\ww) = \Pr(\XX < \ww, Z_0^* < c_0) = \Pr(\ZZ < \ww, Z_0^* < c_0) = \Phi(c_0) + o(1/n),
$$
and, using the integration by parts formula, we find:
\begin{align*}
I_2(\ww) &= \Pr(\XX < \ww, Z_0^* >c_0) = \Pr(\ZZ + (1/\aa)\cdot V_0 < \ww, - Z_0^* < -c_0)\\
&= \int_0^{\infty} \Phi_{d+1}\left(\left(\ww - x/\aa, -c_0\right)^{\top}; \Sigma_0\right)\exp(-x)dx\\
&= \Phi_{d+1}\left((\ww,-c_0)^{\top};\Sigma_0\right) - \sum_{j=1}^dI_{3,j}(\ww) = \Phi(-c_0) - \sum_{j=1}^dI_{3,j}(\ww) + o(1/n),
\end{align*}
where $\Sigma_0$ is the correlation matrix of $(\ZZ^{\top}, -Z_0^*)^{\top}$. By changing the variables, we get:
\begin{align*}
I_{3,j}(\ww) =&\ (1/\alpha_j)\int_0^{\infty}\Phi_{d}\left(\left[\left\{\frac{w_i - x/\alpha_i - \rho_{ij}(w_j - x/\alpha_j)}{\tilde\rho_{ij}}\right\}_{i \in \II_{-j}}, \frac{-c_0 - \rho_{j,0}(w_j - x/\alpha_j)}{\tilde \rho_{j,0}}\right]^{\top}; \Sigma_{0|j}  \right)\\
& \times\phi(w_j - x/\alpha_j)\exp(-x) dx\\
= & \ \frac{\tilde x_j}{n} \int_{-\infty}^{w_j-\alpha_j}\Phi_{d}\left(\left[\left\{\frac{\varrho_{ij} + (\alpha_j/\alpha_i - \rho_{ij})z}{\tilde\rho_{ij}}\right\}_{i \in \II_{-j}}, \frac{-c_0 + \rho_{j,0}\alpha_j + \rho_{j,0}z}{\tilde \rho_{j,0}}\right]^{\top}; \Sigma_{0|j}  \right) \phi(z) dz\\
=&\ \frac{\tilde x_j}{n} \Phi_d \left[\left\{\left(\frac{\lambda_{ij}}{2} + \frac{1}{\lambda_{ij}}\ln \frac{\tilde x_j}{\tilde x_i}\right)_{i \in \II_{-j}};\alpha_j\rho_{j,0} - c_0\right\}^{\top}; \tilde\Sigma^{(j)}\right]  + o(1/n).
\end{align*}
Here, $\Sigma_{0|j}$ is the partial correlation matrix of $(\ZZ^{\top}, -Z_0^*)^{\top}$ conditional on $Z_j$, $\tilde \rho_{ij} = (1-\rho_{ij}^2)^{1/2}$, $\tilde \rho_{j,0} = (1-\rho_{j,0}^2)^{1/2}$, and $\varrho_{ij} = (0.5/\alpha_i)\cdot\lambda_{ij}^2 + (1/\alpha_i)\cdot \ln (\tilde k_j/\tilde k_i)$. The stable tail dependence function is obtained by taking the limit:
\begin{align*}
\ell_{\XX}(\xx) 
&= \lim_{n \to \infty} n \left\{1 - \Pr(\XX < \ww)\right\} \\
&= \sum_{j = 1}^d \tilde x_j \Phi_d \left[\left\{\left(\frac{\lambda_{ij}}{2} + \frac{1}{\lambda_{ij}}\log \frac{\tilde x_j}{\tilde x_i}\right)_{i \in \II_{-j}};\alpha_j\rho_{j,0} - c_0\right\}^{\top}; \tilde\Sigma^{(j)}\right],
\end{align*}
and the result of proposition now follows by recognizing the cdf of the extended Skew-Normal distribution \citet[see][]{Arellano.Genton.2010}.\hfill $\Box$

\subsection{Illustrations of the Marshall-Olkin distribution}
\label{appx-MO}

\begin{figure}[t]
    \centering
    \includegraphics[width=0.32\linewidth]{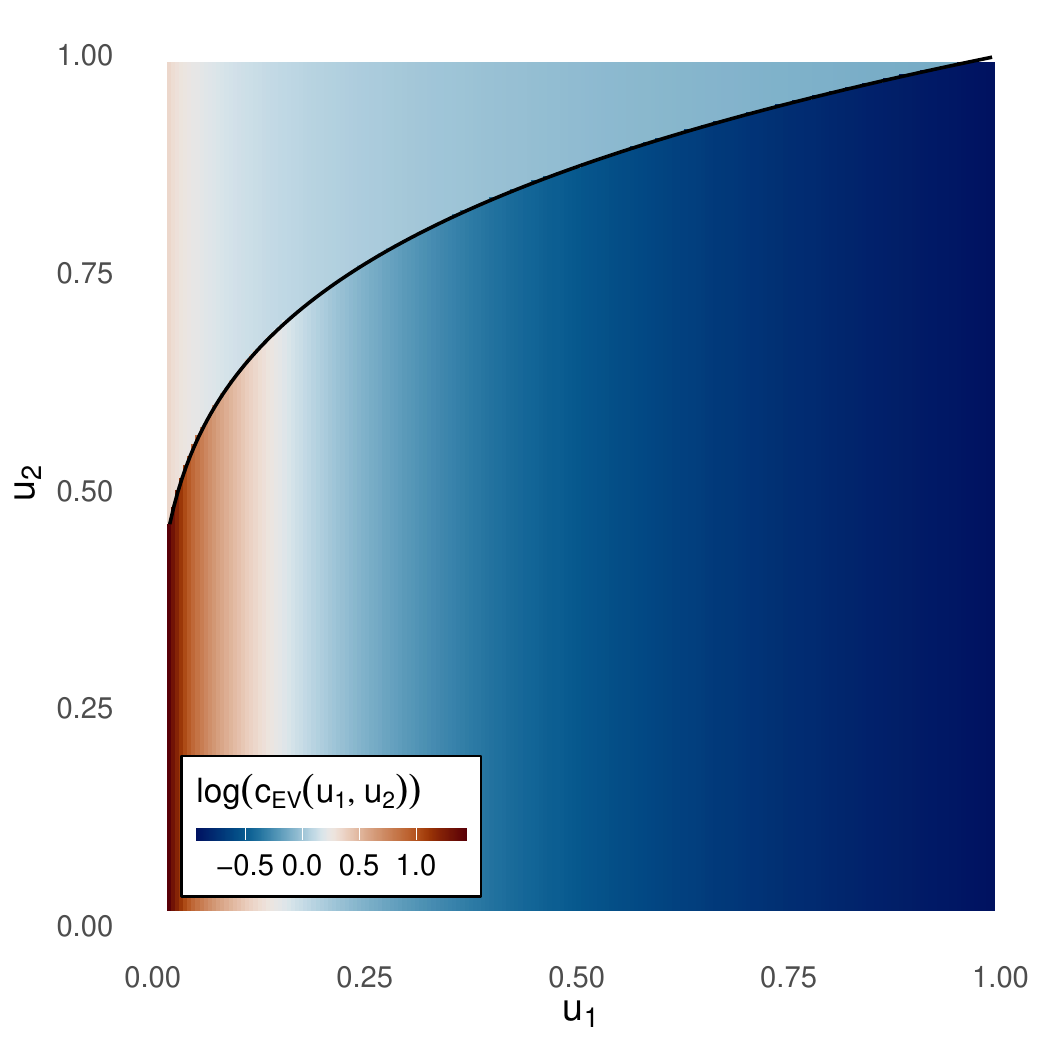}
    \includegraphics[width=0.32\linewidth]{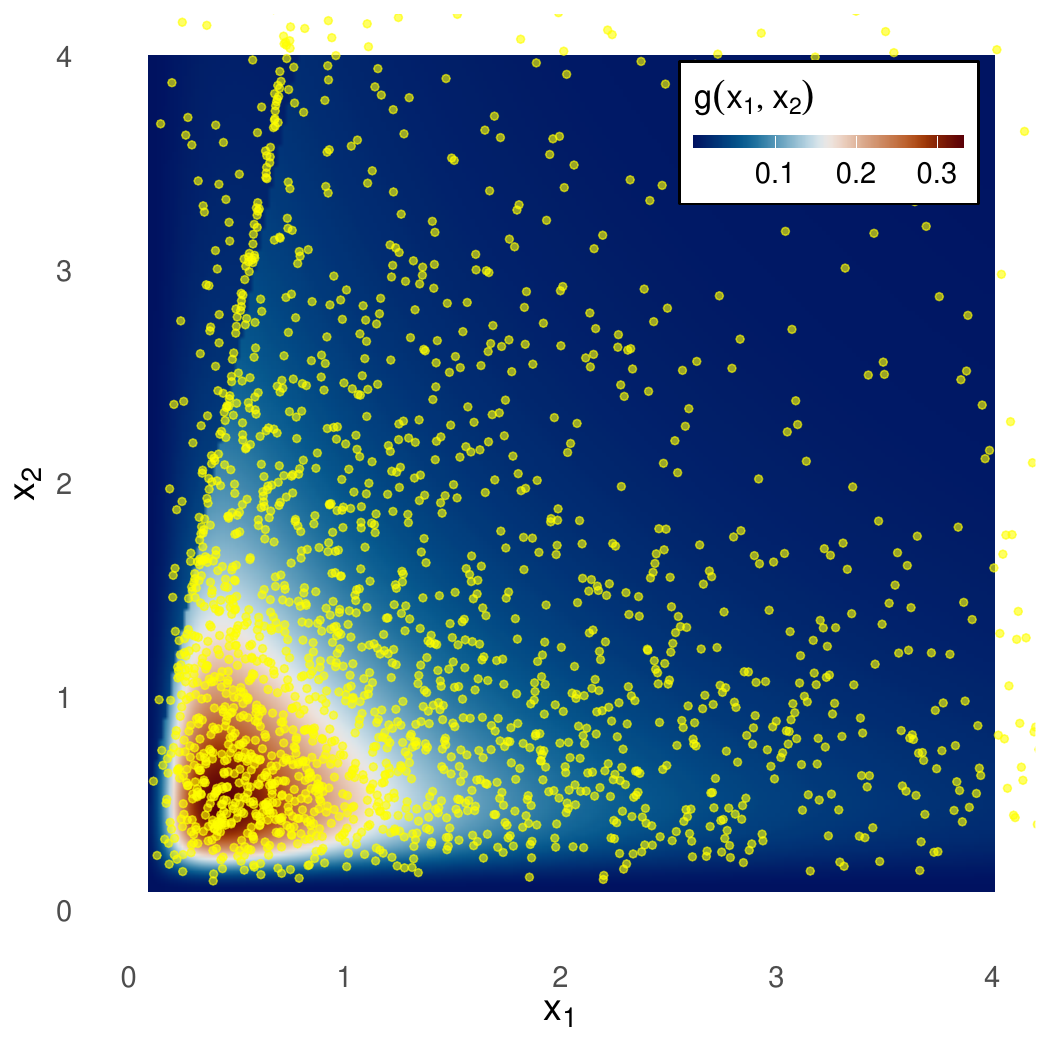}
    \includegraphics[width=0.32\linewidth]{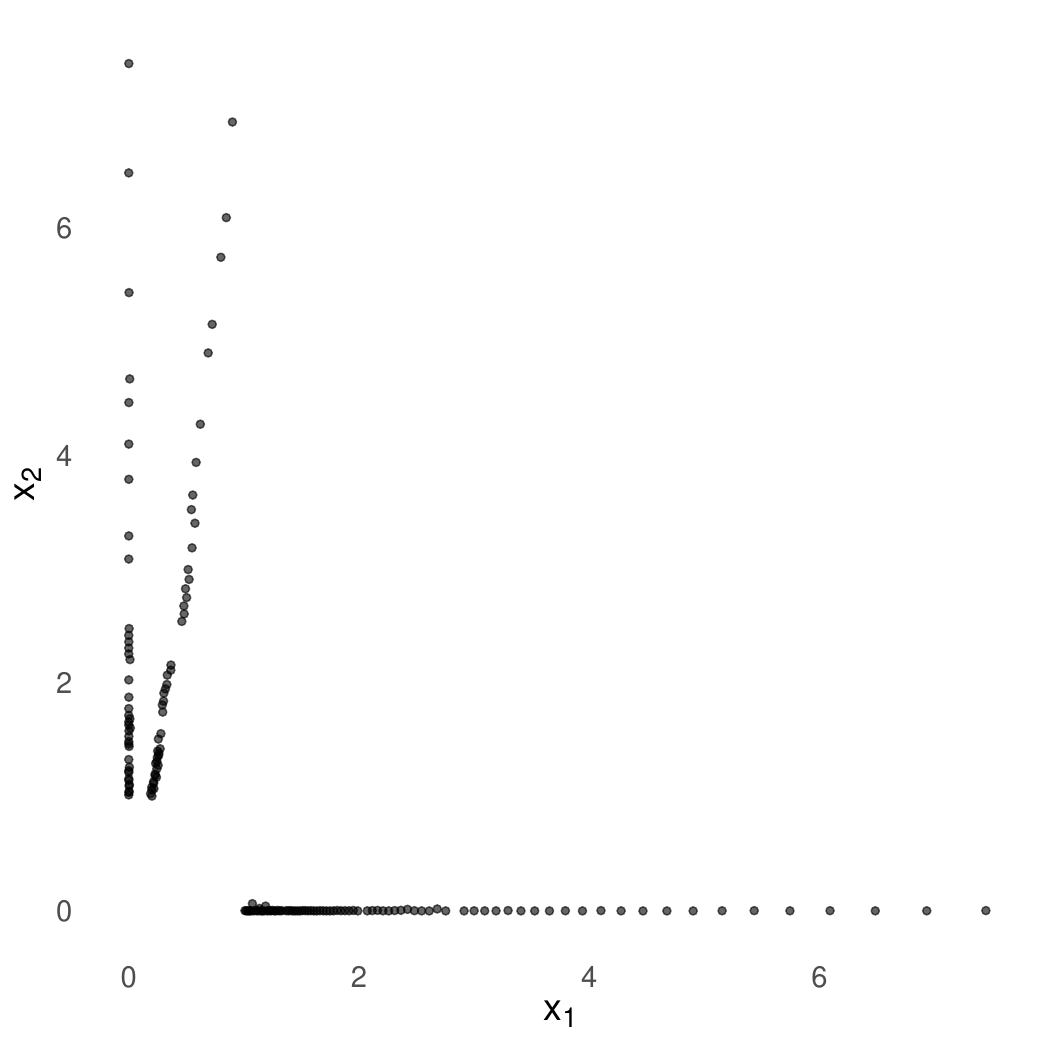} \\
    \includegraphics[width=0.32\linewidth]{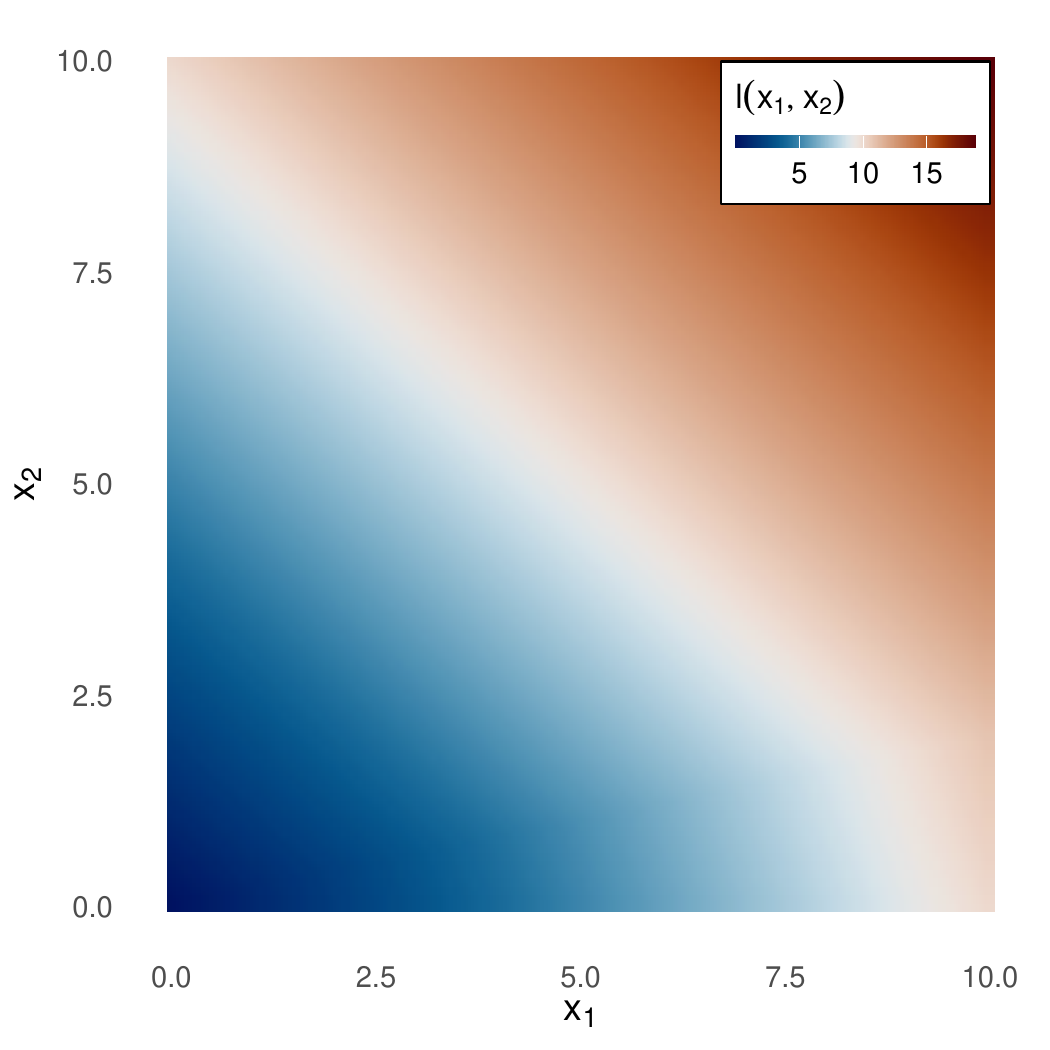}
    \includegraphics[width=0.32\linewidth]{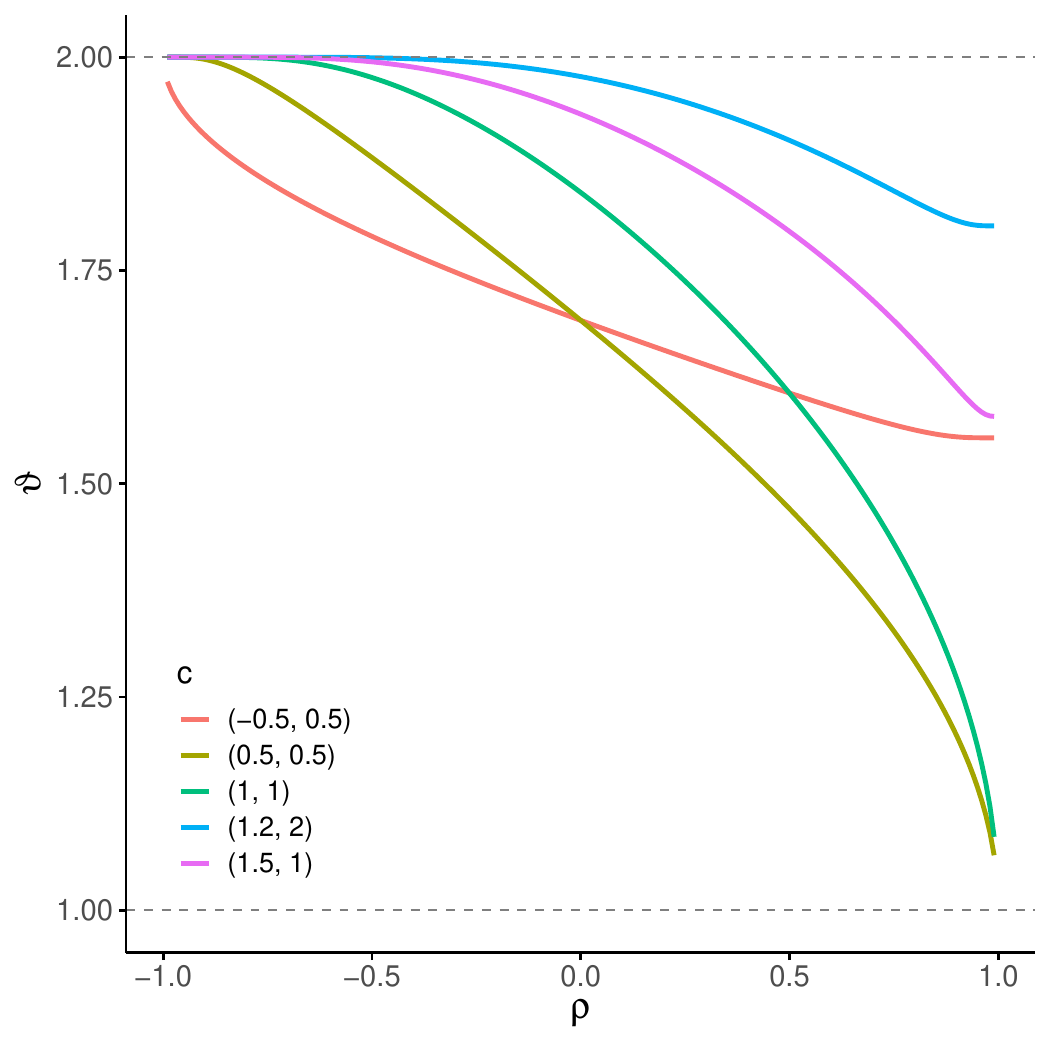}
    \includegraphics[width=0.32\linewidth]{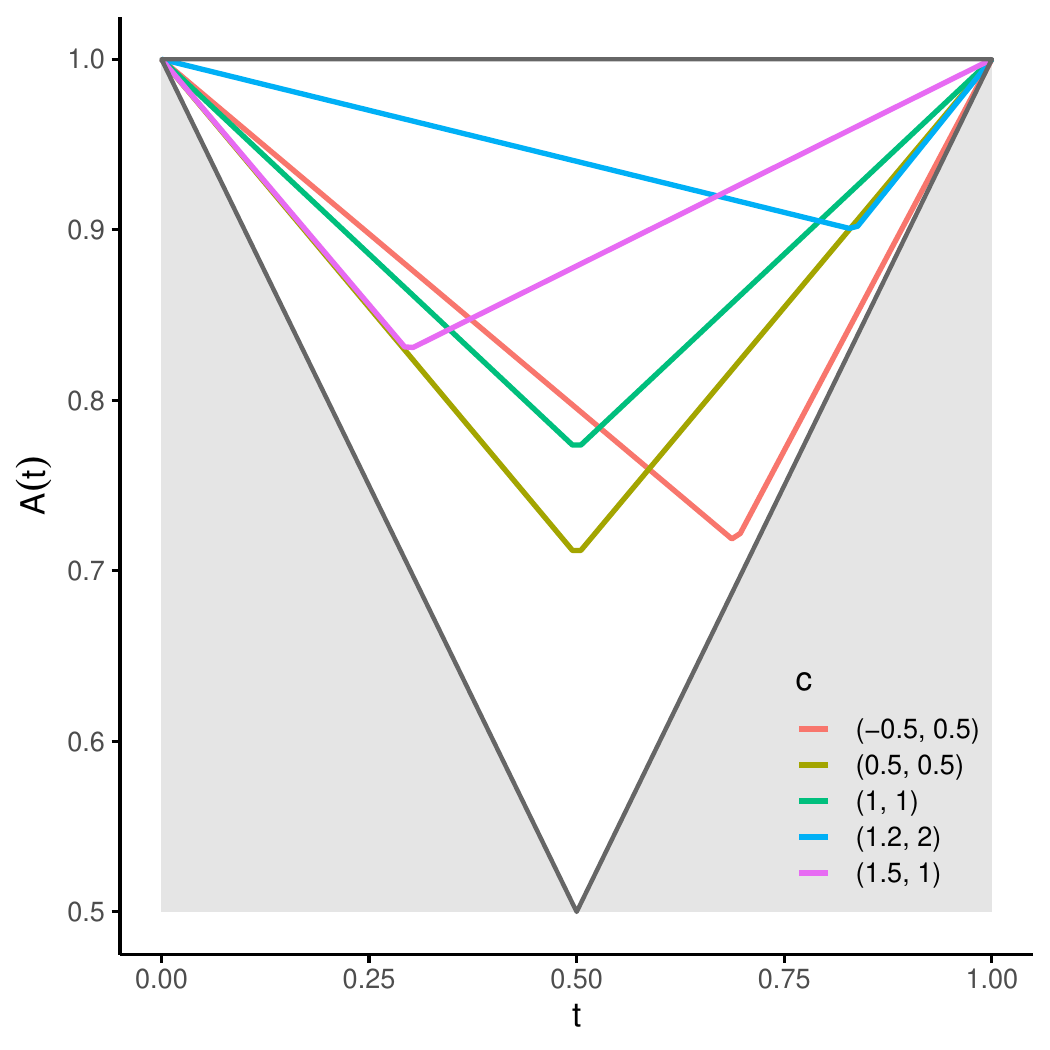}
    \caption{Left to right, top to bottom: density of the extreme value copula (on log scale, $\log c_{EV}$), density of the bivariate extreme value distribution $g(\bm x)$, exceedances above the $(1,1)$ threshold on unit Pareto margins, and the stable tail dependence function $\ell(\bm x)$ of the Marshall-Olkin model with $c = (1.2, 2)$ and correlation $\rho = 0.6$ between $Z_1^*$ and $Z_2^*$. The remaining panels display for several choices of $c$, the extremal coefficient $\vartheta$ as function of $\rho$ and the Pickands dependence function with $\rho = 0.6$.}
    \label{fig:summryMO}
\end{figure}

Consider the Marshall–Olkin model with $c = (1.2, 2)$ and correlation $\rho = 0.6$ between $Z_1^*$ and $Z_2^*$.
The top-left panel of Figure~\ref{fig:summryMO} illustrates the EV copula through its density $$
c_{\textrm{EV}}(u_1, u_2) = \begin{cases} (1-\alpha) u_1^\alpha, & u_1 > u_2^{\beta/\alpha}\\
(1-\beta) u_2^{\beta}, & \text{otherwise,}
\end{cases}
$$
with the black solid line representing the edge case $u_2 = u_1^{\alpha / \beta}$. The top-middle panel shows the multivariate EV density $g(\bm x) = \exp \{ -\ell(1/\bm x) \} (1-\alpha)(1-\beta) / (x_1 x_2)^2$, where the yellow dots correspond to 3,000 randomly generated component-wise maxima. In the top-right panel, the density of the bivariate generalized Pareto is intentionally left blank (except for some simulated points) because the model does not admit a closed-form density. 
The bottom-left panel depicts the stable tail dependence function, while the bottom-middle and bottom-right panels provide intuition about the type of dependence structure captured by this model through the extremal coefficient and the Pickands dependence function.

\newpage

\subsection{Illustrations of Submodel~2}
\label{appx-S2}

\begin{figure}[b!]
    \centering
    \includegraphics[width=0.32\linewidth]{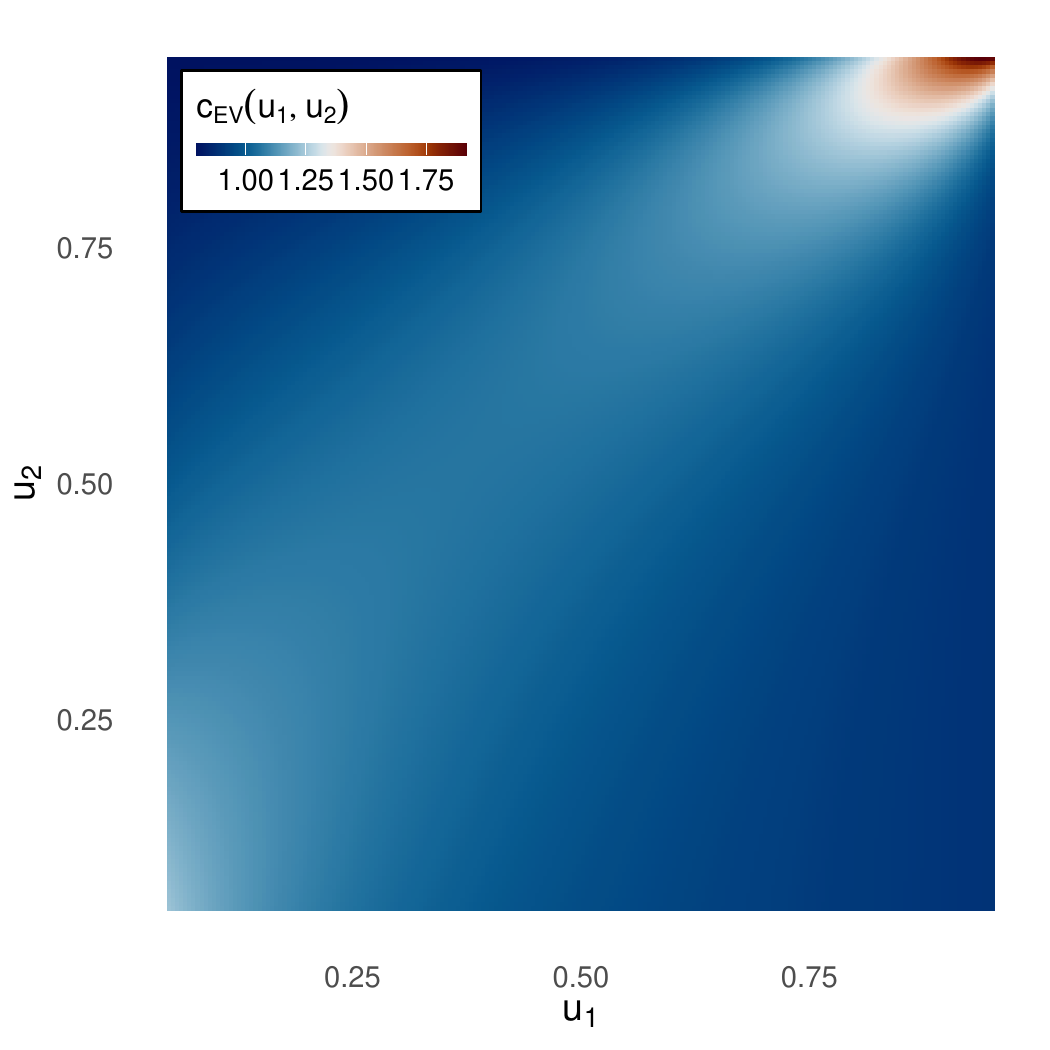}
    \includegraphics[width=0.32\linewidth]{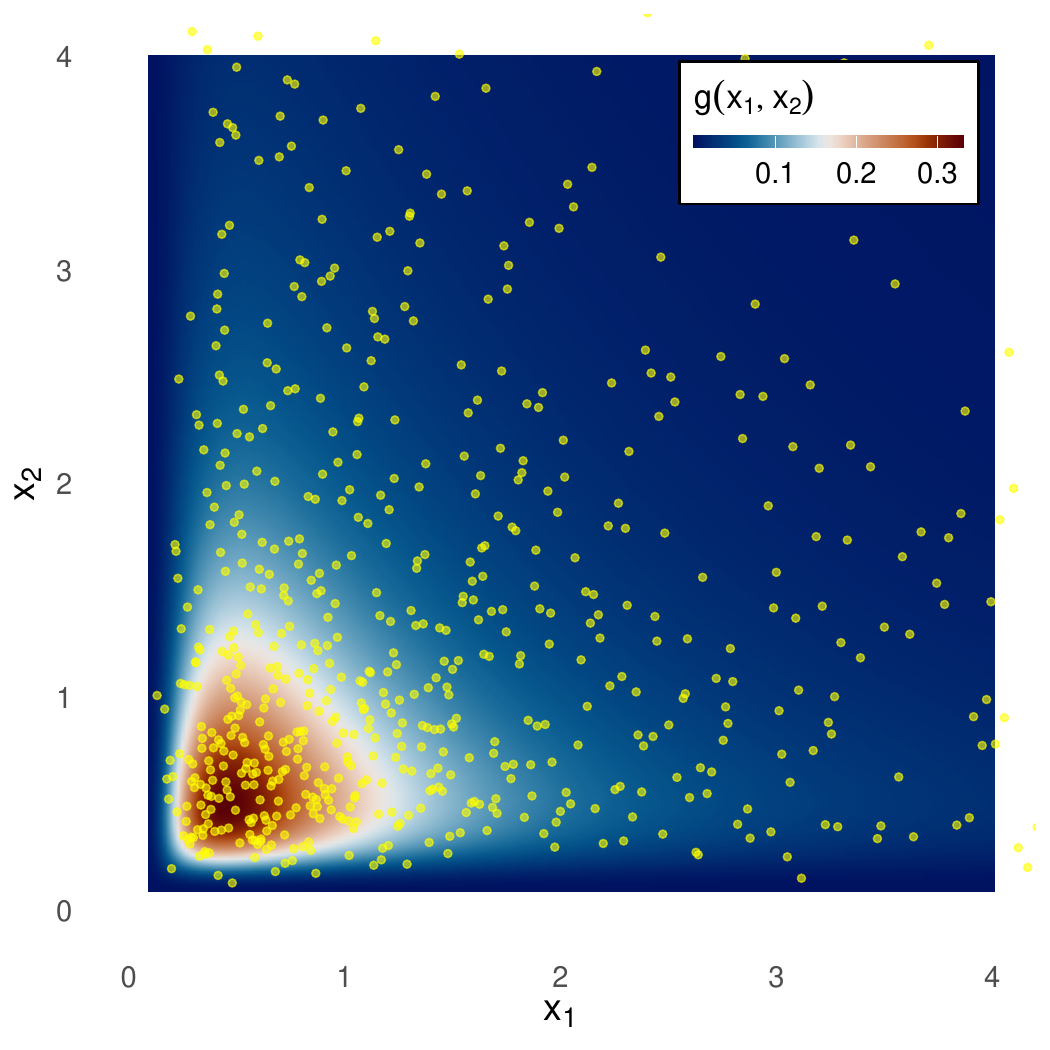}
    \includegraphics[width=0.32\linewidth]{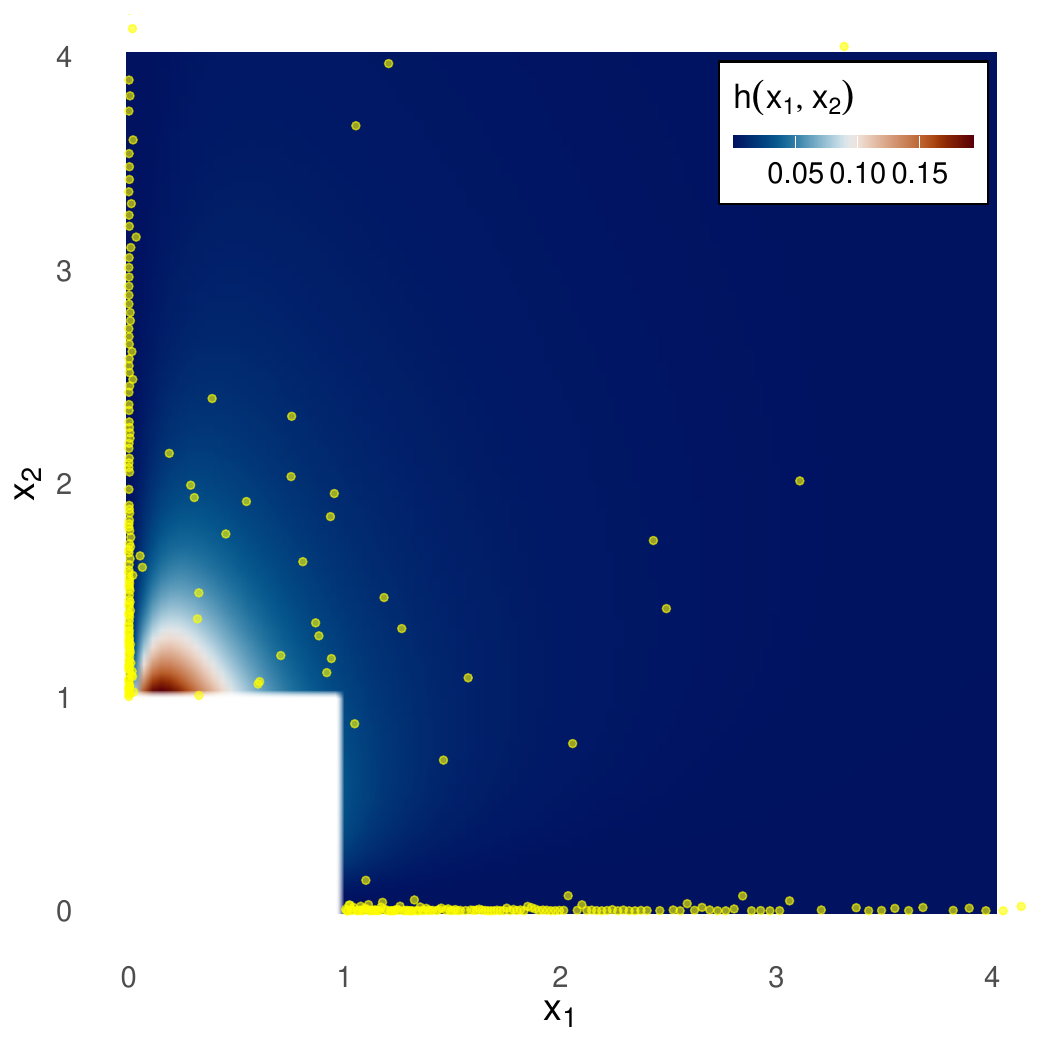} \\
    \includegraphics[width=0.32\linewidth]{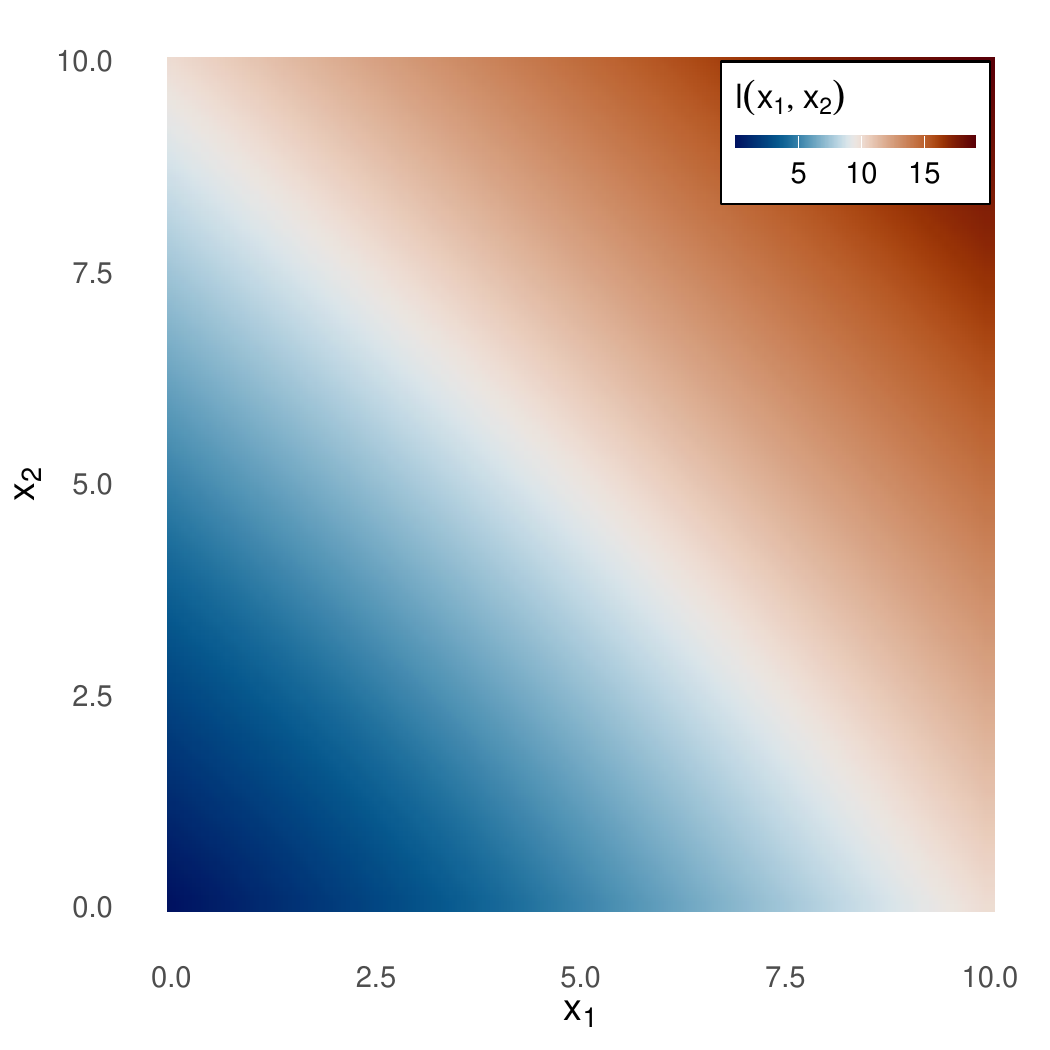}
    \includegraphics[width=0.32\linewidth]{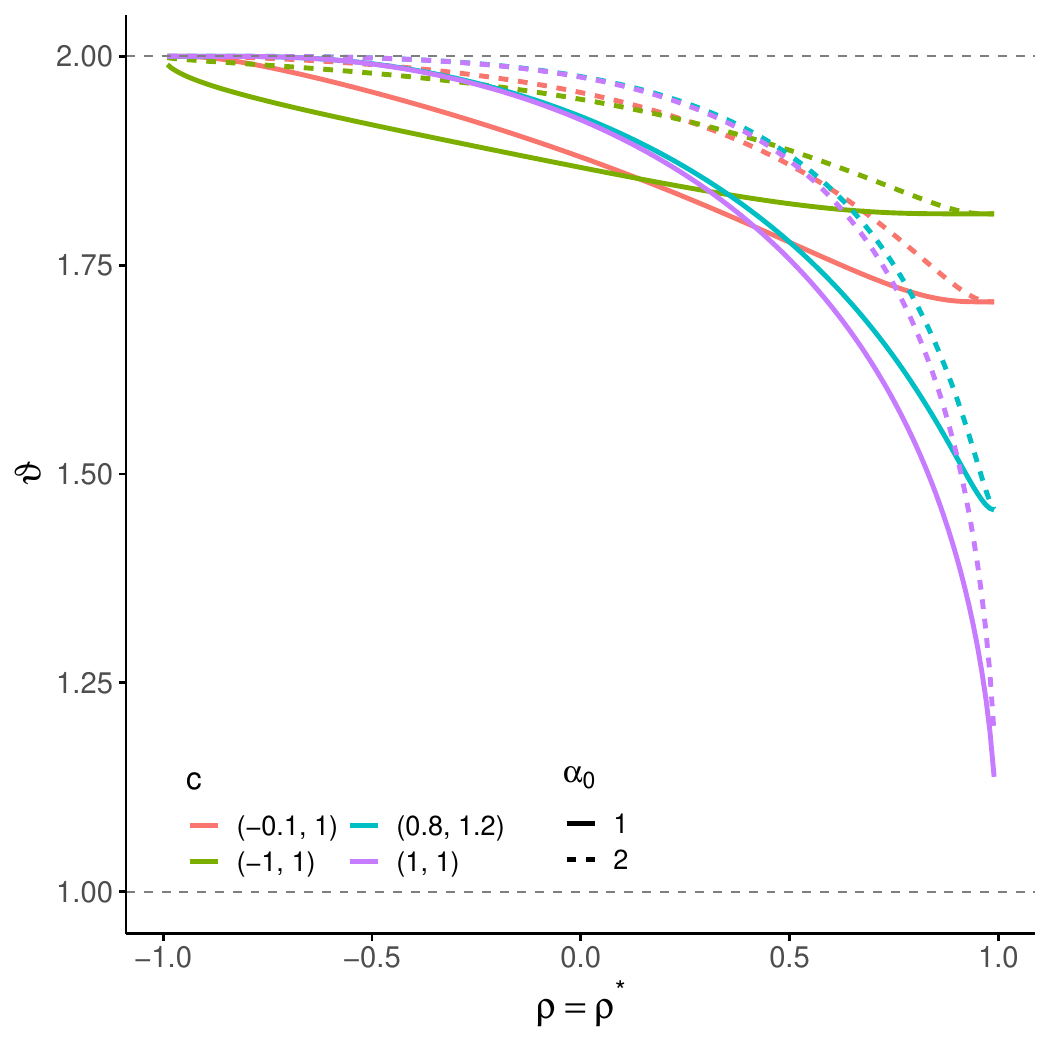}
    \includegraphics[width=0.32\linewidth]{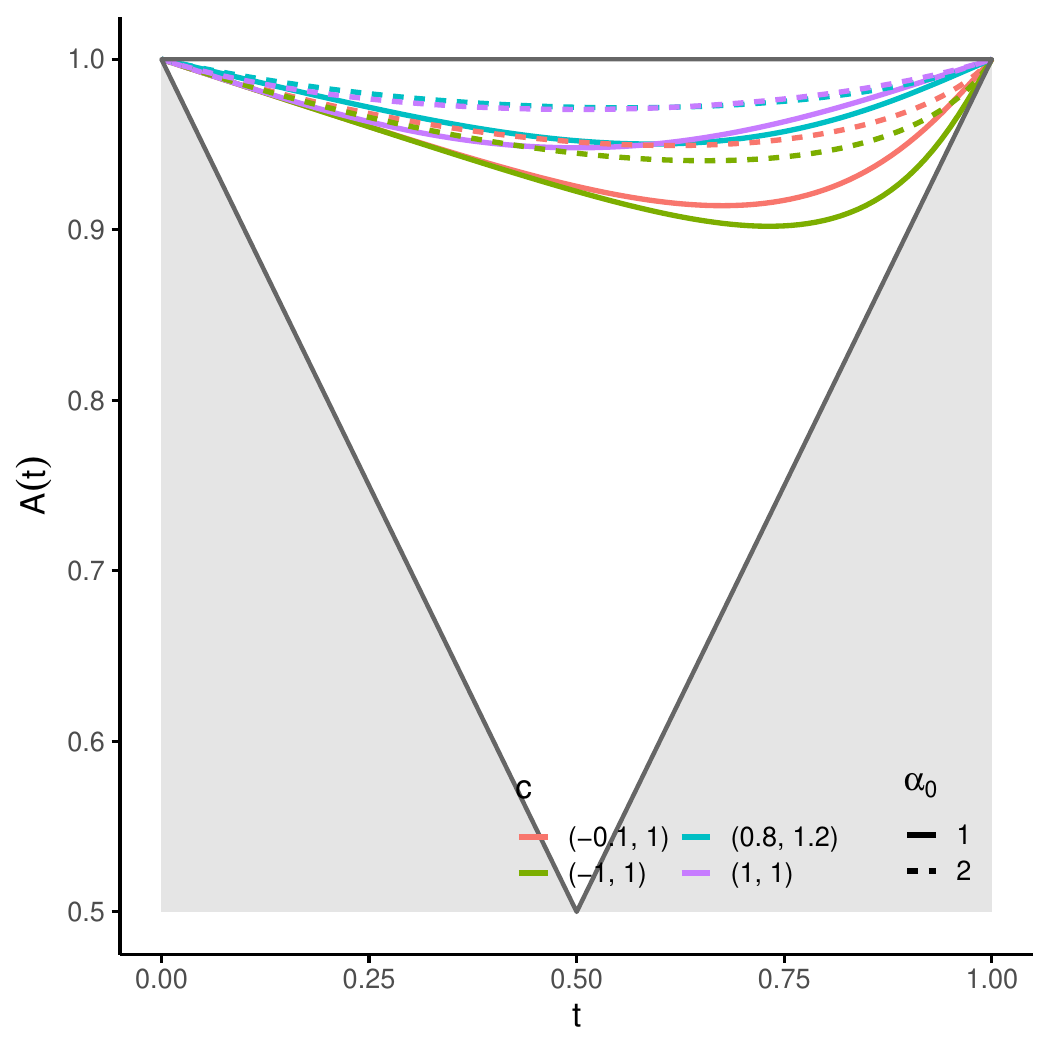}
    \caption{Left to right, top to bottom: density of the extreme value copula (on log scale, $\log c_{EV}$), density of the bivariate extreme value distribution $g(\bm x)$, exceedances above the $(1,1)$ threshold on unit Pareto margins, and the stable tail dependence function $\ell(\bm x)$ of the Submodel~2 with $\bm c = (0.8, 1.2)$, $\alpha_0 = 1$, correlation $\rho = \rho_{1,2} = 0.6$ and $\rho^* =0$. The remaining panels display for several choices of $c$, the extremal coefficient $\vartheta$ as function of $\rho (=\rho^*)$ and the Pickands dependence function with $\rho = 0.6$.}
    \label{fig:summryS2}
\end{figure}

Consider Submodel~2 with $\bm c= (0.8, 1.2)$, $\alpha_0=1$ and correlation $\rho_{1,2} = 0.6$. The key difference with the illustrations of Submodel~1 from Figure~\ref{fig:summryS1} is reduced from $\rho^* = 1$ to $\rho^* = 0$, which leads to a higher degree of independence among the extremes. The six panels of Figure~\ref{fig:summryS2} can be directly compared with those of Figure~\ref{fig:summryS1}. We note that the multivariate generalized Pareto distribution now assigns mass to both boundaries ($x_1=0$ and $x_2=0$) as displayed in the top-right panel.

\pagebreak
\bibliographystyle{model2-names}
\bibliography{fmcop}

\end{document}